\documentclass[10pt,aps,prb,amsfonts,amsmath,amssymb,raggedbottom,longbibliography,reprint,citeautoscript,bibnotes]{revtex4-2}
\usepackage[usenames,dvipsnames]{color}
\usepackage{graphicx}
\usepackage{microtype}
\usepackage{multirow}
\usepackage{longtable}
\usepackage{lmodern}
\usepackage[bookmarks=false,colorlinks]{hyperref}
\hypersetup{
    linkcolor=magenta,        
    citecolor=Plum,     
    filecolor=Plum,      	
    urlcolor=MidnightBlue,           
}

\newcommand{\Autoref}[1]{
  \begingroup%
  \def\sectionautorefname{Appendix}%
  \autoref{#1}%
  \endgroup%
}



\begin{document}

\title{Polar Metals Taxonomy for Materials Classification and Discovery} %

\author{Daniel Hickox-Young}
\email{hickoxyoung@roanoke.edu}
\altaffiliation[Present Address: ]{Department of Mathematics, Computer Science and Physics, Roanoke College, Salem, Virginia 24153, USA}
\affiliation{Department of Materials Science and Engineering, 
Northwestern University, Evanston, Illinois 60208, USA}

\author{Danilo Puggioni}
\email{danilo.puggioni@northwestern.edu }
\affiliation{Department of Materials Science and Engineering, 
Northwestern University, Evanston, Illinois 60208, USA}

\author{James M.\ Rondinelli}
\email{jrondinelli@northwestern.edu}
\affiliation{Department of Materials Science and Engineering, 
Northwestern University, Evanston, Illinois 60208, USA}

\begin{abstract}
Over the past decade, materials that combine broken inversion symmetry with metallic conductivity have gone from a thought experiment to one of the fastest growing research topics. In 2013, the observation of the first uncontested polar transition in a metal, LiOsO$_3$, inspired a surge of theoretical and experimental work on the subject, uncovering a host of materials which combine properties previously thought to be contraindicated [\href{https://doi.org/10.1038/nmat3754}{Nat.\ Mater.\ \textbf{12}, 1024 (2013)}]. As is often the case in a nascent field, the sudden rise in interest has been accompanied by diverse (and sometimes conflicting) terminology. Although ``ferroelectric-like" metals are well-defined in theory, i.e., materials that undergo a symmetry-lowering transition to a polar phase while exhibiting metallic electron transport, real materials find a myriad of ways to push the boundaries of this definition. Here, we review and explore the burgeoning polar metal frontier  from the perspectives of theory, simulation, and experiment while introducing a unified taxonomy. The framework allows one to describe, identify, and classify polar metals; we also use it to discuss some of the fundamental tensions between theory and models of reality inherent in the terms ``ferroelectric" and ``metals.''  In addition, we highlight shortcomings of electrostatic doping simulations in modeling different subclasses of polar metals, noting how the assumptions of this approach depart from experiment. We include a survey of known materials that combine polar symmetry with metallic conductivity, classified according to the mechanisms used to harmonize those two orders and their resulting properties. We conclude by describing opportunities for the discovery of novel polar metals by utilizing our taxonomy. 
\end{abstract}

\date{\today}

\maketitle

\section{Introduction}

The concept of crystalline metals without inversion symmetry, specifically those that lift parity symmetry 
to support a polar crystal structure, has been frequently attributed to a concise 1965 Letter by 
Blount and Anderson  titled ``Symmetry Considerations on Martensitic Transformations: `Ferroelectric' Metals''
\cite{Anderson/Blount:1965PRL}.
What is less well appreciated is that Blount and Anderson's article focuses on how a nominally first-order \emph{ferroelastic}
transition, observed at the time in the metallic  silicide V$_3$Si, could exhibit second-order character, presumably attributed to a displacive 
component akin to the paraelectric-to-ferroelectric transition found in insulating compounds, which exhibit a well-defined polarization 
below the critical temperature. 
At the time, nearly all martensitic (ferroelastic) transformations in metals 
exhibited strong first-order character \cite{salje:2012}; yet, V$_3$Si exhibited second-order behavior. 
The continuous response was rationalized by inferring that the symmetry-break should be \emph{like} that found in second-order 
displacive ferroelectrics, hence the quotes around ``ferroelectric'' in the title. Blount and Anderson did not suppose that ``ferroelectric" metals would possess a switchable polarization. Gauss's Law dictates that no electric field may exist within a metal \cite{de1773attraction,Gauss1877}, so the atomic structure of an ideal polar metal should be immune to perturbation via external applied electric field.

\begin{figure}
\centering
\includegraphics[width=0.49\textwidth]{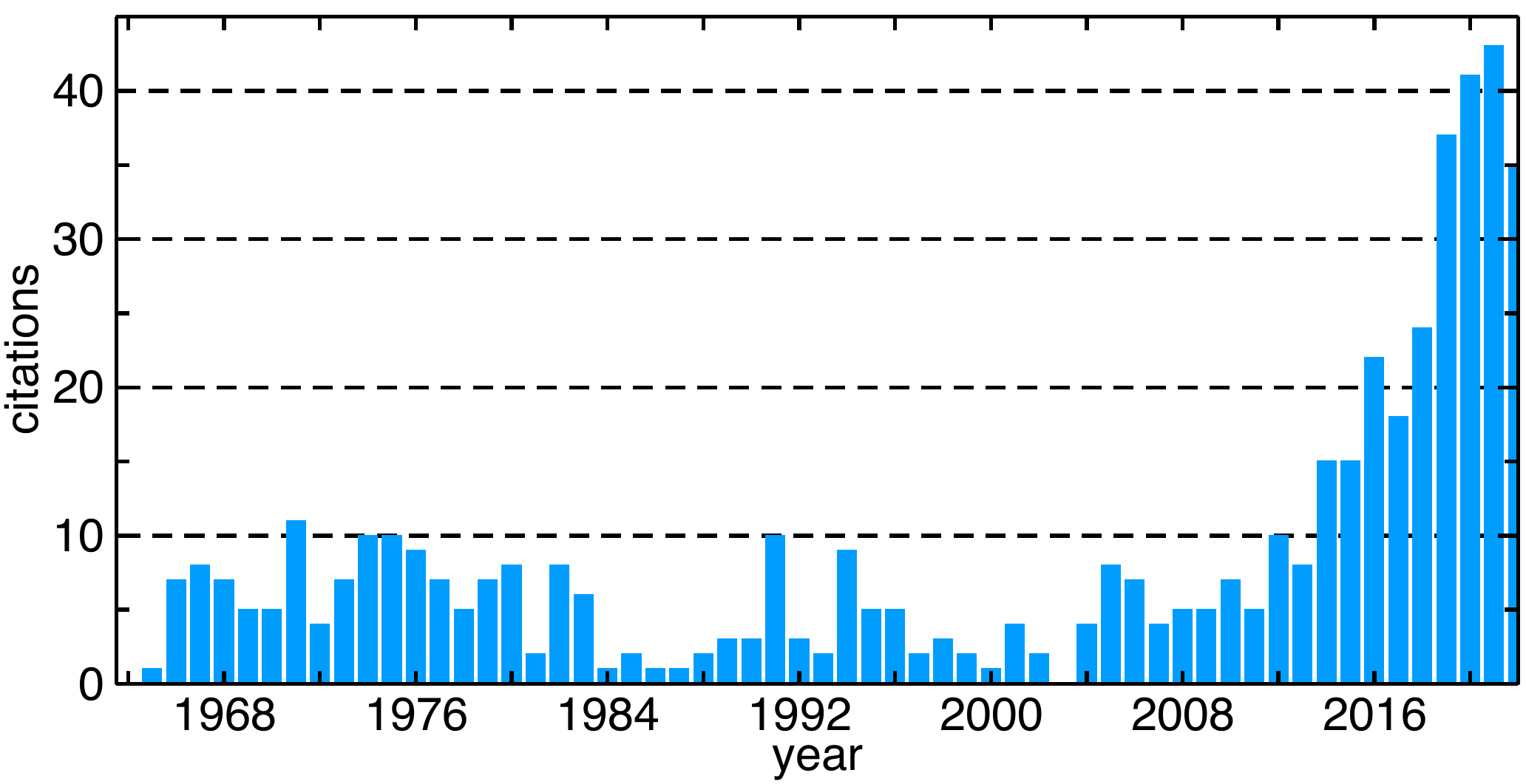}
\caption{Citations per year of Anderson and Blount's 1965 paper on second-order martensitic phase transitions \cite{Anderson/Blount:1965PRL}. Data collected from Google Scholar on October 10th, 2022.}
\label{fig:space}
\end{figure}

Ironically, for years this same physical law seemed to imply that polar metals should not exist at all. In prototypical ferroelectrics like BaTiO$_3$ it was shown that the polar displacement was stabilized by long-range dipole-dipole interactions \cite{Cochran1959,Cohen:1992Nature}. In the presence of free charge carriers, such interactions would be completely screened, favoring the centrosymmetric structure.
In the decades following Anderson and Blount's work, it appeared that Gauss's Law would prevail over the synthesis of a polar metal. There were several candidates in the 2000s which combined polar order and metallicity via compositional ordering, but it was not until 2013 that Shi et al.\ showed the metal LiOsO$_3$ exhibited a displacive transition \cite{Shi:2013Nature}. 
Their work demonstrated that a ``ferroelectric"-like transition need not rely on long-range interactions, but may derive from local structural effects (geometrically-driven Li displacements in the case of LiOsO$_3$).
Since the discovery of LiOsO$_3$ and following the introduction of a `weak-coupling hypothesis' by Puggioni and Rondinelli  outlining mechanisms by which to stabilize polar metals \cite{Puggioni/Rondinelli:2014NatComm},  they have garnered significant interest (\autoref{fig:space}). It was further accelerated by the 
discovery of nontrivial topological metals and their ensuing phenomena and properties \cite{Varjas:2016PRL,Wu2016,Gao2018}.
During the rapid increase in scholarship, diverse methods of combining these previously contraindicated properties have been proposed and executed, ranging from degenerately doped ferroelectrics \cite{He/Jin:2016PRB,ahadi2019enhancing, schooley1964superconductivity,Kolodiazhnyi:2010PRL} to metals with hybrid improper polar distortions \cite{Lei/Gu/Puggioni:2018NanoLett,Yoshida2005} to two-dimensional thin films and interfaces \cite{Fei2018,Zheng2021,Rettie2019}. 
Reviews of the classification of different design strategies for polar metals can be found in Refs.~\onlinecite{Benedek/Birol:2016JMMC,Zhou2020,Bhowal_Spaldin:2022}.

Accompanying these materials are a variety of terms. Polar, ferroelectric, ``ferroelectric,'' ferroelectric-like, and native ferroelectric are all qualifiers used to describe metallic systems with broken inversion symmetry. Furthermore, many so-called ``ferroelectric metals" either push the boundaries of what may be called `metallic' or do not exhibit a switchable polarization. 
Given the advances in dielectric, modern polarization, and soft-mode theories \cite{Baroni1987,Resta1993,King-Smith1993,vanderbilt2018berry,Cohen1990}, we find that 
the ferroelectric-like designation and its derivatives \cite{Gu/Jin:2017PRB, Xiang:2014PRB, Jin2016,Filippetti2016,Fei2018} are 
cumbersome and nonessential descriptions, obfuscating the  physics displayed by very different (and yet, equally interesting) classes of materials.
In addition, we note that both experimental and computational approaches to studying these materials have at times been abused. Computationally, the background-charge approach to electrostatic doping simulations fails to model reality in multiple underappreciated ways \cite{Bruneval2015,Iwazaki2012}. Meanwhile, in experiment, the application of ferroelectric characterization techniques to materials that are not formally ferroelectrics both  
erodes the key differences separating metals from dielectrics and can lead to misinterpretations of 
measured dielectric polarizations, e.g., electric polarization hysteresis \cite{Gu/Jin:2017PRB,Scott2008}.
In this work, we propose using meaningful atomic and electronic structure descriptions to distinguish materials based on conductivity and symmetry considerations. 
We begin by discussing the tensions between theory and experiment, as exemplified by both the various methods for combining polar order and metallicity and the shortcomings of our current terminology.
These issues are further amplified in computational studies using electronic structure methods, e.g., with the popular background-charge approach to electrostatic doping simulations, which can lead to model results inconsistent with experiment.
We then present a survey of polar metals classified using our taxonomy that combines conductivity with broken inversion using terminology based on clearly defined class descriptors.

\section{Tensions in terminology: ``ferroelectric" ``metals"}

Between the realms of theory and experiment, communication is critical in order to foster a productive relationship. Since communication is built on having a common vocabulary, this makes the terminology we use of utmost importance in seeking to advance the current understanding of our field and not just a pedantic dilemma. In the field of polar metals, however, some classifiers that seem obvious in theory are less well-defined in experiment and the result has been confusing and at times unintentionally misleading. We present a few examples of the two most common sources of dissonance between theoretical and experimental labels---namely, ``metallic" and ``ferroelectric"---and suggest methods for relieving the  tension. 

\subsection{Ferroelectric ``metals''}
\textit{How to define metallicity?}---The question of how one defines a `metal' carries significance for many disciplines, but the distinction bears considerable weight when evaluating doped ferroelectrics (FEs), especially as a large number of reports describe doping known FEs as a route to achieve polar or ``ferroelectric" metals. At what point does a doped ferroelectric become a polar metal? Is there an important fundamental difference between a polar structure with intrinsic charge carriers or extrinsic charge carriers? To answer these questions, let us consider fundamental definitions of metallicity.

According to Kohn's theory, while the wave function of a metal in its ground state is delocalized, that of an insulator is localized \cite{Kohn1964}. This distinction strictly defines the difference between metals and insulators.
Electron transport considerations can, in principle, be used to quite clearly separate metals from insulators (dielectrics).
According to Mott, ``\ldots a metal conducts, and a non-metal doesn't'' \cite{Edwards2010,davis1998nevill}. This statement is strictly true at $T=0$\,K and is often 
used as the discriminating factor between a metal and an insulator (or semiconductor) such that:
\begin{equation}
\lim_{T \rightarrow 0} \rho(T)  = 
  \begin{cases}
  	\infty & \textrm{insulator} \\
	\rho_0 & \text{metal}
\end{cases}\,
\end{equation}
where $\rho$ corresponds to the electric resistivity of the material and $\rho_0$ is the residual resistivity due to electron collisions with crystal impurities and imperfections at $T = 0$\,K. 
Importantly, the value of $\rho$ at room-temperature does not matter; thus, although the definition is mathematically well defined, experimental conditions ($T\neq0$\,K) can make the differentiation challenging, especially when the carrier density of a semiconductor is sufficiently high at room temperature. 
These may be intrinsic charge carriers 
or the result of degenerate doping, such that the material is  conductive through extrinsic doping \cite{Takahashi2017, Fujioka2015, Kolodiazhnyi:2010PRL, Kolodiazhnyi2008}. 
These compounds are polar, and for reasonably large temperature ranges there is a positive correlation between resistivity and temperature, even though (strictly speaking) the correlation between resistivity and temperature becomes negative at low temperature. 
This fact may be why the boundary between doped polar semiconductors and polar metals is frequently blurred by experiment, despite the clear theoretical distinction    \cite{Kohn1964,Scalapino/White/Zhang1993,Resta2002}.

\textit{Band Theory}---The distinction between a trivial insulator and a metal can be understood using band theory. For an insulator, the Fermi level is not well-defined, but resides within an energy gap. Therefore, at 0\,K, there are no free charge carriers available and $\rho \rightarrow \infty$. With increasing temperature, the electrons may thermally populate the conduction band and  become available for conduction such that the  resistivity decreases ($d\rho/dT<0$).  
In metals, the Fermi level is located within a band, giving rise to free charge carriers. As temperature increases, so does resistivity ($d\rho/dT>0$). At high temperatures ($T\gg\Theta_D$, where $\Theta_D$ is the Debye temperature), $\rho\sim T$. At low temperatures ($T\ll\Theta_D$) and $\rho$ has four contributions \cite{Ashcroft/Mermin:Book}: 
\[
\rho(T) = \rho_0+AT^2+BT^5+CT^f\exp\left({-\frac{\hbar\omega_{min}}{k_BT}}\right).
\]
The $T^2$ contribution originates from electron-electron scattering. The $T^5$ contribution was evaluated by Bloch and Gr\"uniessen \cite{Bloch1930, Grun1933} and is due to
electron-phonon scattering. The exponential term describes the contribution of electron-phonon umklapp scattering, where $\omega_{min}$ is the minimum phonon frequency below which umklapp process are forbidden and $f$ is an empirical parameter.
To be considered genuinely metallic, polar metals should exhibit a positive correlation between resistivity and temperature for all $T>0$\,K. This is exactly what happens for LiOsO$_3$, which shows Fermi liquid-like behavior ($\rho\sim T^2$) in its polar phase \cite{Shi:2013Nature}. 
With this in mind we can answer the question: At what point does a doped semiconductor become metallic? 
The threshold to achieve metallic conductivity is 
determined by the carrier concentration
required to make $d\rho/dT>0$. 

The ratio $d\rho/dT$ is useful  for differentiating metals from insulators. However, we note that metallic and insulating phases are limiting situations. In transition metal oxides and other strongly correlated systems $\rho(T)$ exhibits a complex behavior. For instance, the polar oxide Ca$_3$Ru$_2$O$_7$ exhibits metallic conductivity above 48\,K, insulating behavior from 48 to 30\,K, and shows again metallic transport below 30\,K \cite{PhysRevB.69.220411}. Even doped ferroelectrics exhibit complex transport properties. In La-doped BaTiO$_3$, $\rho(T)$ shows insulating behavior from 350 to 260\,K, metallic transport between 260 and 70\,K, and again insulating behavior  with weak localization below 70\,K \cite{Fujioka2015}. Therefore, the assessment of metal/non-metal status using $d\rho/dT$ alone is not an easy task and should be performed in conjunction with other descriptors.

\begin{figure}
  \centering
  \hspace*{.6cm}\vspace*{.3cm}\includegraphics[width=0.4\textwidth]{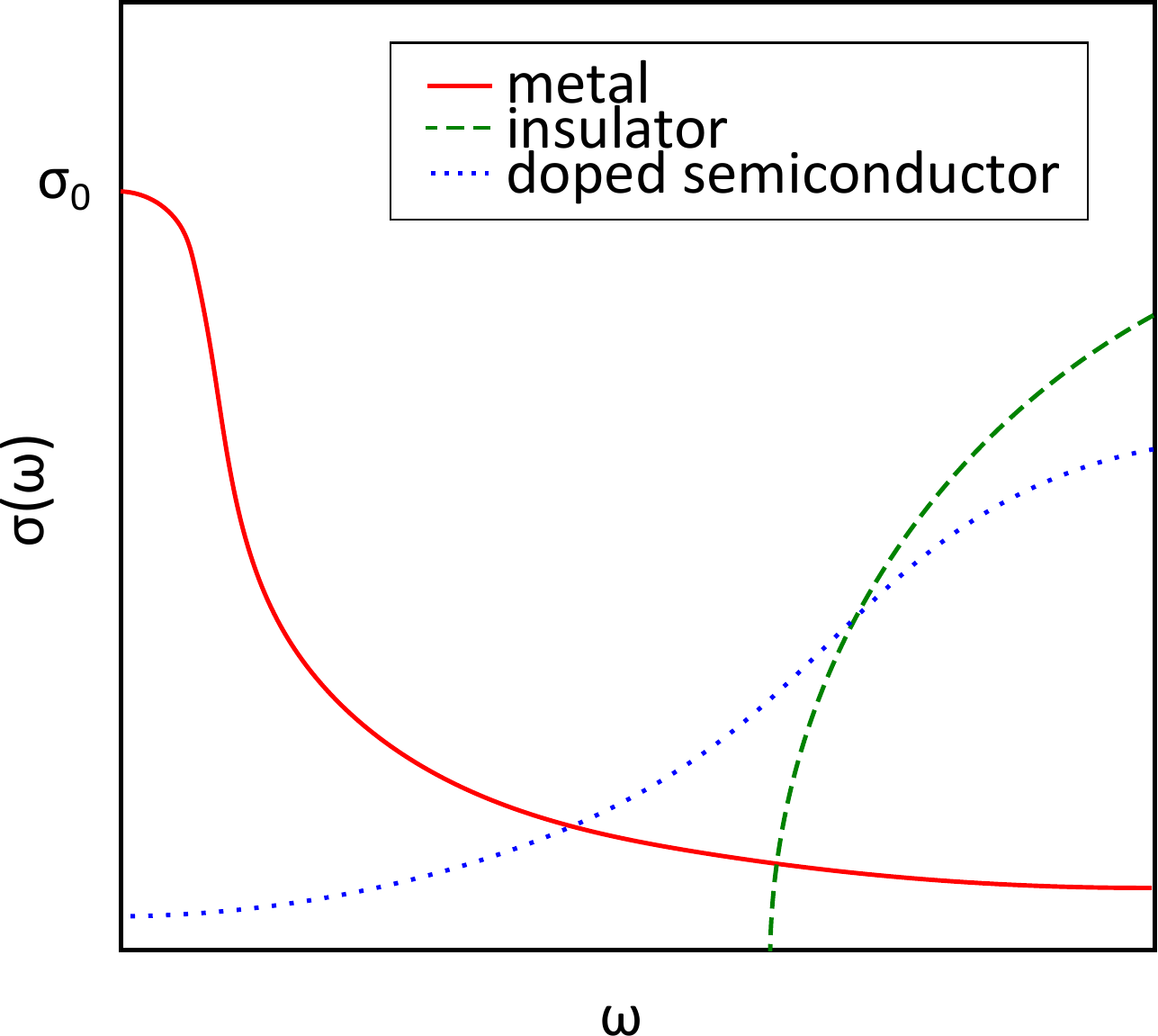}
  \qquad
  \includegraphics[width=0.45\textwidth]{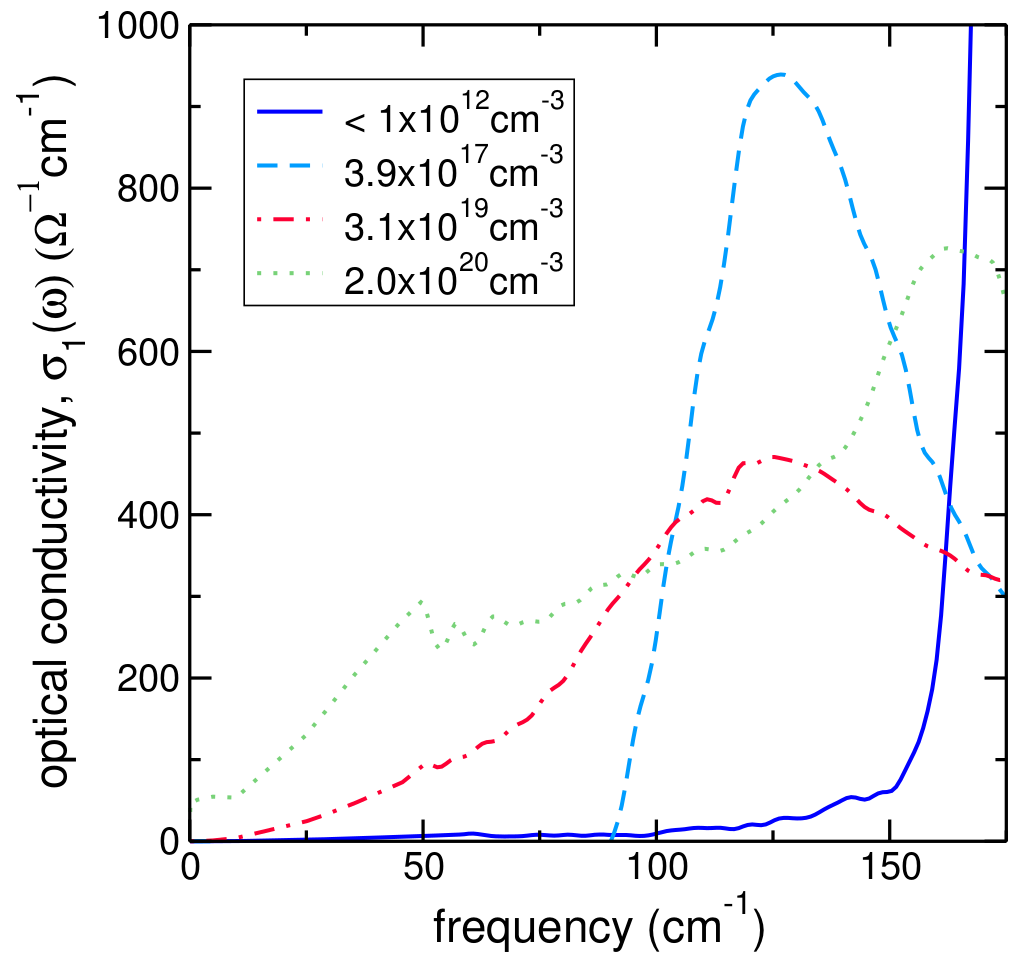}
  \caption{(top) Schematic illustration of the difference in optical conductivity response across materials classes. (bottom) Experimental data from Ref.\ \onlinecite{Hwang2010} illustrating the evolution of the optical conductivity of BaTiO$_{3-\delta}$ under oxygen vacancy doping at $T \approx 30$\,K.}
  \label{fig:sigma}
\end{figure}

\textit{Drude Definition}---Although the DC conductivity would appear to be the natural way to separate metals from insulators, 
in practice assessing the frequency dependent free-carrier response of a material allows one to treat 
metals, doped-semiconductors, and insulators on more equal footing \cite{Millis1991}.
For a perfect crystalline material with non-interacting electrons, the optical conductivity $\sigma$ relates the electrical current 
density due to a spatially uniform transverse electric field as
\[
\sigma(\omega)= \frac{ne^2 \tau}{m (1-i\omega \tau)}\,,
\]
where $m$ is the mass of the carrier (free electron mass or band renormalized mass) and $n$ is 
the density of carriers (electrons or holes). 
\autoref{fig:sigma}a shows that for an ideal metal the dc conductivity $\sigma_0=ne^2\tau/m$ 
appears as a local maximum of $\sigma (\omega)$ at zero frequency (Drude peak) and then decays with a Lorentzian form due to 
finite relaxation time $\tau$.
In contrast, $\sigma(\omega)=0$ for $0\le\omega\le E_g$ in insulators with an optical gap $E_g$. This Drude definition would categorize some doped FEs with very modest room temperature conductivity as fundamentally closer to insulators, providing some clarity. In addition, the nature of the optical conductivity may vary with temperature \cite{Homes1993}, allowing the metal-insulator classification to change with temperature---thereby accounting for metal-insulator transitions.
However, sufficient doping of a ferroelectric can still lead to optical conductivities that remain nonzero as $\omega\rightarrow0$, but do not reach a local maximum at $\omega=0$ (\autoref{fig:sigma}b), which presents a somewhat ambiguous case. 

Another low-frequency dynamical property,  which is of relevance to distinguishing between metals and insulators, is the non-adiabetic Born effective charge (naBEC) \cite{dreyer2021nonadiabatic}. Whereas Born effective charges are typically only well-defined in insulators (as measurements of the changes in dielectric polarization as a function of atomic displacement), naBECs are measurable in metals by considering the current generated in response to atomic motion. 
Atomic motion in this case is produced optically in a regime such that $\omega$ is much greater than the inverse carrier lifetime ($1/\tau$) while still being much smaller than interband resonances. Effectively, i.e.,   for materials with long carrier lifetimes, this amounts to $\omega\rightarrow0$ and the naBECs stabilize as the ``Drude weight" or the density of free charge carriers available for conduction. As with optical conductivity, $\omega\rightarrow0$ in insulators but reaches a non-zero value in metals. The naBECs, as an analog to BECs, are also advantageous as they allow for characterization of polarizability in metals despite polarization itself not being well-defined.

\begin{table}
\centering
\begin{ruledtabular}
\caption{Plasma frequency and resistivity magnitudes as measured at room temperature for band metals (top) and doped semiconductors (bottom).}
\begin{tabular}{lll} 
 Material & $\omega_p (\mbox{cm}^{-1})$  & Resistivity (10\,$^{-6} \Omega$-m) \\ 
 \hline
 Al & 1.19 $\times10^5$ & 2.82$\times10^{-2}$ \\
 Cu & 6.38 $\times10^4$ & 1.70$\times10^{-2}$ \\
 Au & 7.25 $\times10^4$ & 2.44$\times10^{-2}$ \\
 Pb & 6.20 $\times10^4$ & 2.20$\times10^{-1}$ \\
 Ag & 7.25 $\times10^4$ & 1.59$\times10^{-2}$ \\
 \hline
 n-GaAs & 4.94 $\times10^2$ & $\sim10^{3}$ \\
 n-Si & 1.76 $\times10^3$ & 10.7 \\
 p-Si & 2.28 $\times10^3$ & 6.4 \\
 n-InSb & $\sim2.1\times10^2$ & $10^3$--$10^4$ \\
\end{tabular}
\label{tab:wp}
\end{ruledtabular}
\end{table}

\textit{Order of Magnitude}---Although for low-frequencies the optical properties of (doped) semiconductors are qualitatively similar ($\sigma\ne0$), they are quantitatively different, because of the difference in carrier masses and densities. 
At high-frequencies, semiconductors and metals both absorb -- as expected for an insulator with available conduction band 
states -- owing to interband processes that give rise to finite $\sigma(\omega)$. 
The frequency crossover at which the behaviors change is given by the plasma frequency $\omega_p$, corresponding to a zero in 
the real part of the dielectric function. Neglecting any damping effects, the plasma frequency can be expressed as 
$\omega^2_p = {ne^2}/{\varepsilon\epsilon_0m},$ 
where $\varepsilon=\epsilon_\infty$ for a metal in that it includes only electronic contributions from (high energy) interband 
transitions while an insulator includes both electronic and ionic (static) polarization contributions, i.e.,  
$\varepsilon=\epsilon_\infty + \epsilon_\mathrm{ionic}$ and $\epsilon_\mathrm{ionic}>\epsilon_\infty$.
As the carrier density increases, the plasma frequency increases. In conventional 
metals $\omega_p$ is in the UV region which gives rise to the UV reflectivity edge (light of frequency $\omega< \omega_p$ is reflected). 
(In practice, this edge can be difficult to assess experimentally due to 
interband transitions and in some cases can be found just below visible frequencies.) 
By contrast, the carrier density in doped semiconductors places  $\omega_p$ in 
the 100s of meV (5-30\,$\mu$m range) as in $n$-InSb (\autoref{tab:wp}) and is highly 
tunable \cite{Anderson1971}.

\begin{figure}
    \centering
    \includegraphics[width=0.5\textwidth]{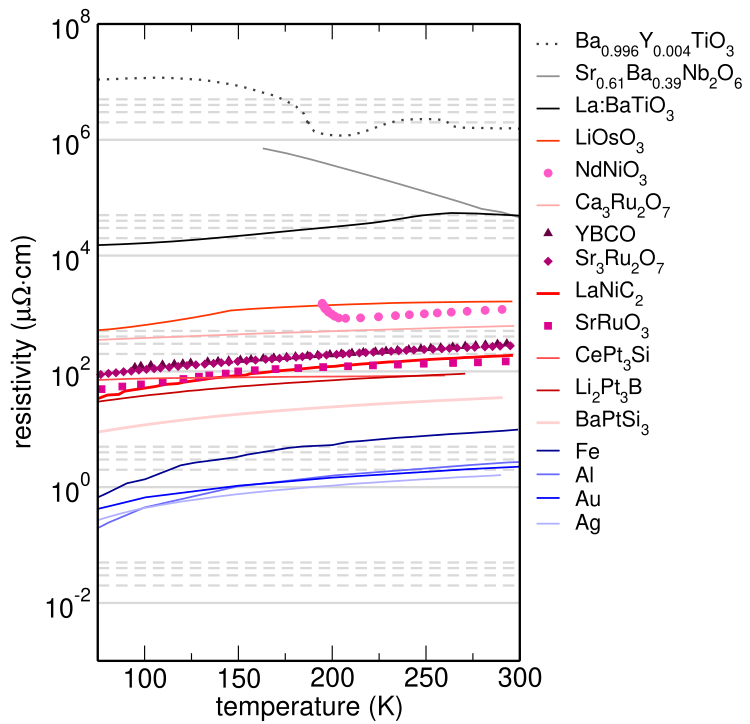}    \caption{Temperature-dependent resistivities of a variety of metals, polar metals, and doped semiconductor. Although the polar metals exhibit electron transport roughly 1-2 orders of magnitude more resistive than typical elemental metals, they are still easily identifiable as significantly more conductive than the doped semiconductors (despite a positive slope in resistivity for La:BaTiO$_3$ throughout the temperature range).}
    \label{fig:transport}
\end{figure}

\autoref{tab:wp} also illustrates the order of magnitude gap in resistivity between doped semiconductors and band metals (as does \autoref{fig:transport}). 
The order of magnitude is a meaningful descriptor to aid classification as it similarly is a useful materials selection when deploying materials under application constraints. This distinction should be used when comparing polar metals and doped FEs. 
However, it is worth noting that some materials would be mislabelled if one uses the order of magnitude of the plasma frequency or resistivity alone. 
Doped SrTiO$_3$ exhibits extremely low resistivity (including a superconducting transition) but the plasma frequency, even at low temperature, is comparable to other doped semiconductors  ($\sim1.5\times10^3$cm$^{-1}$)\cite{Bi2006}. 
Meanwhile, many polar metals  exhibit relatively poor conductivity (\autoref{fig:transport}), belonging to the so-called class of ``bad metals."  LiOsO$_3$, for example, has a room temperature resistivity of $15\times10^{-6}\,\Omega$-m and a plasma frequency on the order of $\sim10^2$\,cm$^{-1}$) \cite{Shi:2013Nature,LoVecchio2016}. 
Therefore, as with all other descriptors discussed thus far, order of magnitude is best used in combination with other criteria when assessing conductivity.

\textit{Doping Sensitivity Analysis}---Although the descriptors above are generally sufficient to clearly differentiate between polar metals and doped FEs, greater clarity may be achieved by considering doping as a perturbation to the initial state of a material and evaluating whether that perturbation has been sufficient to change the material's classification. 
We consider the sensitivity of the electronic and crystallographic structure with respect to the perturbation. The effect of doping on electronic structure is direct and immediately distinguishes intrinsic conductivity from extrinsic conductivity. Doping shifts the Fermi level, which in most metals (i.e., excluding semi-metals) has little effect on the effective mass or concentration of the free charge carriers. In FE insulators, on the other hand, doping has an immediate impact, shifting the Fermi level toward a band edge and often inducing defect states, thereby altering the conduction mechanism. This distinction has practical considerations; the transport properties of doped FEs will be more sensitive to changes in the electron chemical potential than those of polar metals. 

At sufficiently high concentrations, dopant atoms form a partially occupied impurity band which may exhibit metallic conductivity---so-called degenerately doped semiconductors. In this regime, the conductivity of the material is less sensitive to small variations in the concentration of impurity atoms than a traditional doped semiconductor. However, the system should still be considered a perturbation from the pristine state of the semiconductor and can be distinguished from a band metal by both the relatively smaller carrier concentration and the proximity of the Fermi level to a band gap. These distinctions should be clear from electron transport and optical measurements, respectively.

Predicting the effect of doping on crystal structure is less direct and requires an understanding of the structural driving forces. In the case of polar metals or doped FEs, the primary structural concern is the impact of doping on the inversion-lifting mechanism. Once again, different classes of materials will respond differently to doping as a perturbation. 
In doped proper FEs, the asymmetric structure is stabilized by a combination of dipole-dipole interactions and covalent bonds which compete with short-range repulsive forces (which favor a higher symmetry structure) \cite{Bersuker2013:ChemRev,Hickox-Young_Rondinelli2020:PRB,Ghosez:1996EPL}.
Although the addition of charge carriers is not necessarily incompatible with the persistence of broken symmetry, it cannot help but reduce and eventually eliminate the influence of long range dipole-dipole interactions (due to the reduction of the screening length) that cooperatively align the off-centering displacements and may also interfere with bonding, depending on the electronic structure of the material. 
By contrast, long-range interactions in polar metals are always screened and the atoms providing states at the Fermi level typically display weak coupling with the atoms active in the soft phonon(s) driving the symmetry-break \cite{Puggioni/Rondinelli:2014NatComm}.
Nonetheless, for sufficient carrier densities local polar  displacements can persist. Therefore, beyond how one simulates doping in these materials, as discussed in \Autoref{sec:A}, it is also imperative to recognize that \emph{how} we understand the manner in which the atomic structure  responds to doping relies intimately on whether the experimental probe interrogates local or average structure \cite{salmanirezaie2020orderdisorder,Chepkemboi2022}.
In any case, changes in the Fermi level of doped proper FEs will almost always eventually affect the ground state crystal -- local and average --  structure whereas similar changes in the Fermi level of polar metals are more likely to leave the crystal structure unaltered independent of the experimental probe volume.

\begin{figure}
    \centering
    \includegraphics[width=0.47\textwidth]{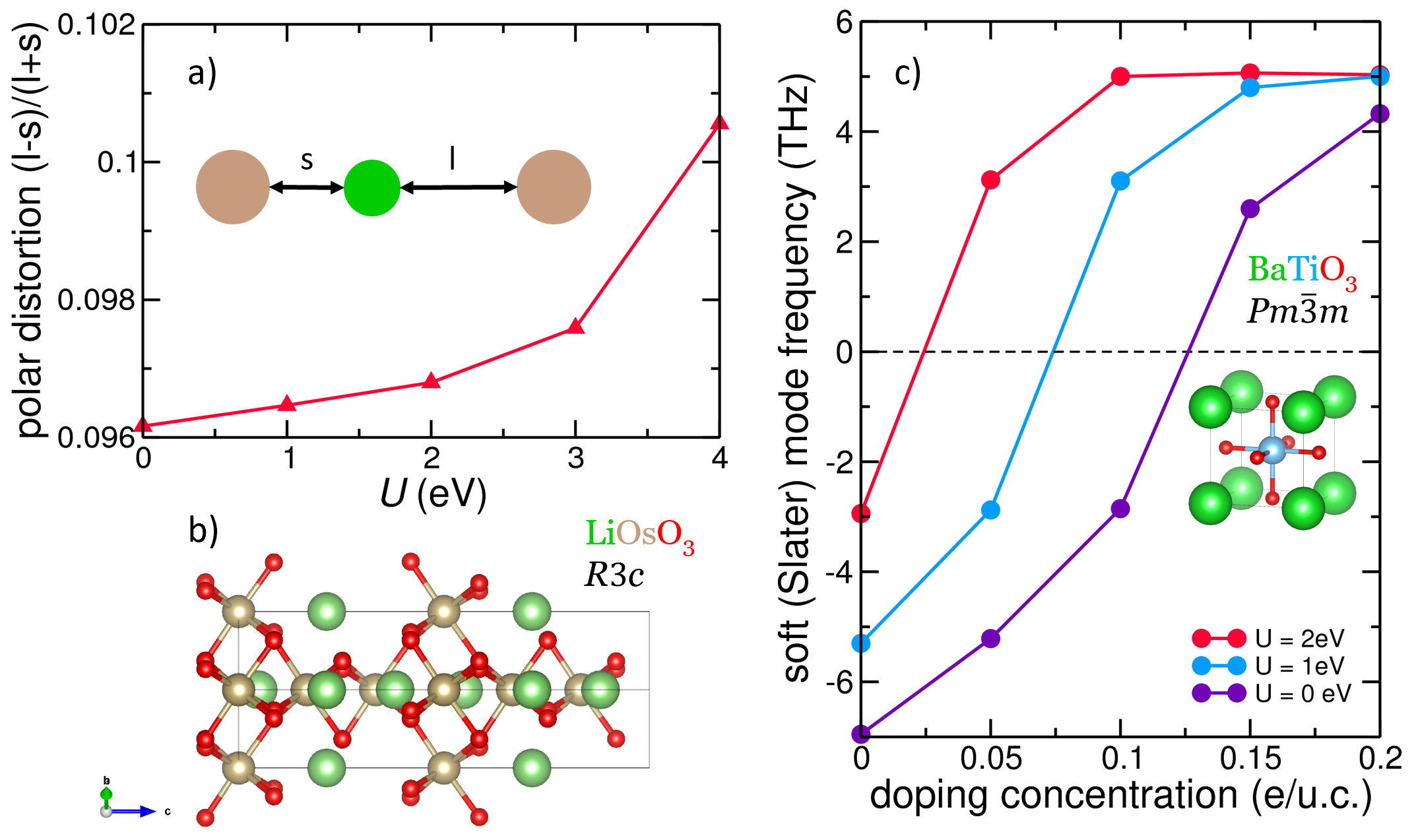}
    \caption{(a) Adding correlation to LiOsO$_3$ via increasing the Hubbard $U$ (applied to Os $d$ states) enhances the amplitude of the polar distortion. (inset) Schematic showing how the polar distortion amplitude is defined by the relative long and short distances between Li and Os along the polar axis. (b) Crystal structure of polar ($R3c$) LiOsO$_3$. (c) Increasing the degree of correlation in cubic BaTiO$_3$ (by applying the Hubbard $U$ to the Ti $d$ states) reduces the critical doping concentration to stabilize the soft $\Gamma$-point phonon mode of the cubic phase. (inset) The crystal structure of cubic ($Pm\overline{3}m$) BaTiO$_3$.}
    \label{fig:corr_effects}
\end{figure}

\textit{Electron Correlation and Magnetism}---%
Correlation also plays a significant role, both in the realization of polar metals generally and in the potential to drive metal-insulator  transitions, often in concert with magnetic ordering. Evidence of the former is found in the number of polar metals which exhibit ``bad" metallic transport from electron-electron interactions (\autoref{fig:transport}). Although it is now well-established that short-range interactions play a dominant role in driving local off-centering in polar metals, reduction of the screening length via correlation may enable longer-range interactions to further enhance the displacement magnitude, or at least allow for long-range coordination of local displacements. It was shown in Ref.\ \onlinecite{Giovannetti2016} that the polar displacements in the predicted polar metal SrEuMo$_2$O$_6$ are enhanced by introduction of additional correlation via a Hubbard $U$.interaction within DFT.  A similar effect is observed when plotting the effective polar amplitude in LiOsO$_3$ as a function of the static $U$ (\autoref{fig:corr_effects}a). However, just as correlation may help to stabilize or enhance polar displacements, when coupled with magnetic ordering it may also drive Mott-type metal insulator transitions, as found in simulations of LiOsO$_3$/LiNbO$_3$ superlattices \cite{Puggioni2015}. 

The effect of correlation on a system is also highly dependent on the distortion mechanism. In the absence of magnetism, correlation was shown to reduce the critical free carrier concentration in doped BaTiO$_3$ (\autoref{fig:corr_effects}c). The more highly correlated Ti $d$ states exhibited weaker covalent bonding with the O $p$ states, favoring the high symmetry structure. Previous studies of correlation in polar metals implied that correlation should assist in decoupling free charges from the symmetry-lowering transition, but the result of BaTiO$_3$ illustrates that the impact of correlation is highly dependent on the driving force behind local off-centering.

\subsection{``Ferroelectric" Metals}

In this section, we consider existing terminology used to denote polar structures in metallic compounds. Most of these labels were developed to describe insulating polar materials and require re-evaluation in a metallic context. We discuss the fundamental contradiction inherent to some labels and highlight others which transfer more easily. Note that we are restricting our discussion to polar metals, rather than noncentrosymmtric metals more broadly. Noncentrosymmetric space groups may be polar (more than one point invariant under all symmetry operations), chiral (symmetry elements contain only proper rotations), polar-chiral, or nonpolar-achiral. Chiral metals in particular (e.g., ferromagnetic MnSi and other B20 transition metal compounds) are of great interest for realizing skyrmions and other emergent physics \cite{Muhlbauer/etal:2009Science,Schulz2012, Yu2010}, but are not explicitly considered herein. A recent perspective on chiral magnets can be found in Ref.\ \onlinecite{Cheong2022}.

\textit{Contraindicated Properties}---
Whereas long-range coordinated off-centering and metallic conductivity were merely thought to be fundamentally incompatible (only to be shown otherwise in the last decade), the switchable polarization of a ferroelectric is, in principle, incapable of coexisting with free charge carriers. A ferroelectric phase is defined as ``one in which the spontaneous electric polarization can be reoriented between possible equilibrium directions (determined by the crystallography of the system) by a realizable, appropriately oriented electric field" \cite{IEEE2003}. 
Unlike some of the conductivity definitions above, ferroelectricity is an engineering definition; whether a material is ferroelectric can only be confirmed via experimental verification of two criteria: (1) That the polar order occurs at zero electric field (i.e.,  \textit{spontaneous} ordering) and (2) the polarization may be \textit{switched} via external electric field according to the symmetry of the crystal structure. These two criteria lead to the appearance of domain microstructures, a well-known feature of ferroic properties which separates ferroelectrics from other noncentrosymmetric materials \cite{doi:10.1143/JPSJ.27.387}. 
Gauss's Law prevents the generation of an internal electric field in a metal, thereby screening any attempt to switch the polarization of a polar metal via external electric field and rendering the possibility of criterion (2) null for any reasonably conducting materials. 

\begin{figure*}
    \centering
    \includegraphics[width=\textwidth]{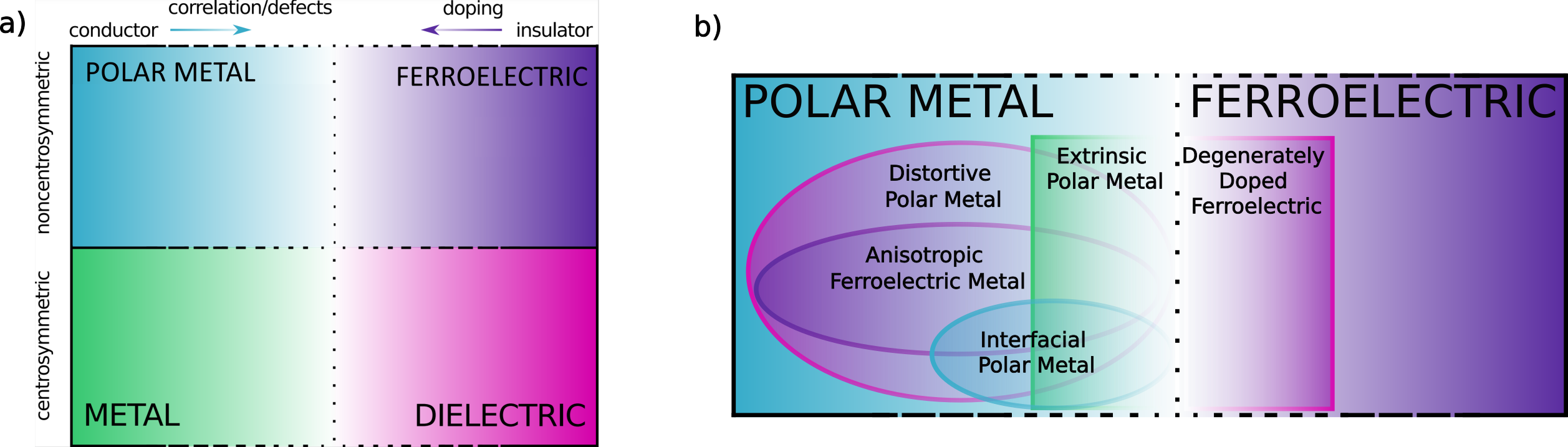}
    \caption{(a) Using crystal symmetry and electron transport as descriptors, the relationship between polar metals, ferroelectrics, metals, and dielectrics is illustrated. The gradient along the conductivity axis (abscissa) is meant to indicate that the line between metals and insulators is somewhat ambiguous in experiment, whereas the solid colors along the crystal symmetry (ordinate) axis indicate that the presence or lack of an inversion center is precisely defined. (b) A more detailed classification portraying the relationship between the various subclasses defined in the text. Transparency of overlapping categories indicates that a material may belong to multiple subclasses.}
    \label{fig:schematic}
\end{figure*}

\textit{``Ferroelectric-like''}---
Nevertheless, ``Ferroelectric-like" or ``ferroelectric" (in quotes) terminology has been used to describe structural phase transitions in polar metals since the first report of its kind in 2013 \cite{Shi:2013Nature}, with many other studies since following suit with this naming convention, either in reference to the material itself or to the inversion-lifting transition (e.g.,  \cite{Fujioka2015,Xiang:2014PRB,Benedek/Birol:2016JMMC,Sakai2016}). Despite the recent history of this convention, describing something as ``ferroelectric-like" is inherently ambiguous, as it leaves unclear which aspect(s) of ferroelectricity are being replicated. The comparison to FEs implies that the polarization is switchable, which---as noted above---is only true in very specific cases. The intended FE attribute being referred to is the presence of a second-order, structural phase transition that removes the spatial parity operation of inversion. Given the possibility for confusion,we recommend using more precise terminology.

\textit{Ferroelectric Bananas}---
Electrical hysteresis measurements assess the change in charge $Q$ on a pair of electrodes in contact with a dielectric material upon reversal of the applied bias, such that 
\begin{equation}
Q = 2 P_r A + \sigma E A t   
\end{equation}
\noindent where $P_r, A, t,$ and $E$ are the remnant electric polarization, area of the electrode contact, thickness of the dielectric, and the applied field, respectively, and $\sigma$ is the conductivity.
For either a degenerately doped ferroelectric or a polar metal, $\sigma\ne0$ and thus the charge that switches is not due to an 
electric polarization but the finite conductivity of the material, which is well-understood to 
be dielectric loss as $P_r\rightarrow0$ \cite{Scott2008}.
Nonetheless, some reports from the literature have claimed both to have synthesized a polar metal and to have measured a polarization \cite{Gu/Jin:2017PRB,Jin2019}.
However, we find these hysteresis loops reported in Ref.\  \onlinecite{Gu/Jin:2017PRB} to be consistent with examples from Reg.\  \onlinecite{Scott2008}, demonstrating that even definitively non-polarizable materials (e.g., bananas) can still exhibit electrical hysteresis.
Indeed, the classification of a material as a metal is fundamentally incompatible with a bulk polarization within the adiabatic regime \cite{Resta2002}.
%
For these reasons, in principle one should not on the one hand call a material a polar metal and then on the other hand report an electric polarization for it.

It is still possible, nonetheless, to experimentally quantify the displacements in polar metals. 
%
To this purpose, appropriate techniques such as scanning transmission electron microscopy, near-edge X-ray absorption fine structure, X-ray diffraction, coherent Bragg rod analysis, Raman scattering, and second harmonic generation measurements could be used \cite{Stone2019,Kim/Puggioni:2016Nature,Jin2016,Lei/Gu/Puggioni:2018NanoLett, Padmanabhan2018,laurita2019evidence}. 
The optical conductivity of noncentrosymmetric metals has been used to evaluate the effect and magnitude of correlation effects \cite{LoVecchio2016,Lei/Gu/Puggioni:2018NanoLett}. Analysis of the spectral weights produced in those measurements could be used to derive non-adiabatic Born effective charges (naBECs) and provide insight into the role of various ions with regard to the inversion-breaking mechanism \cite{dreyer2021nonadiabatic}. 

\textit{Ferroelasticity}---
The mechanical analog of ferroelectricity, a ferroelastic material exhibits two or more domains exhibiting a spontaneous strain, the direction of which may switched by an applied external stress \cite{IEEE2003,Wadhawan1984}. 
While ferroelectric and ferroelastic transitions are often accompanied by one another (as is the case in BaTiO$_3$), they may also occur independently, since all that is required for a ferroelastic transition is a change in unit cell shape, which may occur either with or without loss of inversion symmetry (e.g., sodium trihydrogen selenite \cite{Slosarek1989} and lithium niobate \cite{Gopalan2007,Shapovalov2014} both exhibit nonferroelastic ferroelectric phase transitions). 
Since ferroelasticity is defined without regard to electric fields or charge carriers, there is no fundamental incompatibility between metallicity and ferroelasticity; therefore, ferroelastic metals are an allowed class of materials. However, since they do not necessarily break inversion, we will henceforth restrict our discussion of ferroelastic metals to the scenario where the ferroelastic order is coupled with polar displacements.

\textit{Switchable Polar Metals}---
Nevertheless, recently a number of materials have emerged which use low-dimensionality, anisotropy, or ferroelasticity to demonstrate switchable polar metals \cite{Filippetti2016,Fei2018,Lei/Gu/Puggioni:2018NanoLett}. Although it is still difficult to formally define a polarization ($P\neq0$) for these compounds, it  is possible to reverse the direction of the atomic displacements into an opposite orientation. Perhaps the most widely applicable switching mechanism is ferroelasticity. Although free charges screen external electric fields, strain fields experience no such screening. Although not all polar metals exhibit ferroelastic domains, a change in lattice parameters often accompanies polar distortions, making ferroelastic switching perhaps a viable switching strategy as demonstrated by Lei et al.\  \cite{Lei/Gu/Puggioni:2018NanoLett}.

Switching the polar displacement via electric field was only achieved recently within two-dimensional polar metals that exhibit anisotropic conductivity \cite{Fei2018,Filippetti2016}. With the polar axis oriented perpendicular to the plane, the minimal thickness of the conducting layer limits the number of free charges available to screen an applied electric field. This approach skirts the fundamental contraindication between metallicity and switchable polarization by compromising the conductivity of the polar metal along the polar axis, such that switching is enabled along the more insulating direction. We might argue that the ferroelectric properties associated with these compounds are not intrinsic to the materials themselves but rather a product of the architecture and dimensionality of their synthesis.

We should note that although it has yet to be realized in experiment, there is a third proposed mechanism toward switching a polar metal. Namely, a thin polar metal film is deposited on a FE with low lattice mismatch. An external electric field switches the FE substrate and the resulting strain switches the polar metal \cite{Xiang:2014PRB,Fang2020}. This approach might be considered a variation of the ferroelastic switching mechanism, as both approaches use the unscreened strain field to switch polarization.

\begin{table*}[]
\centering
\begin{ruledtabular}
\caption{\label{tab:classes} Minimal requirements for various classes of polar metals (PM) by subclass according to the number of phases, crystallography, electrical conductivity, and sensitivity to boundary conditions, e.g., geometry and doping. The two phase designation is inclusive of heterojunctions and related nanoscale composite materials. Polar atomic structures in single phase compounds should adopt one of the 10 polar crystal classes: $1, 2, 3, 4, 6, m, mm2, 3m, 4mm, 6mm$. Checkmarks ($\checkmark$) and crosses ($\times$) indicate strict positive or negative requirements for that class, respectively, whereas a $\cdot$ indicates the presence or absence of the feature does not determine the class assignment for the material.
}
\begin{tabular}{llccccccc}
               &                       
               & \multicolumn{1}{c}{\multirow{2}{1in}{\centering polar atomic structure}} 
               & \multirow{2}{1in}{\centering local maximum $\sigma(\omega)$} 
               & \multirow{2}{0.8in}{\centering $~$ phase $~$ transition} 
               & \multicolumn{2}{c}{\centering Conduction} 
               & \multirow{2}{1in}{\centering Switchable Polarization} 
               & \multirow{2}{.6in}{\centering $~~$Doping$~~$  Sensitive} \\ \cline{6-7} 
    Phases         
&   Class / Subclass        
&  
&  
& 
& \multicolumn{1}{c}{3D}             
& \multicolumn{1}{c}{2D}     
&
&    \\
\hline
1              & Polar metal           & $\checkmark$                                                & $\checkmark$                                    &                 $\cdot$                  &        $\cdot$        &       $\checkmark$        &                 $\cdot$ &                  $\times$             \\
1              & --\,Distortive (DPM)           & $\checkmark$                                                & $\checkmark$                                    & $\checkmark$                      & $\cdot$   &     $\checkmark$          &             $\cdot$                    &       $\times$       \\
1              &  --\,Anisotropic  (AFM)         & $\checkmark$                                                & $\checkmark$                                    &       $\cdot$                            &         $\times$       & $\checkmark$  & $\checkmark$                    &        $\times$       \\
1              &  --\,Extrinsic (EPM)             & $\checkmark$                                                & $\checkmark$                                    &        $\cdot$                           & $\cdot$   & $\checkmark$  &              $\cdot$                   &    $\times$    \\
2              &  --\,Interfacial    (IPM)       & $\checkmark$                                                & $\checkmark$                                    &        $\cdot$                           &        $\cdot$        & $\checkmark$  &    $\cdot$                             &        $\times$      \\[0.8em]
1,2 & \multirow{2}{1.2in}{{Degenerately Doped} Ferroelectric (DDF)} & $\checkmark$                                                &                $\cdot$                                 &            $\cdot$                       & $\cdot$   & $\checkmark$  &      $\cdot$                                       & $\checkmark$\\
& & & & & & & & \\ 
\end{tabular}
\end{ruledtabular}
\end{table*}

\textit{Piezoelectric Metals}---
Piezoelectricity is defined as a change in polarization in response to a mechanical stress, the only prerequisite for which is a lack of an inversion center. However, as was observed when discussing ``ferroelectric" metals, polarization in a metal is ill-defined. Nonetheless, the Berry curvature effects observed in noncentrosymmetric insulators give rise to analogous properties in metals. Although a static polarization has no meaning in a conductive material, changes in electric polarization are instead measurable as a bulk current, making piezoelectric metals a meaningful designation. Varjas et al.\ also demonstrated that noncentrosymmetric metals with broken time reversal symmetry should exhibit a magnetopiezoelectric effect (MPE) \cite{Varjas:2016PRL}, as was observed recently in antiferromagnetic EuMnBi$_2$ \cite{Watanabe2017}. It is noteworthy that while all ferroelectrics by definition must exhibit piezoelectricity, not all piezoelectrics will be ferroelectric. This distinction is especially significant in metallic systems as the vast majority of polar metallic system are not switchable, making ``piezoelectric metal" a more accurate and more frequently applicable label than ``ferroelectric metal".

\section{Classifying metallic and polar materials}

\subsection{New terminology}
\begin{figure}
\centering
\includegraphics[width=0.49\textwidth]{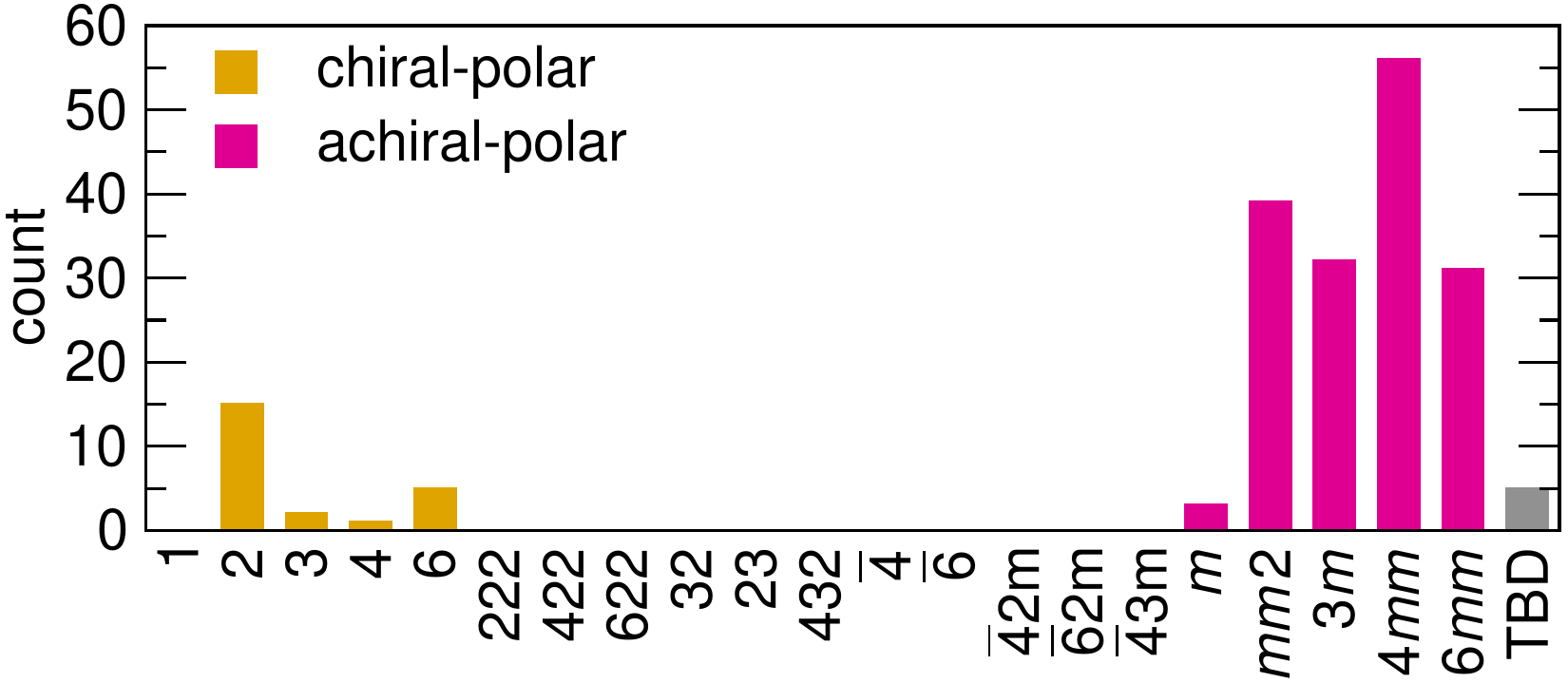}
\caption{Distribution of 189 polar metals in \autoref{tab:compiled_NCSMs} among the 21 noncentrosymmetric crystal classes. Most polar metals are found in achiral-polar symmetries rather than chiral-polar symmetries. The symmetry of compounds labeled `TBD' remain to be determined.}
\label{fig:dist}
\end{figure}

Summarizing the above considerations and descriptors concerning symmetry and conductivity class (\autoref{fig:schematic}a), we have identified four unique subclasses classes of polar metals (\autoref{tab:classes}),  which we describe in detail next and differentiate from degenerately doped ferroelectrics.  
In some cases, a material may be classified by more than one designation as indicated by the overlapping intersections in \autoref{fig:schematic}b.
We then use this schema to surveying known metallic and acentric compounds in the literature (\autoref{tab:compiled_NCSMs}), which is current as of September 2022 and \autoref{fig:dist} shows the distribution of these compounds by polar metal subclass and crystal class.
A dynamic version of  \autoref{tab:compiled_NCSMs} is  maintained at \href{https://mtd.mccormick.northwestern.edu/polar-metals-materials-database/}{https://mtd.mccormick.northwestern.edu/polar-metals-materials-database/}.
We invite researchers to submit entries as new materials are discovered by contacting the corresponding author.  


\textit{Polar Metal}---%
A polar metal (PM) exhibits a polar crystal structure, identifiable via structural characterization techniques, such as X-Ray diffraction, neutron diffraction, or the observation of properties associated with broken inversion symmetry (i.e., second harmonic generation). A polar metal should also conduct electricity, as determined via a local maximum in its optical conductivity as $\omega\rightarrow0$. Finally, the inversion-lifting mechanism and electron transport should be more or less unaffected by small changes in electron concentration, i.e., perturbations to the electron chemical potentiatl. Restated, for a material be considered a polar metal, the polar distortion and metallicity should not be contraindicated. Examples include LiOsO$_3$ \cite{Shi:2013Nature}, Ca$_3$Ru$_2$O$_7$ \cite{Lei/Gu/Puggioni:2018NanoLett,Yoshida2005}, and CePt$_3$Si \cite{Bauer2004}.

\textit{Distortive Polar Metal}---%
A distortive polar metal (DPM) is a subclass of polar metals, meeting all of the above criteria while also exhibiting an inversion-symmetry lifting phase transition. This eliminates the need for misleading ``ferroelectric-like" terminology and its derivatives by highlighting the intended comparison to a ferroelectric phase transition. 

The disortive polar metal subclass also separates polar metals exhibiting phase transitions from polar metals in which inversion is lifted via compositional order and nominally absent for a broad temperature range. 
    This is a meaningful distinction, both from a fundamental physics perspective as well as an application-driven perspective, as the presence of a phase transition implies opportunities to tune physical properties in ways that would not be possible in  polar metals that do not undergo a phase transition. Structural phase transitions in polar metals have been experimentally observed via second-order discontinuities in electronic transport, heat capacity, magnetic susceptibility, dielectric response, second harmonic generation, and with scattering and diffraction techniques \cite{Jiao2020,Shi:2013Nature,Padmanabhan2018,Fujioka2015,Hwang2010,Rischau2017,Berger2021}. Within distortive polar metals---as is true with ferroelectrics---the phase transition may be displacive, order-disorder, or exhibit characteristics of both. Examples of distortive polar metals include LiOsO$_3$ (primarily order-disorder) \cite{Shi:2013Nature,laurita2019evidence,Padmanabhan2018,Jin2016} and Pb$_2$CoOsO$_6$ (primarily displacive) \cite{Jiao2020}. Examples of non-distortive or compositional polar metals include intermetallics such as CePt$_3$Si \cite{Bauer2004} and ErPdBi \cite{Pan2013}.
    
    In differentiating between polar metals that exhibit a structural phase transition and those that do not, a key feature of potential  technological use in distortive polar metals is their domain structure. Metallic domains walls in ferroelectrics like BiFeO$_3$ \cite{Jiang2017} and YMnO$_3$ \cite{Choi2010} are actively investigated for use as non-destructive resistance-based memory devices \cite{jiang2015strain,Holstad2020}. Similarly, in Ca$_3$Ru$_2$O$_7$, both charged and uncharged domain walls were shown to exist---each associated with different electrical conductivity---implying potential application in charge-mediated memory devices as well \cite{Stone2019}. However, under the current classification scheme, Ca$_3$Ru$_2$O$_7$ would not be classified as a distortive polar metal because the temperature at which it would undergo a structural phase transition to recover centrosymmetry exceeds the melting temperature of the compound. 
    
    A number of polar metals both exhibit a local ordering mechanism and fail to exhibit an inversion-lifting structural phase transition. Ca$_3$Ru$_2$O$_7$ is one such example, in which the polar order is derived from trilinear coupling between octahedral rotations but the thermal energy required to overcome this coupling and achieve a centrosymmetric structure is greater than the melting temperature of the compound  
    \cite{Yoshida2005,Lei/Gu/Puggioni:2018NanoLett}. 
    Therefore, materials which exhibit local-ordering mechanisms similar to those observed in distortive polar metals may be called ``pseudo-distortive polar metals." This terminology acknowledges that such materials share important characteristics with distortive metals while remaining distinct due to their lack of an ordering transition.

\textit{Anisotropic Ferroelectric Metal}---%
An anisotropic ferroelectric metal (AFM) meets the structural criteria for a distortive polar metal but exhibits limited conductivity along the polar axis, allowing for observable switching of the polar displacement direction via external electric field. ``Anisotropic" is meant to call attention to the fact that it the anisotropic conductivity tensor and direction of the polar distortions are in transverse orientations. As a consequence, these materials can be deployed in suitable device geometries to sidestep the fundamentally contraindicated relationship between field-switchable polarization and conductivity. Ferroelectricity is not, and by definition cannot, be an intrinsic property of a metal, but low-dimensional conductors provides a design strategy to achieve high electrically  anisotropy such that the polarization becomes switchable by external electric field. Examples include WTe \cite{Fei2018} and Bi$_5$Ti$_5$O$_{17}$ \cite{Filippetti2016}. 
Note that polar metals whose polar order is switched via strain are not classified as anisotropic ferroelectric metals, as this switching mechanism does not require limiting the fundamental relationship between external electric field and free charge carriers.

\begingroup
\setlength\LTleft{0pt}
\setlength\LTright{0pt}
\LTcapwidth=\textwidth
\begin{longtable*}[c]{lllllll}
\caption{\label{tab:compiled_NCSMs}
Published polar compounds exhibiting metallic conductivity. Columns denote space group (SG), presence of a polar structural phase transition ($T_c$), presence of a superconducting transition ($T_{sc}$), and presence of a magnetic ordering transition ($T_M$), respectively. Under SG: * = only the point group is known; ``unknown" = the specific symmetry elements are unknown but there is evidence of broken inversion symmetry. Under $T_c$: P = compound is predicted but has not yet been synthesized. Under Class, the following abbreviations are used:  Polar metal (PM), Distortive polar metal (DPM), Anisotropic ferroelectric metal (AFM), Extrinsic polar metal (EPM), interfacial polar metal (IPM), and Degenerately doped ferroelectric (DDF).}
{}\\

 \hline\hline\\[0.1em]
    Composition & SG & $T_c$ & $T_{sc}$ & $T_M$ & Class & Digital Object Identifier (DOI) \\[0.1em]
    \hline\\[0.05em]
    \endfirsthead

    \multicolumn{7}{c}%
{{\bfseries \tablename\ \thetable{} -- continued from previous page}} \\
\hline\hline\\[0.1em]
  Composition & SG & $T_c$ & $T_{sc}$ & $T_M$ & Class & Digital Object Identifier (DOI) \\[0.1em]
  \hline\\[0.05em] 
  \endhead
  
  \hline\\[0.1em] \multicolumn{7}{c}{{\textit{Continued on next page}}} \\ 
\endfoot

\hline \hline
\endlastfoot
BaTiO$_3$	&	$P4mm$	&	yes	&		&		&	DDF	&	10.1038/s41598-017-04635-3	\\
BaTiO$_{3-\delta}$	&	$P4mm$	&	yes	&		&		&	DDF	&	10.1103/PhysRevLett.104.147602	\\
PbTiO$_{3-\delta}$	&	$P4mm$	&	P	&		&		&	EPM	&	10.1103/PhysRevB.94.224107	\\
PbTi$_{1-x}$Nb$_x$O$_3$	&	$P4mm$	&		&		&		&	EPM	&	10.1103/PhysRevB.96.165206	\\
BiFeO$_{3-\delta}$	&	$R3c$	&	P	&		&	yes	&	EPM	&	10.1103/PhysRevB.93.174110	\\
Sr$_{1-x}$Ca$_x$TiO$_{3-\delta}$	&	$P4mm$	&	yes	&	yes	&		&	DPM	&	10.1038/NPHYS4085	\\
BaMnO$_3$	&	$Amm2$	&	P	&		&		&	EPM	&	10.1103/PhysRevB.97.054107	\\
BiAlO$_3$	&	$P4mm$	&	P	&		&		&	EPM	&	10.1103/PhysRevB.97.054107	\\
BaTiO$_3$/SrTiO$_3$/LaTiO$_3$	&	$mm4*$	&		&		&		&	IPM	&	10.1038/s41467-018-03964-9	\\
LaAlO$_3$/BaSr$_{0.8}$TiO$_3$/SrTiO$_3$	&	unknown	&		&		&		&	IPM	&	10.1038/s42005-019-0227-4	\\
LaFeO$_3$/YFeO$_3$	&	$Pmc2_1$	&	P	&		&		&	IPM	&	10.1103/PhysRevB.97.054107	\\
NdNiO$_3$	&	$Pc$	&		&		&		&	PM	&	10.1038/nature17628	\\
LaNiO$_3$	&	$Pc$	&		&		&		&	PM	&	10.1038/nature17628	\\
LiOsO$_3$	&	$R3c$	&	yes	&		&		&	DPM	&	10.1038/NMAT3754	\\
LiNbO$_3$	&	$R3c$	&	P	&		&		&	DDF	&	10.1103/PhysRevMaterials.3.054405	\\
MgReO$_3$	&	$R3c$	&	P	&		&		&	PM	&	10.1103/PhysRevB.90.094108	\\
TiGaO$_3$	&	$R3c$	&	P	&		&		&	PM	&	10.1103/PhysRevB.90.094108	\\
Ca$_3$Ru$_2$O$_7$	&	$Bb2_1m$	&		&		&	yes	&	PM	&	10.1021/acs.nanolett.8b00633	\\
(Sr,Ca)Ru$_2$O$_6$	&	$Pmc2_1$	&	P	&		&	yes	&	PM	&	10.1088/0953-8984/26/26/265501	\\
Bi$_5$Ti$_5$O$_{17}$	&	$Pm2_1n$	&	P	&		&		&	AFM	&	10.1038/ncomms11211	\\
BiPbTi$_2$O$_6$	&	$Pmm2$	&	P	&		&		&	PM	&	10.1038/s43246-019-0005-6	\\
La$_2$Ti$_2$O$_7$	&	$P2_1$	&	P	&		&		&	EPM	&	10.1103/PhysRevB.97.054107	\\
Sr$_2$Nb$_2$O$_7$	&	$Cmc2_1$	&	P	&		&		&	EPM	&	10.1103/PhysRevB.97.054107	\\
MgCNi$_3$	&	$P4mm$	&	P	&		&		&	PM	&	10.1103/PhysRevMaterials.2.125004	\\
ZnCNi$_3$	&	$P4mm$	&	P	&		&		&	PM	&	10.1103/PhysRevMaterials.2.125004	\\
CdCNi$_3$	&	$P4mm$	&	P	&		&		&	PM	&	10.1103/PhysRevMaterials.2.125004	\\
CeSiPt$_3$	&	$P4mm$	&		&	yes	&	yes	&	PM	&	10.1103/PhysRevLett.92.027003	\\
LiGaGe	&	$P6_3mc$	&		&		&		&	PM	&	10.1103/PhysRevB.99.195154	\\
SrHgPb	&	$P6_3mc$	&	P	&		&		&	PM	&	10.1103/PhysRevLett.121.106404	\\
SrHgSn	&	$P6_3mc$	&	P	&		&		&	PM	&	10.1103/PhysRevLett.121.106404	\\
CaHgSn	&	$P6_3mc$	&	P	&		&		&	PM	&	10.1103/PhysRevLett.121.106404	\\
KMgSb$_{0.2}$Bi$_{0.8}$	&	$P6_3mc$	&	P	&		&		&	PM	&	10.1103/PhysRevLett.117.076401	\\
CaAgBi	&	$P6_3mc$	&	P	&		&		&	PM	&	10.1103/PhysRevMaterials.1.044201	\\
LiZnBi	&	$P6_3mc$	&	P	&		&		&	PM	&	10.1103/PhysRevB.96.115203	\\
LaAuGe	&	$P6_3mc$	&		&		&		&	PM	&	10.1063/1.5132339	\\
LaPtSb	&	$P6_3mc$	&		&		&		&	PM	&	10.1063/1.5132339	\\
WTe$_2$	&	$Pnm2_1$	&	yes	&	yes	&		&	AFM	&	10.1038/s41586-018-0336-3	\\
MoTe$_2$	&	$Pnm2_1$	&	yes	&		&		&	PM	&	10.1126/sciadv.1601378	\\
CrN	&	$6mm*$	&		&		&	yes	&	AFM	&	10.1103/PhysRevB.96.235415	\\
CrB$_2$	&	$6mm*$	&	P	&		&	yes	&	AFM	&	10.1103/PhysRevB.96.235415	\\
FeB$_2$	&	$6mm*$	&		&		&		&	AFM	&	10.1021/acs.nanolett.6b02335	\\
P	&	$P6_3mc$	&	P	&		&		&	PM	&	10.1088/1361-648X/aadeaa	\\
As	&	$P6_3mc$	&	P	&		&		&	PM	&	10.1088/1361-648X/aadeaa	\\
Sb	&	$P6_3mc$	&	P	&		&		&	PM	&	10.1088/1361-648X/aadeaa	\\
Bi	&	$P6_3mc$	&	P	&		&		&	PM	&	10.1088/1361-648X/aadeaa	\\
SnP	&	$I4mm$	&	yes	&	yes	&		&	DPM	&	10.1103/PhysRevLett.119.207001	\\
PdBi	&	$P2_1$	&		&	yes	&		&	PM	&	10.1016/j.phpro.2013.04.062 	\\
UIr	&	$P2_1$	&		&	yes	&	yes	&	PM	&	10.1007/978-3-642-24624-1\_2	\\
LaNiC$_2$	&	$Amm2$	&		&	yes	&	yes	&	PM	&	10.1016/j.physc.2014.01.008	\\
NdRhC$_2$	&	$Amm2$	&		&		&	yes	&	PM	&	10.1021/cm00006a007	\\
PrRhC$_2$	&	$Amm2$	&		&		&	yes	&	PM	&	10.1021/cm00006a007	\\
LaSr$_2$Cu$_2$GaO$_7$	&	$Ima2$	&		&	yes	&	yes	&	PM	&	10.1021/cm00017a032	\\
CeSr$_2$Cu$_2$GaO$_7$	&	$Ima2$	&		&	yes	&	yes	&	PM	&	10.1021/cm00017a032	\\
PrSr$_2$Cu$_2$GaO$_7$	&	$Ima2$	&		&	yes	&	yes	&	PM	&	10.1021/cm00017a032	\\
NdSr$_2$Cu$_2$GaO$_7$	&	$Ima2$	&		&	yes	&	yes	&	PM	&	10.1021/cm00017a032	\\
PmSr$_2$Cu$_2$GaO$_7$	&	$Ima2$	&		&	yes	&	yes	&	PM	&	10.1021/cm00017a032	\\
SmSr$_2$Cu$_2$GaO$_7$	&	$Ima2$	&		&	yes	&	yes	&	PM	&	10.1021/cm00017a032	\\
EuSr$_2$Cu$_2$GaO$_7$	&	$Ima2$	&		&	yes	&	yes	&	PM	&	10.1021/cm00017a032	\\
GdSr$_2$Cu$_2$GaO$_7$	&	$Ima2$	&		&	yes	&	yes	&	PM	&	10.1021/cm00017a032	\\
TbSr$_2$Cu$_2$GaO$_7$	&	$Ima2$	&		&	yes	&	yes	&	PM	&	10.1021/cm00017a032	\\
DySr$_2$Cu$_2$GaO$_7$	&	$Ima2$	&		&	yes	&	yes	&	PM	&	10.1021/cm00017a032	\\
HoSr$_2$Cu$_2$GaO$_7$	&	$Ima2$	&		&	yes	&	yes	&	PM	&	10.1021/cm00017a032	\\
ErSr$_2$Cu$_2$GaO$_7$	&	$Ima2$	&		&	yes	&	yes	&	PM	&	10.1021/cm00017a032	\\
TmSr$_2$Cu$_2$GaO$_7$	&	$Ima2$	&		&	yes	&	yes	&	PM	&	10.1021/cm00017a032	\\
YbSr$_2$Cu$_2$GaO$_7$	&	$Ima2$	&		&	yes	&	yes	&	PM	&	10.1021/cm00017a032	\\
YSr$_2$Cu$_2$GaO$_7$	&	$Ima2$	&		&	yes	&	yes	&	PM	&	10.1021/cm00017a032	\\
V$_2$Hf	&	$Imm2$	&	yes	&	yes	&		&	DPM	&	10.1103/PhysRevB.17.1136	\\
Li$_2$IrSi$_3$	&	$P3_1c$	&		&	yes	&		&	PM	&	10.7566/JPSJ.83.093706	\\
Mg$_2$Al$_3$	&	$R3m$	&	yes	&	yes	&		&	DPM	&	10.1103/PhysRevB.76.014528	\\
La$_5$B$_2$C$_6$	&	$P4$	&		&	yes	&		&	PM	&	10.1016/0022-5088(83)90520-9	\\
BaPtSi$_3$	&	$I4mm$	&		&	yes	&		&	PM	&	10.1103/PhysRevB.80.064504	\\
CeIrSi$_3$	&	$I4mm$	&		&	yes	&	yes	&	PM	&	10.1143/JPSJS.77SA.37	\\
CeRhSi$_3$	&	$I4mm$	&		&	yes	&	yes	&	PM	&	10.1143/JPSJ.76.051010	\\
LaPtSi$_3$	&	$I4mm$	&		&	yes	&		&	PM	&	10.1103/PhysRevB.89.094509	\\
LaPdSi$_3$	&	$I4mm$	&		&	yes	&		&	PM	&	10.1103/PhysRevB.89.094509	\\
CeIrGe$_3$	&	$I4mm$	&		&	yes	&	yes	&	PM	&	10.1016/j.jmmm.2006.10.151	\\
EuNiGe$_3$	&	$I4mm$	&		&		&	yes	&	PM	&	10.1103/PhysRevB.87.064406	\\
LaRhGe$_3$	&	$I4mm$	&		&		&		&	PM	&	10.1063/5.0042924	\\
IrRhGe$_3$	&	$I4mm$	&		&		&		&	PM	&	10.1063/5.0042924	\\
PdRhGe$_3$	&	$I4mm$	&		&		&		&	PM	&	10.1063/5.0042924	\\
SrPdGe$_3$	&	$I4mm$	&		&	yes	&		&	PM	&	10.1088/1742-6596/273/1/012078	\\
SrPtGe$_3$	&	$I4mm$	&		&	yes	&		&	PM	&	10.1088/1742-6596/273/1/012078	\\
PrPdIn$_2$	&	$I4mm$	&		&		&	yes	&	PM	&	10.1021/cm031139m	\\
CePt$_3$Si	&	$I4mm$	&		&	yes	&	yes	&	PM	&	10.1103/PhysRevLett.92.027003	\\
Li$_2$(Pd$_{1-x}$Pt$_x$)$_3$B	&	$I4mm$	&		&	yes	&	yes	&	PM	&	10.1088/1742-6596/400/2/022096	\\
KCu$_7$P$_3$	&	$P31m$	&		&		&		&	PM	&	10.1021/acs.inorgchem.9b01336	\\
Bi$_2$FeCrO$_6$	&	$R3$	&	P	&		&	yes	&	PM	&	10.1103/PhysRevLett.123.107201	\\
CeAuGe	&	$P6_3mc$	&		&		&	yes	&	PM	&	10.1016/0304-8853(95)00430-0	\\
LuAuGe	&	$P6_3mc$	&		&		&		&	PM	&	10.1016/0925-8388(95)02069-1	\\
ScAuGe	&	$P6_3mc$	&		&		&		&	PM	&	10.1016/0925-8388(95)02069-1	\\
HoAuGe	&	$P6_3mc$	&		&		&	yes	&	PM	&	10.1088/0953-8984/13/11/315	\\
CeCuSn	&	$P6_3mc$	&		&		&		&	PM	&	10.1016/j.jallcom.2004.09.086	\\
La$_{15}$Ge$_9$C	&	$P6_3mc$	&		&		&		&	PM	&	10.1016/j.jallcom.2011.03.092	\\
La$_{15}$Ge$_9$Fe	&	$P6_3mc$	&		&		&	yes	&	PM	&	10.1021/ic9515158	\\
La$_{15}$Ge$_9$Co	&	$P6_3mc$	&		&		&		&	PM	&	10.1021/ic9515158	\\
La$_{15}$Ge$_9$Ni	&	$P6_3mc$	&		&		&		&	PM	&	10.1021/ic9515158	\\
Sr$_3$Cu$_8$Sn$_4$	&	$P6_3mc$	&		&		&		&	PM	&	10.1016/j.intermet.2011.02.018	\\
IrMg$_{2.03}$In$_{.97}$	&	$P6_3mc$	&		&		&		&	PM	&	10.1016/j.intermet.2003.12.001	\\
IrMg$_{2.20}$In$_{.80}$	&	$P6_3mc$	&		&		&		&	PM	&	10.1016/j.intermet.2003.12.001	\\
CaAlSi	&	$P6_3$	&		&	yes	&		&	PM	&	10.1143/JPSJ.75.043713	\\
TlV$_6$S$_8$	&	$P6_3$	&		&	yes	&		&	PM	&	10.1016/S0038-1098(01)00333-7	\\
KV$_6$S$_8$	&	$P6_3$	&		&	yes	&		&	PM	&	10.1016/S0038-1098(01)00333-7	\\
RbV$_6$S$_8$	&	$P6_3$	&		&	yes	&		&	PM	&	10.1016/S0038-1098(01)00333-7	\\
CsV$_6$S$_8$	&	$P6_3$	&		&	yes	&		&	PM	&	10.1016/S0038-1098(01)00333-7	\\
LaPt$_3$B	&	$P4mm$	&		&		&	yes	&	PM	&	10.1016/S0925-8388(03)00373-6	\\
PrPt$_3$B	&	$P4mm$	&		&		&	yes	&	PM	&	10.1016/S0925-8388(03)00373-6	\\
NdPt$_3$B	&	$P4mm$	&		&		&	yes	&	PM	&	10.1016/S0925-8388(03)00373-6	\\
LaRhSi$_3$	&	$I4mm$	&		&	yes	&		&	PM	&	10.1016/0025-5408(84)90017-5	\\
LaIrSi$_3$	&	$I4mm$	&		&	yes	&		&	PM	&	10.1016/0025-5408(84)90017-5	\\
CeCoGe$_3$	&	$I4mm$	&		&	yes	&	yes	&	PM	&	10.1016/j.jmmm.2006.10.717	\\
LaCoGe$_3$	&	$I4mm$	&		&		&	yes	&	PM	&	10.1143/JPSJ.75.044711	\\
CeRhGe$_3$	&	$I4mm$	&		&	yes	&	yes	&	PM	&	10.1143/JPSJ.77.064716	\\
CeRuSi$_3$	&	$I4mm$	&		&		&		&	PM	&	10.1143/JPSJ.77.064716	\\
LaIrGe$_3$	&	$I4mm$	&		&		&		&	PM	&	10.1143/JPSJ.77.064717	\\
LaFeGe$_3$	&	$I4mm$	&		&		&		&	PM	&	10.1143/JPSJ.77.064717	\\
PrCoGe$_3$	&	$I4mm$	&		&		&		&	PM	&	10.1143/JPSJ.77.064717	\\
CaIrSi$_3$	&	$I4mm$	&		&	yes	&		&	PM	&	10.1016/j.physc.2009.10.120	\\
CaPtSi$_3$	&	$I4mm$	&		&	yes	&		&	PM	&	10.1134/S0021364010170157	\\
SrAuSi$_3$	&	$I4mm$	&		&	yes	&		&	PM	&	10.1021/cm500032u	\\
EuPdGe$_3$	&	$I4mm$	&		&		&	yes	&	PM	&	10.1016/j.ssc.2012.02.022	\\
EuPtSi$_3$	&	$I4mm$	&		&		&	yes	&	PM	&	10.1103/PhysRevB.81.144414	\\
NdPdIn$_2$	&	$I4mm$	&		&		&		&	PM	&	10.1021/cm031139m	\\
SmPdIn$_2$	&	$I4mm$	&		&		&		&	PM	&	10.1021/cm031139m	\\
GdPdIn$_2$	&	$I4mm$	&		&		&		&	PM	&	10.1021/cm031139m	\\
ErPdIn$_2$	&	$I4mm$	&		&		&		&	PM	&	10.1021/cm031139m	\\
TmPdIn$_2$	&	$I4mm$	&		&		&		&	PM	&	10.1021/cm031139m	\\
LuPdIn$_2$	&	$I4mm$	&		&		&		&	PM	&	10.1021/cm031139m	\\
La$_2$NiAl$_7$	&	$I4mm$	&		&		&		&	PM	&	10.1021/cm050513a	\\
SnP	&	$I4mm$	&		&		&		&	PM	&	10.1021/ic50084a032	\\
GeP	&	$I4mm$	&		&	yes	&		&	PM	&	10.1016/0022-4596(70)90005-8	\\
Ir$_9$Al$_{28}$	&	$P3_1c$	&		&		&		&	PM	&	10.1016/j.jallcom.2005.06.027	\\
$\gamma$-Bi$_2$Pt	&	$P3_1m$	&		&		&		&	PM	&	10.1002/zaac.201400331	\\
Au$_{6.05}$Zn$_{12.51}$	&	$P3_1m$	&		&		&		&	PM	&	10.1021/ic301933a	\\
Ba$_{21}$Al$_{40}$	&	$P3_1m$	&		&		&		&	PM	&	10.1021/ic0400235	\\
Li$_{17}$Ag$_3$Sn$_6$	&	$P3_1m$	&		&		&		&	PM	&	10.1021/ja038868n	\\
Cr$_5$Al$_8$	&	$R3m$	&		&		&		&	PM	&	10.1107/S0567740874004997	\\
Mn$_5$Al$_8$	&	$R3m$	&	yes	&		&		&	DPM	&	10.1007/BF02672582	\\
Cu$_{7.8}$Al$_5$	&	$R3m$	&		&		&		&	PM	&	10.1107/S0108768191005694	\\
Cu$_7$Hg$_6$	&	$R3m$	&		&		&		&	PM	&	10.3891/acta.chem.scand.23-1181	\\
NbS$_2$	&	$R3m$	&		&	yes	&		&	PM	&	10.1107/S0567740874003220	\\
Pr$_2$Fe$_{17}$	&	$R3m$	&		&		&	yes	&	PM	&	10.1103/PhysRevB.68.054424	\\
Pr$_2$Co$_{17}$	&	$R3m$	&		&		&		&	PM	&	10.1103/PhysRevB.68.054424	\\
Sn$_4$As$_3$	&	$R3m$	&		&	yes	&		&	PM	&	10.1002/zaac.19683630102	\\
Sn$_4$P$_3$	&	$R3m$	&		&	yes	&		&	PM	&	10.1002/zaac.19683630102	\\
LiOsO$_3$	&	$R3c$	&	yes	&		&		&	DPM	&	10.1038/NMAT3754 A	\\
La$_4$Mg$_5$Ge$_6$	&	$Cmc2_1$	&		&		&		&	PM	&	10.1021/ic2014732	\\
La$_4$Mg$_7$Ge$_6$	&	$Cmc2_1$	&		&		&		&	PM	&	10.1021/ic2014732	\\
Yb$_2$Ga$_4$Ge$_6$	&	$Cmc2_1$	&		&		&		&	PM	&	10.1002/chem.200305755	\\
Ce$_2$Rh$_3$(Pb,Bi)$_5$	&	$Cmc2_1$	&		&		&	yes	&	PM	&	10.1016/j.jssc.2007.06.012	\\
Eu$_2$Pt$_3$Sn$_5$	&	$Cmc2_1$	&		&		&	yes	&	PM	&	10.1524/zkri.2009.1160	\\
Lu$_4$Zn$_5$Ge$_6$	&	$Cmc2_1$	&		&		&		&	PM	&	10.1016/j.intermet.2013.02.018	\\
Hg$_3$Te$_2$Br$_2$	&	$R3$	&	yes	&		&		&	PM	&	10.1038/s41467-021-21836-7	\\
In$_2$S$_3$	&	$R3m$	&	P	&		&		&	AFM	&	10.1039/D1MH01556G	\\
In$_2$Se$_3$	&	$R3m$	&	P	&		&		&	AFM	&	10.1039/D1MH01556G	\\
In$_2$Te$_3$	&	$R3m$	&	P	&		&		&	AFM	&	10.1039/D1MH01556G	\\
NaYMnReO$_3$	&	$P2_1$	&	P	&		&	yes	&	PM	&	10.1021/acs.chemmater.0c02976	\\
NaYFeReO$_3$	&	$P2_1$	&	P	&		&	yes	&	PM	&	10.1021/acs.chemmater.0c02976	\\
NaYCoReO$_3$	&	$P2_1$	&	P	&		&	yes	&	PM	&	10.1021/acs.chemmater.0c02976	\\
NaYNiReO$_3$	&	$P2_1$	&	P	&		&	yes	&	PM	&	10.1021/acs.chemmater.0c02976	\\
NaYMnOsO$_3$	&	$P2_1$	&	P	&		&	yes	&	PM	&	10.1021/acs.chemmater.0c02976	\\
NaYCoOsO$_3$	&	$P2_1$	&	P	&		&	yes	&	PM	&	10.1021/acs.chemmater.0c02976	\\
NaYNiOsO$_3$	&	$P2_1$	&	P	&		&	yes	&	PM	&	10.1021/acs.chemmater.0c02976	\\
NaYFeWO$_3$	&	$P2_1$	&	P	&		&	yes	&	PM	&	10.1021/acs.chemmater.0c02976	\\
YAl$_2$	&	$P6mm$	&	P	&		&		&	AFM	&	10.1021/acs.jpclett.0c03136	\\
CaRh$_2$	&	$P6mm$	&	P	&		&		&	AFM	&	10.1021/acs.jpclett.0c03136	\\
doped SiGe	&	$P3m1$	&	P	&		&		&	AFM	&	10.1088/1361-648X/abdce9/pdf	\\
doped SiSn	&	$P3m1$	&	P	&		&		&	AFM	&	10.1088/1361-648X/abdce9/pdf	\\
doped GeSn	&	$P3m1$	&	P	&		&		&	AFM	&	10.1088/1361-648X/abdce9/pdf	\\
KNbO$_3$/BaTiO$_3$	&	$P4mm$	&	P	&		&	yes	&	AFM	&	10.1016/j.commatsci.2020.110235	\\
(SrRuO$_3$)$_1$/(BaTiO$_3$)$_{10}$	&	$mm2*$	&		&		&	yes	&	AFM	&	10.1021/acs.nanolett.0c03417	\\
strained EuTiO$_{3-x}$H$_x$	&	$Pmm2$	&	P	&		&	yes	&	EPM	&	10.1103/PhysRevB.102.224102	\\
KTiO$_2$H	&	unknown	&	P	&	yes	&		&	DDF	&	10.1103/PhysRevMaterials.5.054802	\\
RbTiO$_2$H	&	unknown	&	P	&	yes	&		&	DDF	&	10.1103/PhysRevMaterials.5.054802	\\
CsTiO$_2$H	&	unknown	&	P	&	yes	&		&	DDF	&	10.1103/PhysRevMaterials.5.054802	\\
doped PbZrO$_3$	&	unknown	&	P	&		&		&	DDF	&	10.1103/PhysRevB.102.134118	\\
LaFeAsO$_{1-x}$H$_x$	&	$mm2*$	&		&	yes	&	yes	&	PM	&	10.21203/rs.3.rs-77544/v1	\\
Pb$_2$CoOsO$_6$	&	$Pc$	&	yes	&		&	yes	&	DPM	&	10.1103/PhysRevB.102.144418	\\
PrAlGe	&	$I4_1md$	&		&		&	yes	&	PM	&	10.1038/s41467-020-16879-1	\\
doped PbTe monolayer	&	$P3m1$	&	P	&		&		&	EPM, AFM	&	10.1039/d0nh00188k	\\
en-CoS	&	$Pca2_1$	&		&		&	yes	&	PM	&	10.1021/acs.chemmater.1c00540	\\
Eu(Ti$_{0.875}$Nb$_{0.125}$)$_3$	&	$P4mm$	&	P	&	yes	&	E	&	EPM	&	10.48550/arXiv.2203.10646	\\
Bi$_5$Mn$_5$O$_{17}$	&	$Pmn2_1/Pm2_1n$	&	P	&	yes	&	A	&	AFM	&	10.1038/s41467-020-18664-6	\\
2D In	&	$3m$	&		&		&		&	PM	&	10.1021/acsnano.1c05944	\\
2D Ga	&	$3m$	&		&		&		&	PM	&	10.1021/acsnano.1c05944	\\
2D In/Ga	&	$3m$	&		&		&		&	PM	&	10.1021/acsnano.1c05944	\\
(Fe$_{0.5}$Co$_{0.5}$)$_5$GeTe$_2$	&	$P6_3mc$	&		&		&	yes	&	PM	&	 10.1126/sciadv.abm7103	\\ 
GdCaMnNiO$_6$	&	$P2_1$	&	P	&	yes	&		&	PM	&	10.2139/ssrn.4147202	\\
LaCaMnNiO$_6$	&	$P2_1$	&	P	&	yes	&		&	PM	&	10.2139/ssrn.4147202	\\
SmCaMnNiO$_6$	&	$P2_1$	&	P	&	yes	&		&	PM	&	10.2139/ssrn.4147202	\\
TmCaMnNiO$_6$	&	$P2_1$	&	P	&	yes	&		&	PM	&	10.2139/ssrn.4147202	\\

\end{longtable*}
\endgroup

\textit{Extrinsic Polar Metal}---%
An extrinsic polar metal (EPM) must meet each of the structural and electronic criteria of a polar metal (i.e., polar structure, metallic optical conductivity, robust polar distortion in the presence of perturbations to the Fermi level). However, a polar metal is deemed extrinsic if the metallic electron transport is a result of perturbations to the pristine state of the material. The number of charge carriers or conductivity of an extrinsic polar metal is determined by the doping mechanisms, such as chemical substitution, interstitials, vacancies, photodoping, and electrostatic gating. This distinction has consequences for application contexts in which sensitivity to small changes in chemical potential either positively or detrimentally impacts performance. In any case, the broken inversion symmetry in an extrinsic polar metal persists over the doping ranges explored.  Examples include Nb-doped PbTiO$_3$ \cite{He/Jin:2016PRB} and doped SrTiO$_3$ \cite{ahadi2019enhancing, schooley1964superconductivity}.

\textit{Interfacial Polar Metal}---%
When the two fundamental criteria for a polar metal exist only at the interface between two compounds, we define the resulting composite material an interfacial polar metal (IPM). Interfacial polar metals exhibit some similarity to both anisotropic ferroelectric metals and extrinsic polar metals, but can be fundamentally distinguished from both. 
    Whereas anisotropic ferroelectric metals often exhibit low-dimensional electronic structures distinct from bulk polar metals, interfacial polar metals emerge out of the heterojunction created by interfacing two materials which in the bulk are either non-metallic, non-polar, or both. 
    In this sense, either the conductivity, the broken inversion symmetry, or both should be exclusively limited to the interface. In addition, whereas anisotropic ferroelectric metals must be switchable, interfacial polar metals may or may not be switchable.
    
    When the interfacial polar metal forms at the interface of two doped semiconductors, it may be considered a subcategory of extrinsic polar metals. However, if no doping is required or if one of the two materials is a bulk conductor, then the conductivity of the interface is no longer considered to be extrinsically derived. The most well-known interfacial polar metals are grown as perovskite superlattices, e.g., BaTiO$_3$/SrTiO$_3$/LaTiO$_3$ \cite{Cao2018}, LaAlO$_3$/BaSr$_{0.8}$TiO$_3$/SrTiO$_3$ \cite{Zhou2019}, and doped LaFeO$_3$/YFeO$_3$ \cite{Zhao_Iniguez2018:PRB}.
    
\textit{Degenerately Doped Ferroelectric}---%
A degenerately doped ferroelectric (DDF) meets the same structural criteria as the other categories, but the structural transition is detrimentally impacted by perturbations to the Fermi level from a pristine insulating state. In such a material, the polar order and the conduction mechanism are contraindicated, which both implies a fundamentally different relationship between the polar structure and the conduction mechanism than in the above categories.
These differences will affect how  degenerately doped ferroelectrics are used in devices, e.g., some are proposed for thermoelectric applications \cite{Lee2010,Lee2012,Banik2019,Dangi2020}, whereas the higher carrier density in polar metals would make them less well suited for these energy devices. Furthermore, the pristine undoped material is a good dielectric that exhibits a spontaneous and switchable polarization. Nonetheless, if degenerately doping a ferroelectric is seen as a pathway to accessing the many desirable properties of a polar metal, it should be performed in a context where careful control of the chemical potential is possible and practical. The conductivity threshold required to be deemed ``degenerately doped" rather than ``metallic" is somewhat ambiguous, but a minimum criteria should be $\sigma\neq0$ as $\omega\rightarrow0$. The primary example of a degenerately doped ferroelectric is doped BaTiO$_3$ \cite{Hwang2010, Hickox-Young_Rondinelli2020:PRB,Kolodiazhnyi:2010PRL}. 
    

\subsection{Classification Caveats}
Despite the consideration put into the above classification scheme, there remain materials which prove difficult to categorize into a single class. The majority of such cases involve materials which undergo significant changes in conductivity or displacive mode dynamics as a result of changes in their environment. In most cases, it is simplest to describe a material as moving between categories as a function of some perturbation. For example, a polar material exhibiting a metal-insulator transition would transition from a polar metal to  ferroelectric \cite{Puggioni2015,PhysRevLett.115.106401,Zhang_pssr.201900436}. 

Another unique case is found in materials which exhibit a polar displacement that is \emph{enhanced} as a consequence of doping an initially insulating material, as predicted in Sr$_3$Sn$_2$O$_7$ and various binary metal oxides \cite{li2021free, Cao2022arxiv}. Strictly speaking, assuming doping occurs at such level so as to pass the Drude criterion for conductivity, it would be tempting to categorize Sr$_3$Sn$_2$O$_7$ as a degenerately doped ferroelectric, since the polar distortion is sensitive to changes at the Fermi level and the pristine state is ferroelectric. However, since polar order and conductivity are clearly not fundamentally contraindicated in this compound, it may be wise to reclassify as an extrinsic polar metal. Such a classification would be made easier if the polar mode amplitude plateaus under doping to reach a steady state, relatively insensitive to changes at the Fermi level. It is also worth noting that the material has not yet been synthesized and characterized in experiment and we may discover that the real material behaves differently than anticipated.

Finally, although we referenced 2D polar metals in the ``Anisotropic Ferroelectric Metals" category, it should be noted that a low-dimensional polar metal may have applications beyond switchability under external electric field. For example, 2D polar metallic In and Ga thin films exhibit remarkable second harmonic generation due to their unique bonding environment and fine control of the crystal lattice \cite{Nisi2020,Steves2020}. These materials cannot be described as ferroelectrics, and would be considered simply polar metals under our current scheme, although the strong coupling between thin film morphology and crystal structure implies a similarity to ``Interfacial Polar Metals" and may warrant a separate category entirely.

\section{Outlook}

The outlook for polar metals research has never held more possibility. The number of publications in the research space continues to climb and we have made great strides in determining the necessary relationship between electronic and crystallographic structure to enable their coexistence. Although there remain some unanswered questions (e.g. is `weak-coupling' a necessary condition or merely
a useful design principle?), there are also opportunities to significantly broaden the polar metallic design space:
\begin{itemize}
    \item Polar metals requiring new subclass designations may combine multiple orders or enhanced functions, e.g., magnetism and strong electron-electron interactions can enable magnetochiral anisotropy with  nonreciprocal electrical transport. 
    \item Polar metals violating the weak-coupling principle would broaden the materials landscape and bring control routes to physical properties using non-conjugate fields.
\end{itemize}

Beginning with magnetic polar metals, there are many known acentric conductors with magnetic ordering temperature (across all of the above categories), including the following: en-CoS \cite{Zheng2021}, Pb$_2$CoOsO$_6$ \cite{Jiao2020}, Pb$_2$NiOsO$_6$ \cite{Feng2021}, tri-color superlattices BaTiO$_3$/SrTiO$_3$/LaTiO$_3$ \cite{Cao2018}, Ca$_3$Ru$_2$O$_7$ \cite{Yoshida2005}, and more. The relationship between magnetic ordering in these materials and their designation as polar metals varies. 
For example,  Jiao et al.\ describe separating polar metals with magnetic ordering temperatures into ``type-I'' and ``type-II'' categories depending on the strength of the coupling between magnetic and polar orders, with Pb$_2$CoOsO$_6$ being an example of a ``type-II'' magnetic polar metal \cite{Jiao2020}. By contrast, the magnetic ordering predicted in SrCaRu$_2$O$_6$ is not expected to couple to the inversion-lifting distortion \cite{Puggioni/Rondinelli:2014NatComm}, or one might note the insensitivity of the polar transition in doped SrTiO$_3$ to the presence of magnetic doping and isovalent substitution as an example of ``type-I" behavior \cite{SalmaniRezaie2021}. 
Still, we may describe other categories, including magnetically-driven metal-insulator transitions in Pb$_2$CaOsO$_6$ \cite{Jacobsen2020} and B-site substituted Ca$_3$Ru$_2$O$_7$\cite{Zhu2017,Lei2019:phasediagram,Tsuda2013} (in which case the magnetic and transport properties are highly sensitive to dilute concentrations and dependent on the substituting species---Mn, Ti, or Fe).  Such sub-categorization according to magnetic properties helps describe fundamentally different physical phenomena, but may be well-suited to more detailed investigations which may coexist with the classification scheme presented above. 

The `weak-coupling hypothesis' states that in order for broken inversion and metallic conductivity to coexist there must be limited coupling between electrons at the Fermi level and the phonon(s) responsible for driving the symmetry-breaking transition  \cite{Puggioni/Rondinelli:2014NatComm}.
If polar metals exhibiting strong coupling between the Fermi level and the structural distortion were to be discovered, they may deserve separate categorization as well. Although TiGaO$_3$ has been predicted as a compound that might violate the weak-coupling hypothesis \cite{Xiang:2014PRB}, it be would hasty to separately categorize such a material prior to experimental synthesis and characterization. We recently computed the decomposition enthalpy of TiGaO$_3$ to be about -1\,eV/f.u., making it unlikely to be realized in experiment. Studying a series of related ABO$_3$ compounds (A\,=\,Ti,\,Zr; B\,=\,Al,\,Ga,\,In) revealed that while each exhibits a dynamically stable polar phase driven by displacements of the A-site cation (which also contributes to the Fermi level), none have a decomposition enthalpy smaller in magnitude than -0.8\,eV/f.u. As was the case with Sr$_3$Sn$_2$O$_7$, it seems prudent to primarily concern ourselves with the classification of known compounds, as the synthesis of exotic predicted compounds may reveal features that aid in their classification. Alternatively, we may yet discover factors that prevent the synthesis of acentric conductors that challenge our classification scheme. In the mean time, it is incumbent on theoretical and computational materials scientists to examine the stability of their predicted compounds before sharing their results \cite{Malyi2020}.
At the same time, the high number of superconducting polar metals (\autoref{tab:compiled_NCSMs}) implies that perhaps the description of weak coupling between crystallographic and electronic orders in polar metals is not entirely accurate. There is certainly evidence of a coupled relationship in degenerately doped superconducting ferroelectrics SrTiO$_3$ \cite{ahadi2019enhancing, schooley1964superconductivity} and BaTiO$_3$ \cite{Ma2021}. The apparent connection between superconductivity and polar metals (and the associated implications for coupling between the two orders) warrants further investigation.

Accompanying the advent of 2D materials has been a recent interest in the emergent phenomena that occur when two sheets of material are stacked on top of one another; Moiré  heterostructures like those in the famous “magic angle” graphene are being explored in a variety of van der Waals materials \cite{Kennes2021,Cao2018,Susner2017,Mak2019,Gong2019}. 
However, work thus far has exclusively focused on small-bandgap semiconductors and semimetals. In addition, “twisting” typically occurs mechanically and continuously. Concurrently, in the past year we have seen a significant increase in the number of predicted and synthesized two-dimensional polar metals \cite{Fei2018,Filippetti2016,Xu2020,ye2019observation,Luo2017,Lu2019}. We see an opportunity to significantly broaden the phase space of both topologically non-trivial materials and acentric conductors by considering two dimensional metals and scalable, discrete stacking mechanisms. 

Metallic correlated transition metal dichalcogenides and halides have yet to be investigated in Moiré  heterostructures. Noting that interlayer twisting at most angles breaks inversion symmetry, this presents a pathway for a new class of noncentrosymmetric metals, which may facilitate the generation of tunable charge density waves as well as topological conduction mechanisms and create exotic opportunities to support new physics. However, the number of metallic 2D transition metal dichalcogenides is limited, motivating an interest in looking beyond van der Waals stacking (which requires mechanical stacking amidst a continuum of possible twist angles) toward hybrid organic-inorganic materials that can be directly assembled \cite{Bostrm2018,Aubrey2021}.

As evidenced by \autoref{tab:compiled_NCSMs}, there are a great many predicted compounds; experimental verification of predicted polar metallic dynamics is a pressing need. In addition, the search for compounds that violate weak-coupling continues to be of interest and may require expanding the field into new structure families. Finally, the properties and applications of polar metals are not yet fully appreciated. Recent work on topological phonons indicates that polar metals are ideal candidates to host this recently described phenomenon \cite{Peng2020}, while the potential for polar metals to impact the fields of photovoltaics \cite{D1TC02213J}, energy harvesting \cite{doi:10.1021/acs.jpclett.8b03654, doi:10.1021/acsnano.7b02756}, catalysis \cite{Mann_2022},   microelectronic \cite{Puggioni_electrodes, puggioni_rondinelli_2019_pat,C9NR05404A}, optoelectronic, photonic and plasmonic devices \cite{plasmonics1, plasmonics2,hotelectron,photocat,Juan2009,Juan:2011,Ndukaife:2016,Nisi2020,Steves2020}, and quantum information systems \cite{Basov2017}  remains largely untapped.

\begin{acknowledgments}

This work was supported by the Army Research Office (ARO) under Grant No.\ W911NF-15-1-0017. The authors thank their numerous colleagues and collaborators for enlightening discussions on polar metals and sharing their laboratory discoveries with us over the last decade.

\end{acknowledgments}

\appendix
\renewcommand{\thefigure}{A\arabic{figure}}
\renewcommand{\thetable}{A\arabic{table}}
\setcounter{figure}{0}
\setcounter{table}{0}

\section{Assumptions of the background-charge approximation}\label{sec:A}

The background charge approach has become the de facto tool for use in studying the effects of electrostatic doping on solid state materials \cite{Zhao_Iniguez2018:PRB,Wang/Tsymbal:2012PRL,Hickox-Young_Rondinelli2020:PRB,Bruneval2015,Michel2021}. 
This method works by changing the number of electrons present in an  electronic structure  simulation, while providing an additional homogeneous background charge to maintain charge-balance.
The background charge approximation has become popular in doping simulations because it allows one to use smaller unit cells than in substitutional or vacancy-doping simulations and because it allows one to separate the effects of changes in electronic structure from the distortions which accompany real chemical dopants.
This is an attractive combination of attributes when studying the interplay between free charge carriers and the polar distortion active in ferroelectrics.

The background-charge method is frequently used both to investigate fundamental relationships and to design  polar metals \cite{Zhao_Iniguez2018:PRB,Wang/Tsymbal:2012PRL,Michel2021,He/Jin:2016PRB}.
This approximation, however, has consequences which do not always translate to physical systems. These include, but are not limited to:
(i) Changes in volume; 
(ii) Abrupt changes in conductivity; and 
(iii) Homogeneous changes to electronic and crystallographic structure.
%
Some of these problems are linked to the use of periodic density functional theory (DFT) as a simulation method in general, e.g., (3) can be challenging to avoid while using periodic boundary conditions, and all have been previously observed, but their shortcomings have been amplified by widespread application of the background charge technique to doped FEs.
Despite its limitations, the background-charge approximation remains an extremely valuable method of exploring the effects of doping on electronic and crystallographic structure. 
%
The consequences of using this approximation instead of more realistic substitution or vacancy-based doping simulations, however, should be acknowledged to avoid arriving at inappropriate conclusions about the effect of doping in semiconductors and in doped ferroelectrics in particular. 

\subsection{Changes in volume}

The homogeneous background charge introduced in electrostatic doping simulations produces unintended consequences for the volume of the semiconductor being studied.
The effective `pressure' resulting from the additional electrons and associated background is ill-defined, resulting in inconsistent consequences depending on the choice of DFT software \cite{Bruneval2015}. 
It follows naturally that this arbitrary treatment of volume leads to results that deviate significantly from experiment. For example, Iwazaki et al.\ showed that volume increase due to electrostatic doping simulations via the background charge method in BaTiO$_3$ dramatically overpredicted volume changes as compared to experimental substitutional doping \cite{Iwazaki2012}.
Volume discrepancies dramatically impact properties, including the stability of the crystal structure, as volume is closely tied to the strength of interatomic forces and hence phonon frequencies.  Simulations of soft phonon modes in BaTiO$_3$ under electrostatic doping with and without volume relaxation produce significantly different critical doping thresholds \cite{Hickox-Young_Rondinelli2020:PRB}. 
If the volume difference between polar and non-polar phases is substantial, then the volume changes induced by this method will intrinsically favor high or low symmetry structures depending on the relative volume \cite{Zhao_Iniguez2018:PRB,Iwazaki2012}, obfuscating the changes induced by doping.
Therefore, it is recommended that electrostatic doping via the compensating background charge approximation be performed with fixed volume. Volume relaxation may only be used when simulating substitutional, interstitial, or vacancy-based doping. Even in those cases, the dopant concentration should be considered  when performing volume relaxations.

\subsection{Abrupt changes in conductivity}
In the pristine case, ferroelectric insulators are easily differentiated from metals, both computationally (i.e.\ gapped band structure) and experimentally (optical conductivity approaches zero at low frequency).
Next, the addition of charge carries can be achieved through chemical means (dopants) or electrostatic gating. 
Upon chemical substitution, defect states are introduced into the electronic structure of a semiconductor, allowing for modest conductivity due to thermally excited charge carriers occupying the defect states. This may lead to the material exhibiting metal-like temperature-dependent resistivity (i.e.\ positive slope) away from 0 K, but as temperature is reduced the thermally activated charge carriers are eventually frozen out and resistivity increases. This is reflected in the optical conductivity ($\sigma_0$), which still heads to 0 at low frequency (\autoref{fig:sigma}a).

\begin{figure}
  \centering
  \includegraphics[width=0.4\textwidth]{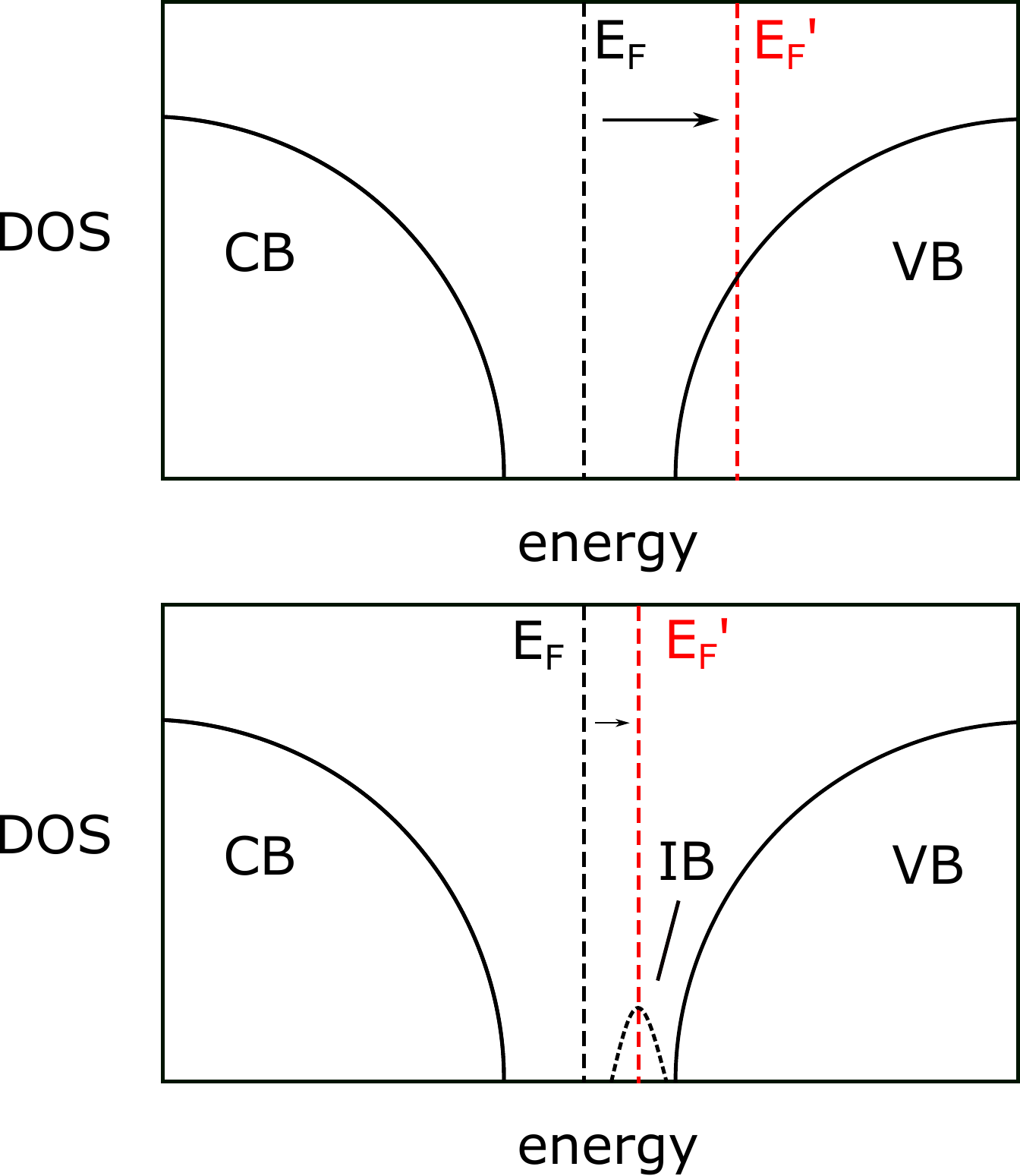}
  \caption{Schematic illustration of the effect of doping on the electronic structure under the background-charge technique (top) and under more realistic substitutional doping mechanisms involving the formation of an impurity band (bottom) }
  \label{fig:dos_doping}
\end{figure}

Upon additional doping, an impurity band may form, transforming a degenerately doped semiconductor into a bad metal with low mobility from impurity scattering \cite{Serre1983,Efros1979}, as shown schematically in \autoref{fig:dos_doping}.
In this case, the material will exhibit exclusively metallic characteristics, as the impurity band produces a finite $\sigma_0$ as $\omega\rightarrow0$, 
which can lead to the apparent closure of the gap with sufficient doping as low frequency spectra weight appears (\autoref{fig:sigma}a). 

This gradual evolution in conductivity under doping, accompanied by the formation of an impurity band, is not well-described using the background charge method to approximate doping.
In real semiconductors, doping shifts the Fermi level up or down within the band gap, but often not to a degree sufficient to push the Fermi level into either the valence or conduction band (\autoref{fig:dos_doping}). 
With the background charge method, even a highly dilute concentration of additional electrons (or holes) is sufficient to place the Fermi level within a band, making the system metallic far below even the electron concentration given by the Mott criterion ($n_c^{1/3} a_0 \approx 0.25$) for free electrons.
Furthermore, the undoped band structure remains more or less intact, precluding the possibility of changes to the electronic structure due to impurity bands or polaron formation (\autoref{fig:dos_doping}).
This simplification obscures the complex reality of changes to the electronic structure that accompany chemical dopants.

These changes are highly dependent on choice of starting material and doping method (substitution, vacancy, electrostatic, etc.).
The critical charge carrier concentration ($n^*$) for conductivity in BaTiO$_3$ alone varies considerably across several doping mechanisms \cite{Raghavan2016,Fujioka2015,Kolodiazhnyi:2010PRL}, all of which are significantly higher than $n^*$ for SrTiO$_3$ \cite{Page2008}.
The background charge method is not able to account for these differences and therefore may lead to misleading conclusions about the ease of altering the conductivity of a material. It may come closest to approximating electrostatic gating, as this field-driven method typically does not generate an impurity band.
However, it still ignores charge localization mechanisms, as will be discussed in the next section.

\subsection{Homogeneous changes to crystallographic and electronic structure}

Below the impurity band formation limit, or under electrostatic gating, the carriers will either be homogeneously or inhomogeneously distributed at the nanoscale. 
At present, nearly all DFT calculations make the assumption that carriers are homogeneously distributed, and by construction the simulations result in a metallic electronic structure. 
Whether or not this model accurately captures the experimental situation requires attention, as recent extensions of semiconductor doping principles to transition metal compounds show that homogeneous distributions are often the exception rather than the rule \cite{Walsh2017}.
For the inhomogeneous case, we anticipate nanoscale phase segregation to occur in close analogy to what happens across metal-insulator phase transitions in complex oxides, e.g., VO$_2$ \cite{Marcelli2017}.
Here doping leads to domains that are either metallic or insulating and local structures that are centric and noncentrosymmetric, respectively. 
In other words, the regions that are metallic recover inversion while inversion symmetry remains lifted in the insulating regions. 
This electronic- and parity-symmetry phase separation is supported by recent studies on BaTiO$_3$ and SrTiO$_3$ \cite{Page2008,Raghavan2016}.
Indeed, BaTiO$_3$ is much more difficult to make metallic than SrTiO$_3$ \cite{Page2008} with the key difference being the lattice dynamical properties; BaTiO$_3$ is a soft-mode semiconducting ferroelectric whereas SrTiO$_3$ is an incipient ferroelectric with a highly dilute superconducting transition. 
In practice this means that long range order quickly breaks down in doped BaTiO$_3$, as the driving force for local off-centering persists leading to a network of disordered insulating octahedra separated by centrosymmetric metallic clusters hosting pseudo-localized free charge carriers.
By contrast, under similar doping conditions, the more regular octahedra of SrTiO$_3$ enter a metallic---and subsequently superconducting---state. SrTiO$_3$ is not immune to nanodomain formation \cite{SalmaniRezaie2020}, but charge does not localize as easily. 
This highlights the fact that the experimental structure may exhibit electronic structure heterogeneities in real-space with regard to either the electronic structure and/or crystal structure.

Meanwhile, nearly all DFT calculations assume homogeneous doping. This is unavoidable to a certain extent, due to the prohibitive cost of simulating the large supercells necessary to capture nanodomain structures on the mesoscale. 
However, the background charge method further enforces uniformity, as the added charge carriers are distributed among the bands closest to the Fermi level while a homogeneous compensating background charge is added for charge balance. 
This makes it very challenging to simulate localization or polaron formation, let alone account for multiple phases.

\bibliography{review}

\providecommand{\noopsort}[1]{}\providecommand{\singleletter}[1]{#1}%
\begin{thebibliography}{144}%
\makeatletter
\providecommand \@ifxundefined [1]{%
 \@ifx{#1\undefined}
}%
\providecommand \@ifnum [1]{%
 \ifnum #1\expandafter \@firstoftwo
 \else \expandafter \@secondoftwo
 \fi
}%
\providecommand \@ifx [1]{%
 \ifx #1\expandafter \@firstoftwo
 \else \expandafter \@secondoftwo
 \fi
}%
\providecommand \natexlab [1]{#1}%
\providecommand \enquote  [1]{``#1''}%
\providecommand \bibnamefont  [1]{#1}%
\providecommand \bibfnamefont [1]{#1}%
\providecommand \citenamefont [1]{#1}%
\providecommand \href@noop [0]{\@secondoftwo}%
\providecommand \href [0]{\begingroup \@sanitize@url \@href}%
\providecommand \@href[1]{\@@startlink{#1}\@@href}%
\providecommand \@@href[1]{\endgroup#1\@@endlink}%
\providecommand \@sanitize@url [0]{\catcode `\\12\catcode `\$12\catcode
  `\&12\catcode `\#12\catcode `\^12\catcode `\_12\catcode `\%12\relax}%
\providecommand \@@startlink[1]{}%
\providecommand \@@endlink[0]{}%
\providecommand \url  [0]{\begingroup\@sanitize@url \@url }%
\providecommand \@url [1]{\endgroup\@href {#1}{\urlprefix }}%
\providecommand \urlprefix  [0]{URL }%
\providecommand \Eprint [0]{\href }%
\providecommand \doibase [0]{https://doi.org/}%
\providecommand \selectlanguage [0]{\@gobble}%
\providecommand \bibinfo  [0]{\@secondoftwo}%
\providecommand \bibfield  [0]{\@secondoftwo}%
\providecommand \translation [1]{[#1]}%
\providecommand \BibitemOpen [0]{}%
\providecommand \bibitemStop [0]{}%
\providecommand \bibitemNoStop [0]{.\EOS\space}%
\providecommand \EOS [0]{\spacefactor3000\relax}%
\providecommand \BibitemShut  [1]{\csname bibitem#1\endcsname}%
\let\auto@bib@innerbib\@empty
\bibitem [{\citenamefont {Anderson}\ and\ \citenamefont
  {Blount}(1965)}]{Anderson/Blount:1965PRL}%
  \BibitemOpen
  \bibfield  {author} {\bibinfo {author} {\bibfnamefont {P.~W.}\ \bibnamefont
  {Anderson}}\ and\ \bibinfo {author} {\bibfnamefont {E.~I.}\ \bibnamefont
  {Blount}},\ }\bibfield  {title} {\bibinfo {title} {Symmetry considerations on
  martensitic transformations: `ferroelectric' metals?},\ }\href
  {https://doi.org/10.1103/physrevlett.14.217} {\bibfield  {journal} {\bibinfo
  {journal} {Physical Review Letters}\ }\textbf {\bibinfo {volume} {14}},\
  \bibinfo {pages} {217} (\bibinfo {year} {1965})}\BibitemShut {NoStop}%
\bibitem [{\citenamefont {Salje}(2012)}]{salje:2012}%
  \BibitemOpen
  \bibfield  {author} {\bibinfo {author} {\bibfnamefont {E.~K.}\ \bibnamefont
  {Salje}},\ }\bibfield  {title} {\bibinfo {title} {Ferroelastic materials},\
  }\href {https://doi.org/10.1146/annurev-matsci-070511-155022} {\bibfield
  {journal} {\bibinfo  {journal} {Annual Review of Materials Research}\
  }\textbf {\bibinfo {volume} {42}},\ \bibinfo {pages} {265} (\bibinfo {year}
  {2012})}\BibitemShut {NoStop}%
\bibitem [{\citenamefont {de~Lagrange}(1773)}]{de1773attraction}%
  \BibitemOpen
  \bibfield  {author} {\bibinfo {author} {\bibfnamefont {J.~L.}\ \bibnamefont
  {de~Lagrange}},\ }\bibfield  {title} {\bibinfo {title} {Sur l’attraction
  des sph{\'e}ro{\"\i}des elliptiques},\ }\href@noop {} {\bibfield  {journal}
  {\bibinfo  {journal} {Nouv. M{\'e}m. Acad. royale Berlin ann{\'e}e}\ ,\
  \bibinfo {pages} {619}} (\bibinfo {year} {1773})}\BibitemShut {NoStop}%
\bibitem [{\citenamefont {Gauss}(1877)}]{Gauss1877}%
  \BibitemOpen
  \bibfield  {author} {\bibinfo {author} {\bibfnamefont {C.~F.}\ \bibnamefont
  {Gauss}},\ }\bibfield  {title} {\bibinfo {title} {Theoria attractionis
  corporum sphaeroidicorum ellipticorum homogeneorum},\ }in\ \href
  {https://doi.org/10.1007/978-3-642-49319-5_1} {\emph {\bibinfo {booktitle}
  {Werke}}}\ (\bibinfo  {publisher} {Springer Berlin Heidelberg},\ \bibinfo
  {year} {1877})\ pp.\ \bibinfo {pages} {3--22}\BibitemShut {NoStop}%
\bibitem [{\citenamefont {Cochran}(1959)}]{Cochran1959}%
  \BibitemOpen
  \bibfield  {author} {\bibinfo {author} {\bibfnamefont {W.}~\bibnamefont
  {Cochran}},\ }\bibfield  {title} {\bibinfo {title} {Crystal stability and the
  theory of ferroelectricity},\ }\href
  {https://doi.org/10.1103/physrevlett.3.412} {\bibfield  {journal} {\bibinfo
  {journal} {Physical Review Letters}\ }\textbf {\bibinfo {volume} {3}},\
  \bibinfo {pages} {412} (\bibinfo {year} {1959})}\BibitemShut {NoStop}%
\bibitem [{\citenamefont {Cohen}(1992)}]{Cohen:1992Nature}%
  \BibitemOpen
  \bibfield  {author} {\bibinfo {author} {\bibfnamefont {R.~E.}\ \bibnamefont
  {Cohen}},\ }\bibfield  {title} {\bibinfo {title} {Origin of ferroelectricity
  in perovskite oxides},\ }\href {https://doi.org/10.1038/358136a0} {\bibfield
  {journal} {\bibinfo  {journal} {Nature}\ }\textbf {\bibinfo {volume} {358}},\
  \bibinfo {pages} {136} (\bibinfo {year} {1992})}\BibitemShut {NoStop}%
\bibitem [{\citenamefont {Shi}\ \emph {et~al.}(2013)\citenamefont {Shi},
  \citenamefont {Guo}, \citenamefont {Wang}, \citenamefont {Princep},
  \citenamefont {Khalyavin}, \citenamefont {Manuel}, \citenamefont {Michiue},
  \citenamefont {Sato}, \citenamefont {Tsuda}, \citenamefont {Yu},
  \citenamefont {Arai}, \citenamefont {Shirako}, \citenamefont {Akaogi},
  \citenamefont {Wang}, \citenamefont {Yamaura},\ and\ \citenamefont
  {Boothroyd}}]{Shi:2013Nature}%
  \BibitemOpen
  \bibfield  {author} {\bibinfo {author} {\bibfnamefont {Y.}~\bibnamefont
  {Shi}}, \bibinfo {author} {\bibfnamefont {Y.}~\bibnamefont {Guo}}, \bibinfo
  {author} {\bibfnamefont {X.}~\bibnamefont {Wang}}, \bibinfo {author}
  {\bibfnamefont {A.~J.}\ \bibnamefont {Princep}}, \bibinfo {author}
  {\bibfnamefont {D.}~\bibnamefont {Khalyavin}}, \bibinfo {author}
  {\bibfnamefont {P.}~\bibnamefont {Manuel}}, \bibinfo {author} {\bibfnamefont
  {Y.}~\bibnamefont {Michiue}}, \bibinfo {author} {\bibfnamefont
  {A.}~\bibnamefont {Sato}}, \bibinfo {author} {\bibfnamefont {K.}~\bibnamefont
  {Tsuda}}, \bibinfo {author} {\bibfnamefont {S.}~\bibnamefont {Yu}}, \bibinfo
  {author} {\bibfnamefont {M.}~\bibnamefont {Arai}}, \bibinfo {author}
  {\bibfnamefont {Y.}~\bibnamefont {Shirako}}, \bibinfo {author} {\bibfnamefont
  {M.}~\bibnamefont {Akaogi}}, \bibinfo {author} {\bibfnamefont
  {N.}~\bibnamefont {Wang}}, \bibinfo {author} {\bibfnamefont {K.}~\bibnamefont
  {Yamaura}},\ and\ \bibinfo {author} {\bibfnamefont {A.~T.}\ \bibnamefont
  {Boothroyd}},\ }\bibfield  {title} {\bibinfo {title} {A ferroelectric-like
  structural transition in a metal},\ }\href {https://doi.org/10.1038/nmat3754}
  {\bibfield  {journal} {\bibinfo  {journal} {Nature Materials}\ }\textbf
  {\bibinfo {volume} {12}},\ \bibinfo {pages} {1024} (\bibinfo {year}
  {2013})}\BibitemShut {NoStop}%
\bibitem [{\citenamefont {Puggioni}\ and\ \citenamefont
  {Rondinelli}(2014)}]{Puggioni/Rondinelli:2014NatComm}%
  \BibitemOpen
  \bibfield  {author} {\bibinfo {author} {\bibfnamefont {D.}~\bibnamefont
  {Puggioni}}\ and\ \bibinfo {author} {\bibfnamefont {J.~M.}\ \bibnamefont
  {Rondinelli}},\ }\bibfield  {title} {\bibinfo {title} {Designing a robustly
  metallic noncenstrosymmetric ruthenate oxide with large thermopower
  anisotropy},\ }\href {https://doi.org/10.1038/ncomms4432} {\bibfield
  {journal} {\bibinfo  {journal} {Nature Communications}\ }\textbf {\bibinfo
  {volume} {5}},\ \bibinfo {pages} {3432} (\bibinfo {year} {2014})}\BibitemShut
  {NoStop}%
\bibitem [{\citenamefont {Varjas}\ \emph {et~al.}(2016)\citenamefont {Varjas},
  \citenamefont {Grushin}, \citenamefont {Ilan},\ and\ \citenamefont
  {Moore}}]{Varjas:2016PRL}%
  \BibitemOpen
  \bibfield  {author} {\bibinfo {author} {\bibfnamefont {D.}~\bibnamefont
  {Varjas}}, \bibinfo {author} {\bibfnamefont {A.~G.}\ \bibnamefont {Grushin}},
  \bibinfo {author} {\bibfnamefont {R.}~\bibnamefont {Ilan}},\ and\ \bibinfo
  {author} {\bibfnamefont {J.~E.}\ \bibnamefont {Moore}},\ }\bibfield  {title}
  {\bibinfo {title} {Dynamical piezoelectric and magnetopiezoelectric effects
  in polar metals from {Berry} phases and orbital moments},\ }\href
  {https://doi.org/10.1103/PhysRevLett.117.257601} {\bibfield  {journal}
  {\bibinfo  {journal} {Physical Review Letters}\ }\textbf {\bibinfo {volume}
  {117}},\ \bibinfo {pages} {257601} (\bibinfo {year} {2016})},\ \Eprint
  {https://arxiv.org/abs/1607.05278} {arXiv:1607.05278} \BibitemShut {NoStop}%
\bibitem [{\citenamefont {Wu}\ \emph {et~al.}(2016)\citenamefont {Wu},
  \citenamefont {Patankar}, \citenamefont {Morimoto}, \citenamefont {Nair},
  \citenamefont {Thewalt}, \citenamefont {Little}, \citenamefont {Analytis},
  \citenamefont {Moore},\ and\ \citenamefont {Orenstein}}]{Wu2016}%
  \BibitemOpen
  \bibfield  {author} {\bibinfo {author} {\bibfnamefont {L.}~\bibnamefont
  {Wu}}, \bibinfo {author} {\bibfnamefont {S.}~\bibnamefont {Patankar}},
  \bibinfo {author} {\bibfnamefont {T.}~\bibnamefont {Morimoto}}, \bibinfo
  {author} {\bibfnamefont {N.~L.}\ \bibnamefont {Nair}}, \bibinfo {author}
  {\bibfnamefont {E.}~\bibnamefont {Thewalt}}, \bibinfo {author} {\bibfnamefont
  {A.}~\bibnamefont {Little}}, \bibinfo {author} {\bibfnamefont {J.~G.}\
  \bibnamefont {Analytis}}, \bibinfo {author} {\bibfnamefont {J.~E.}\
  \bibnamefont {Moore}},\ and\ \bibinfo {author} {\bibfnamefont
  {J.}~\bibnamefont {Orenstein}},\ }\bibfield  {title} {\bibinfo {title} {Giant
  anisotropic nonlinear optical response in transition metal monopnictide
  {Weyl} semimetals},\ }\href {https://doi.org/10.1038/nphys3969} {\bibfield
  {journal} {\bibinfo  {journal} {Nature Physics}\ }\textbf {\bibinfo {volume}
  {13}},\ \bibinfo {pages} {350} (\bibinfo {year} {2016})}\BibitemShut
  {NoStop}%
\bibitem [{\citenamefont {Gao}\ \emph {et~al.}(2018)\citenamefont {Gao},
  \citenamefont {Kim}, \citenamefont {Venderbos}, \citenamefont {Kane},
  \citenamefont {Mele}, \citenamefont {Rappe},\ and\ \citenamefont
  {Ren}}]{Gao2018}%
  \BibitemOpen
  \bibfield  {author} {\bibinfo {author} {\bibfnamefont {H.}~\bibnamefont
  {Gao}}, \bibinfo {author} {\bibfnamefont {Y.}~\bibnamefont {Kim}}, \bibinfo
  {author} {\bibfnamefont {J.~W.}\ \bibnamefont {Venderbos}}, \bibinfo {author}
  {\bibfnamefont {C.}~\bibnamefont {Kane}}, \bibinfo {author} {\bibfnamefont
  {E.}~\bibnamefont {Mele}}, \bibinfo {author} {\bibfnamefont {A.~M.}\
  \bibnamefont {Rappe}},\ and\ \bibinfo {author} {\bibfnamefont
  {W.}~\bibnamefont {Ren}},\ }\bibfield  {title} {\bibinfo {title}
  {Dirac-{Weyl} semimetal: Coexistence of {Dirac} and {Weyl} fermions in polar
  hexagonal {ABC} crystals},\ }\bibfield  {journal} {\bibinfo  {journal}
  {Physical Review Letters}\ }\textbf {\bibinfo {volume} {121}},\ \href
  {https://doi.org/10.1103/physrevlett.121.106404}
  {10.1103/physrevlett.121.106404} (\bibinfo {year} {2018})\BibitemShut
  {NoStop}%
\bibitem [{\citenamefont {He}\ and\ \citenamefont {juan
  Jin}(2016)}]{He/Jin:2016PRB}%
  \BibitemOpen
  \bibfield  {author} {\bibinfo {author} {\bibfnamefont {X.}~\bibnamefont
  {He}}\ and\ \bibinfo {author} {\bibfnamefont {K.}~\bibnamefont {juan Jin}},\
  }\bibfield  {title} {\bibinfo {title} {Persistence of polar distortion with
  electron doping in lone-pair driven ferroelectrics},\ }\bibfield  {journal}
  {\bibinfo  {journal} {Physical Review B}\ }\textbf {\bibinfo {volume} {94}},\
  \href {https://doi.org/10.1103/physrevb.94.224107}
  {10.1103/physrevb.94.224107} (\bibinfo {year} {2016})\BibitemShut {NoStop}%
\bibitem [{\citenamefont {Ahadi}\ \emph {et~al.}(2019)\citenamefont {Ahadi},
  \citenamefont {Galletti}, \citenamefont {Li}, \citenamefont {Salmani-Rezaie},
  \citenamefont {Wu},\ and\ \citenamefont {Stemmer}}]{ahadi2019enhancing}%
  \BibitemOpen
  \bibfield  {author} {\bibinfo {author} {\bibfnamefont {K.}~\bibnamefont
  {Ahadi}}, \bibinfo {author} {\bibfnamefont {L.}~\bibnamefont {Galletti}},
  \bibinfo {author} {\bibfnamefont {Y.}~\bibnamefont {Li}}, \bibinfo {author}
  {\bibfnamefont {S.}~\bibnamefont {Salmani-Rezaie}}, \bibinfo {author}
  {\bibfnamefont {W.}~\bibnamefont {Wu}},\ and\ \bibinfo {author}
  {\bibfnamefont {S.}~\bibnamefont {Stemmer}},\ }\bibfield  {title} {\bibinfo
  {title} {Enhancing superconductivity in {SrTiO$_3$} films with strain},\
  }\bibfield  {journal} {\bibinfo  {journal} {Science Advances}\ }\textbf
  {\bibinfo {volume} {5}},\ \href {https://doi.org/10.1126/sciadv.aaw0120}
  {10.1126/sciadv.aaw0120} (\bibinfo {year} {2019})\BibitemShut {NoStop}%
\bibitem [{\citenamefont {Schooley}\ \emph {et~al.}(1964)\citenamefont
  {Schooley}, \citenamefont {Hosler},\ and\ \citenamefont
  {Cohen}}]{schooley1964superconductivity}%
  \BibitemOpen
  \bibfield  {author} {\bibinfo {author} {\bibfnamefont {J.~F.}\ \bibnamefont
  {Schooley}}, \bibinfo {author} {\bibfnamefont {W.~R.}\ \bibnamefont
  {Hosler}},\ and\ \bibinfo {author} {\bibfnamefont {M.~L.}\ \bibnamefont
  {Cohen}},\ }\bibfield  {title} {\bibinfo {title} {Superconductivity in
  semiconducting {SrTiO}$_3$},\ }\href
  {https://doi.org/10.1103/physrevlett.12.474} {\bibfield  {journal} {\bibinfo
  {journal} {Physical Review Letters}\ }\textbf {\bibinfo {volume} {12}},\
  \bibinfo {pages} {474} (\bibinfo {year} {1964})}\BibitemShut {NoStop}%
\bibitem [{\citenamefont {Kolodiazhnyi}\ \emph {et~al.}(2010)\citenamefont
  {Kolodiazhnyi}, \citenamefont {Tachibana}, \citenamefont {Kawaji},
  \citenamefont {Hwang},\ and\ \citenamefont
  {Takayama-Muromachi}}]{Kolodiazhnyi:2010PRL}%
  \BibitemOpen
  \bibfield  {author} {\bibinfo {author} {\bibfnamefont {T.}~\bibnamefont
  {Kolodiazhnyi}}, \bibinfo {author} {\bibfnamefont {M.}~\bibnamefont
  {Tachibana}}, \bibinfo {author} {\bibfnamefont {H.}~\bibnamefont {Kawaji}},
  \bibinfo {author} {\bibfnamefont {J.}~\bibnamefont {Hwang}},\ and\ \bibinfo
  {author} {\bibfnamefont {E.}~\bibnamefont {Takayama-Muromachi}},\ }\bibfield
  {title} {\bibinfo {title} {{Persistence of ferroelectricity in BaTiO$_3$
  through the insulator-metal transition}},\ }\href
  {https://doi.org/10.1103/PhysRevLett.104.147602} {\bibfield  {journal}
  {\bibinfo  {journal} {Physical Review Letters}\ }\textbf {\bibinfo {volume}
  {104}},\ \bibinfo {pages} {147602} (\bibinfo {year} {2010})}\BibitemShut
  {NoStop}%
\bibitem [{\citenamefont {Lei}\ \emph {et~al.}(2018)\citenamefont {Lei},
  \citenamefont {Gu}, \citenamefont {Puggioni}, \citenamefont {Stone},
  \citenamefont {Peng}, \citenamefont {Ge}, \citenamefont {Wang}, \citenamefont
  {Wang}, \citenamefont {Yuan}, \citenamefont {Wang}, \citenamefont {Mao},
  \citenamefont {Rondinelli},\ and\ \citenamefont
  {Gopalan}}]{Lei/Gu/Puggioni:2018NanoLett}%
  \BibitemOpen
  \bibfield  {author} {\bibinfo {author} {\bibfnamefont {S.}~\bibnamefont
  {Lei}}, \bibinfo {author} {\bibfnamefont {M.}~\bibnamefont {Gu}}, \bibinfo
  {author} {\bibfnamefont {D.}~\bibnamefont {Puggioni}}, \bibinfo {author}
  {\bibfnamefont {G.}~\bibnamefont {Stone}}, \bibinfo {author} {\bibfnamefont
  {J.}~\bibnamefont {Peng}}, \bibinfo {author} {\bibfnamefont {J.}~\bibnamefont
  {Ge}}, \bibinfo {author} {\bibfnamefont {Y.}~\bibnamefont {Wang}}, \bibinfo
  {author} {\bibfnamefont {B.}~\bibnamefont {Wang}}, \bibinfo {author}
  {\bibfnamefont {Y.}~\bibnamefont {Yuan}}, \bibinfo {author} {\bibfnamefont
  {K.}~\bibnamefont {Wang}}, \bibinfo {author} {\bibfnamefont {Z.}~\bibnamefont
  {Mao}}, \bibinfo {author} {\bibfnamefont {J.~M.}\ \bibnamefont
  {Rondinelli}},\ and\ \bibinfo {author} {\bibfnamefont {V.}~\bibnamefont
  {Gopalan}},\ }\bibfield  {title} {\bibinfo {title} {Observation of
  quasi-two-dimensional polar domains and ferroelastic switching in a metal,
  {Ca$_3$Ru$_2$O$_7$}},\ }\href {https://doi.org/10.1021/acs.nanolett.8b00633}
  {\bibfield  {journal} {\bibinfo  {journal} {Nano Letters}\ }\textbf {\bibinfo
  {volume} {18}},\ \bibinfo {pages} {3088} (\bibinfo {year}
  {2018})}\BibitemShut {NoStop}%
\bibitem [{\citenamefont {Yoshida}\ \emph {et~al.}(2005)\citenamefont
  {Yoshida}, \citenamefont {Ikeda}, \citenamefont {Matsuhata}, \citenamefont
  {Shirakawa}, \citenamefont {Lee},\ and\ \citenamefont
  {Katano}}]{Yoshida2005}%
  \BibitemOpen
  \bibfield  {author} {\bibinfo {author} {\bibfnamefont {Y.}~\bibnamefont
  {Yoshida}}, \bibinfo {author} {\bibfnamefont {S.-I.}\ \bibnamefont {Ikeda}},
  \bibinfo {author} {\bibfnamefont {H.}~\bibnamefont {Matsuhata}}, \bibinfo
  {author} {\bibfnamefont {N.}~\bibnamefont {Shirakawa}}, \bibinfo {author}
  {\bibfnamefont {C.~H.}\ \bibnamefont {Lee}},\ and\ \bibinfo {author}
  {\bibfnamefont {S.}~\bibnamefont {Katano}},\ }\bibfield  {title} {\bibinfo
  {title} {Crystal and magnetic structure of {Ca$_3$Ru$_2$O$_7$}},\ }\bibfield
  {journal} {\bibinfo  {journal} {Physical Review B}\ }\textbf {\bibinfo
  {volume} {72}},\ \href {https://doi.org/10.1103/physrevb.72.054412}
  {10.1103/physrevb.72.054412} (\bibinfo {year} {2005})\BibitemShut {NoStop}%
\bibitem [{\citenamefont {Fei}\ \emph {et~al.}(2018)\citenamefont {Fei},
  \citenamefont {Zhao}, \citenamefont {Palomaki}, \citenamefont {Sun},
  \citenamefont {Miller}, \citenamefont {Zhao}, \citenamefont {Yan},
  \citenamefont {Xu},\ and\ \citenamefont {Cobden}}]{Fei2018}%
  \BibitemOpen
  \bibfield  {author} {\bibinfo {author} {\bibfnamefont {Z.}~\bibnamefont
  {Fei}}, \bibinfo {author} {\bibfnamefont {W.}~\bibnamefont {Zhao}}, \bibinfo
  {author} {\bibfnamefont {T.~A.}\ \bibnamefont {Palomaki}}, \bibinfo {author}
  {\bibfnamefont {B.}~\bibnamefont {Sun}}, \bibinfo {author} {\bibfnamefont
  {M.~K.}\ \bibnamefont {Miller}}, \bibinfo {author} {\bibfnamefont
  {Z.}~\bibnamefont {Zhao}}, \bibinfo {author} {\bibfnamefont {J.}~\bibnamefont
  {Yan}}, \bibinfo {author} {\bibfnamefont {X.}~\bibnamefont {Xu}},\ and\
  \bibinfo {author} {\bibfnamefont {D.~H.}\ \bibnamefont {Cobden}},\ }\bibfield
   {title} {\bibinfo {title} {Ferroelectric switching of a two-dimensional
  metal},\ }\href {https://doi.org/10.1038/s41586-018-0336-3} {\bibfield
  {journal} {\bibinfo  {journal} {Nature}\ }\textbf {\bibinfo {volume} {560}},\
  \bibinfo {pages} {336} (\bibinfo {year} {2018})}\BibitemShut {NoStop}%
\bibitem [{\citenamefont {Zheng}\ \emph {et~al.}(2021)\citenamefont {Zheng},
  \citenamefont {Wilfong}, \citenamefont {Hickox-Young}, \citenamefont
  {Rondinelli}, \citenamefont {Zavalij},\ and\ \citenamefont
  {Rodriguez}}]{Zheng2021}%
  \BibitemOpen
  \bibfield  {author} {\bibinfo {author} {\bibfnamefont {H.}~\bibnamefont
  {Zheng}}, \bibinfo {author} {\bibfnamefont {B.~C.}\ \bibnamefont {Wilfong}},
  \bibinfo {author} {\bibfnamefont {D.}~\bibnamefont {Hickox-Young}}, \bibinfo
  {author} {\bibfnamefont {J.~M.}\ \bibnamefont {Rondinelli}}, \bibinfo
  {author} {\bibfnamefont {P.~Y.}\ \bibnamefont {Zavalij}},\ and\ \bibinfo
  {author} {\bibfnamefont {E.~E.}\ \bibnamefont {Rodriguez}},\ }\bibfield
  {title} {\bibinfo {title} {Polar ferromagnetic metal by intercalation of
  metal{\textendash}amine complexes},\ }\href
  {https://doi.org/10.1021/acs.chemmater.1c00540} {\bibfield  {journal}
  {\bibinfo  {journal} {Chemistry of Materials}\ }\textbf {\bibinfo {volume}
  {33}},\ \bibinfo {pages} {4936} (\bibinfo {year} {2021})}\BibitemShut
  {NoStop}%
\bibitem [{\citenamefont {Rettie}\ \emph {et~al.}(2019)\citenamefont {Rettie},
  \citenamefont {Malliakas}, \citenamefont {Botana}, \citenamefont {Bao},
  \citenamefont {Chung},\ and\ \citenamefont {Kanatzidis}}]{Rettie2019}%
  \BibitemOpen
  \bibfield  {author} {\bibinfo {author} {\bibfnamefont {A.~J.~E.}\
  \bibnamefont {Rettie}}, \bibinfo {author} {\bibfnamefont {C.~D.}\
  \bibnamefont {Malliakas}}, \bibinfo {author} {\bibfnamefont {A.~S.}\
  \bibnamefont {Botana}}, \bibinfo {author} {\bibfnamefont {J.-K.}\
  \bibnamefont {Bao}}, \bibinfo {author} {\bibfnamefont {D.~Y.}\ \bibnamefont
  {Chung}},\ and\ \bibinfo {author} {\bibfnamefont {M.~G.}\ \bibnamefont
  {Kanatzidis}},\ }\bibfield  {title} {\bibinfo {title} {{KCu}$_7${P}$_3$: A
  two-dimensional noncentrosymmetric metallic pnictide},\ }\href
  {https://doi.org/10.1021/acs.inorgchem.9b01336} {\bibfield  {journal}
  {\bibinfo  {journal} {Inorganic Chemistry}\ }\textbf {\bibinfo {volume}
  {58}},\ \bibinfo {pages} {10201} (\bibinfo {year} {2019})}\BibitemShut
  {NoStop}%
\bibitem [{\citenamefont {Benedek}\ and\ \citenamefont
  {Birol}(2016)}]{Benedek/Birol:2016JMMC}%
  \BibitemOpen
  \bibfield  {author} {\bibinfo {author} {\bibfnamefont {N.~A.}\ \bibnamefont
  {Benedek}}\ and\ \bibinfo {author} {\bibfnamefont {T.}~\bibnamefont
  {Birol}},\ }\bibfield  {title} {\bibinfo {title} {`ferroelectric' metals
  reexamined: fundamental mechanisms and design considerations for new
  materials},\ }\href {https://doi.org/10.1039/c5tc03856a} {\bibfield
  {journal} {\bibinfo  {journal} {Journal of Materials Chemistry C}\ }\textbf
  {\bibinfo {volume} {4}},\ \bibinfo {pages} {4000} (\bibinfo {year}
  {2016})}\BibitemShut {NoStop}%
\bibitem [{\citenamefont {Zhou}\ and\ \citenamefont
  {Ariando}(2020)}]{Zhou2020}%
  \BibitemOpen
  \bibfield  {author} {\bibinfo {author} {\bibfnamefont {W.~X.}\ \bibnamefont
  {Zhou}}\ and\ \bibinfo {author} {\bibfnamefont {A.}~\bibnamefont {Ariando}},\
  }\bibfield  {title} {\bibinfo {title} {Review on ferroelectric/polar
  metals},\ }\href {https://doi.org/10.35848/1347-4065/ab8bbf} {\bibfield
  {journal} {\bibinfo  {journal} {Japanese Journal of Applied Physics}\
  }\textbf {\bibinfo {volume} {59}},\ \bibinfo {pages} {SI0802} (\bibinfo
  {year} {2020})}\BibitemShut {NoStop}%
\bibitem [{\citenamefont {Bhowal}\ and\ \citenamefont
  {Spaldin}(2022)}]{Bhowal_Spaldin:2022}%
  \BibitemOpen
  \bibfield  {author} {\bibinfo {author} {\bibfnamefont {S.}~\bibnamefont
  {Bhowal}}\ and\ \bibinfo {author} {\bibfnamefont {N.~A.}\ \bibnamefont
  {Spaldin}},\ }\href {https://doi.org/10.48550/ARXIV.2210.02993} {\bibinfo
  {title} {Polar metals: Principles and prospects}} (\bibinfo {year}
  {2022})\BibitemShut {NoStop}%
\bibitem [{\citenamefont {Baroni}\ \emph {et~al.}(1987)\citenamefont {Baroni},
  \citenamefont {Giannozzi},\ and\ \citenamefont {Testa}}]{Baroni1987}%
  \BibitemOpen
  \bibfield  {author} {\bibinfo {author} {\bibfnamefont {S.}~\bibnamefont
  {Baroni}}, \bibinfo {author} {\bibfnamefont {P.}~\bibnamefont {Giannozzi}},\
  and\ \bibinfo {author} {\bibfnamefont {A.}~\bibnamefont {Testa}},\ }\bibfield
   {title} {\bibinfo {title} {{Green's-Function Approach to Linear Response in
  Solids}},\ }\href {https://doi.org/10.1103/PhysRevLett.58.1861} {\bibfield
  {journal} {\bibinfo  {journal} {Physical Review Letters}\ }\textbf {\bibinfo
  {volume} {58}},\ \bibinfo {pages} {1861} (\bibinfo {year}
  {1987})}\BibitemShut {NoStop}%
\bibitem [{\citenamefont {Resta}(1993)}]{Resta1993}%
  \BibitemOpen
  \bibfield  {author} {\bibinfo {author} {\bibfnamefont {R.}~\bibnamefont
  {Resta}},\ }\bibfield  {title} {\bibinfo {title} {{Macroscopie electric
  polarization as a geometric quantum phase}},\ }\href
  {https://doi.org/10.1209/0295-5075/22/2/010} {\bibfield  {journal} {\bibinfo
  {journal} {EPL (Europhysics Letters)}\ }\textbf {\bibinfo {volume} {22}},\
  \bibinfo {pages} {133} (\bibinfo {year} {1993})}\BibitemShut {NoStop}%
\bibitem [{\citenamefont {King-Smith}\ and\ \citenamefont
  {Vanderbilt}(1993)}]{King-Smith1993}%
  \BibitemOpen
  \bibfield  {author} {\bibinfo {author} {\bibfnamefont {R.~D.}\ \bibnamefont
  {King-Smith}}\ and\ \bibinfo {author} {\bibfnamefont {D.}~\bibnamefont
  {Vanderbilt}},\ }\bibfield  {title} {\bibinfo {title} {Theory of polarization
  of crystalline solids},\ }\href {https://doi.org/10.1103/physrevb.47.1651}
  {\bibfield  {journal} {\bibinfo  {journal} {Physical Review B}\ }\textbf
  {\bibinfo {volume} {47}},\ \bibinfo {pages} {1651} (\bibinfo {year}
  {1993})}\BibitemShut {NoStop}%
\bibitem [{\citenamefont {Vanderbilt}(2018)}]{vanderbilt2018berry}%
  \BibitemOpen
  \bibfield  {author} {\bibinfo {author} {\bibfnamefont {D.}~\bibnamefont
  {Vanderbilt}},\ }\href@noop {} {\emph {\bibinfo {title} {Berry Phases in
  Electronic Structure Theory: Electric Polarization, Orbital Magnetization and
  Topological Insulators}}}\ (\bibinfo  {publisher} {Cambridge University
  Press},\ \bibinfo {year} {2018})\BibitemShut {NoStop}%
\bibitem [{\citenamefont {Cohen}\ and\ \citenamefont
  {Krakauer}(1990)}]{Cohen1990}%
  \BibitemOpen
  \bibfield  {author} {\bibinfo {author} {\bibfnamefont {R.~E.}\ \bibnamefont
  {Cohen}}\ and\ \bibinfo {author} {\bibfnamefont {H.}~\bibnamefont
  {Krakauer}},\ }\bibfield  {title} {\bibinfo {title} {{Lattice dynamics and
  origin of ferroelectricity in BaTiO3: Linearized-augmented-plane-wave
  total-energy calculations}},\ }\href
  {https://doi.org/10.1103/PhysRevB.42.6416} {\bibfield  {journal} {\bibinfo
  {journal} {Physical Review B}\ }\textbf {\bibinfo {volume} {42}},\ \bibinfo
  {pages} {6416} (\bibinfo {year} {1990})}\BibitemShut {NoStop}%
\bibitem [{\citenamefont {{Jun-xing Gu and Kui-juan Jin and Chao Ma and
  Qing-hua Zhang and Lin Gu and Chen Ge and Jie-su Wang and Can Wang and
  Hai-zhong Guo and Guo-zhen Yang}}(2017)}]{Gu/Jin:2017PRB}%
  \BibitemOpen
  \bibfield  {author} {\bibinfo {author} {\bibnamefont {{Jun-xing Gu and
  Kui-juan Jin and Chao Ma and Qing-hua Zhang and Lin Gu and Chen Ge and Jie-su
  Wang and Can Wang and Hai-zhong Guo and Guo-zhen Yang}}},\ }\bibfield
  {title} {\bibinfo {title} {Coexistence of polar distortion and metallicity in
  {PbTi$_{1-x}$Nb$_x$O}$_3$},\ }\bibfield  {journal} {\bibinfo  {journal}
  {Physical Review B}\ }\textbf {\bibinfo {volume} {96}},\ \href
  {https://doi.org/10.1103/physrevb.96.165206} {10.1103/physrevb.96.165206}
  (\bibinfo {year} {2017})\BibitemShut {NoStop}%
\bibitem [{\citenamefont {Xiang}(2014)}]{Xiang:2014PRB}%
  \BibitemOpen
  \bibfield  {author} {\bibinfo {author} {\bibfnamefont {H.~J.}\ \bibnamefont
  {Xiang}},\ }\bibfield  {title} {\bibinfo {title} {Origin of polar distortion
  in {LiNbO}$_3$-type {\textquotedblleft}ferroelectric{\textquotedblright}
  metals: Role of {A}-site instability and short-range interactions},\
  }\bibfield  {journal} {\bibinfo  {journal} {Physical Review B}\ }\textbf
  {\bibinfo {volume} {90}},\ \href {https://doi.org/10.1103/physrevb.90.094108}
  {10.1103/physrevb.90.094108} (\bibinfo {year} {2014})\BibitemShut {NoStop}%
\bibitem [{\citenamefont {Jin}\ \emph {et~al.}(2016)\citenamefont {Jin},
  \citenamefont {Zhang}, \citenamefont {Ji}, \citenamefont {Liu}, \citenamefont
  {Wang}, \citenamefont {Shi}, \citenamefont {Tian}, \citenamefont {Ma},\ and\
  \citenamefont {Zhang}}]{Jin2016}%
  \BibitemOpen
  \bibfield  {author} {\bibinfo {author} {\bibfnamefont {F.}~\bibnamefont
  {Jin}}, \bibinfo {author} {\bibfnamefont {A.}~\bibnamefont {Zhang}}, \bibinfo
  {author} {\bibfnamefont {J.}~\bibnamefont {Ji}}, \bibinfo {author}
  {\bibfnamefont {K.}~\bibnamefont {Liu}}, \bibinfo {author} {\bibfnamefont
  {L.}~\bibnamefont {Wang}}, \bibinfo {author} {\bibfnamefont {Y.}~\bibnamefont
  {Shi}}, \bibinfo {author} {\bibfnamefont {Y.}~\bibnamefont {Tian}}, \bibinfo
  {author} {\bibfnamefont {X.}~\bibnamefont {Ma}},\ and\ \bibinfo {author}
  {\bibfnamefont {Q.}~\bibnamefont {Zhang}},\ }\bibfield  {title} {\bibinfo
  {title} {Raman phonons in the ferroelectric-like metal {LiOsO}$_3$},\
  }\bibfield  {journal} {\bibinfo  {journal} {Physical Review B}\ }\textbf
  {\bibinfo {volume} {93}},\ \href {https://doi.org/10.1103/physrevb.93.064303}
  {10.1103/physrevb.93.064303} (\bibinfo {year} {2016})\BibitemShut {NoStop}%
\bibitem [{\citenamefont {Filippetti}\ \emph {et~al.}(2016)\citenamefont
  {Filippetti}, \citenamefont {Fiorentini}, \citenamefont {Ricci},
  \citenamefont {Delugas},\ and\ \citenamefont
  {{\'{I}}{\~{n}}iguez}}]{Filippetti2016}%
  \BibitemOpen
  \bibfield  {author} {\bibinfo {author} {\bibfnamefont {A.}~\bibnamefont
  {Filippetti}}, \bibinfo {author} {\bibfnamefont {V.}~\bibnamefont
  {Fiorentini}}, \bibinfo {author} {\bibfnamefont {F.}~\bibnamefont {Ricci}},
  \bibinfo {author} {\bibfnamefont {P.}~\bibnamefont {Delugas}},\ and\ \bibinfo
  {author} {\bibfnamefont {J.}~\bibnamefont {{\'{I}}{\~{n}}iguez}},\ }\bibfield
   {title} {\bibinfo {title} {Prediction of a native ferroelectric metal},\
  }\bibfield  {journal} {\bibinfo  {journal} {Nature Communications}\ }\textbf
  {\bibinfo {volume} {7}},\ \href {https://doi.org/10.1038/ncomms11211}
  {10.1038/ncomms11211} (\bibinfo {year} {2016})\BibitemShut {NoStop}%
\bibitem [{\citenamefont {Bruneval}\ \emph {et~al.}(2015)\citenamefont
  {Bruneval}, \citenamefont {Varvenne}, \citenamefont {Crocombette},\ and\
  \citenamefont {Clouet}}]{Bruneval2015}%
  \BibitemOpen
  \bibfield  {author} {\bibinfo {author} {\bibfnamefont {F.}~\bibnamefont
  {Bruneval}}, \bibinfo {author} {\bibfnamefont {C.}~\bibnamefont {Varvenne}},
  \bibinfo {author} {\bibfnamefont {J.-P.}\ \bibnamefont {Crocombette}},\ and\
  \bibinfo {author} {\bibfnamefont {E.}~\bibnamefont {Clouet}},\ }\bibfield
  {title} {\bibinfo {title} {Pressure, relaxation volume, and elastic
  interactions in charged simulation cells},\ }\bibfield  {journal} {\bibinfo
  {journal} {Physical Review B}\ }\textbf {\bibinfo {volume} {91}},\ \href
  {https://doi.org/10.1103/physrevb.91.024107} {10.1103/physrevb.91.024107}
  (\bibinfo {year} {2015})\BibitemShut {NoStop}%
\bibitem [{\citenamefont {Iwazaki}\ \emph {et~al.}(2012)\citenamefont
  {Iwazaki}, \citenamefont {Suzuki}, \citenamefont {Mizuno},\ and\
  \citenamefont {Tsuneyuki}}]{Iwazaki2012}%
  \BibitemOpen
  \bibfield  {author} {\bibinfo {author} {\bibfnamefont {Y.}~\bibnamefont
  {Iwazaki}}, \bibinfo {author} {\bibfnamefont {T.}~\bibnamefont {Suzuki}},
  \bibinfo {author} {\bibfnamefont {Y.}~\bibnamefont {Mizuno}},\ and\ \bibinfo
  {author} {\bibfnamefont {S.}~\bibnamefont {Tsuneyuki}},\ }\bibfield  {title}
  {\bibinfo {title} {Doping-induced phase transitions in ferroelectric
  {BaTiO}$_3$ from first-principles calculations},\ }\bibfield  {journal}
  {\bibinfo  {journal} {Physical Review B}\ }\textbf {\bibinfo {volume} {86}},\
  \href {https://doi.org/10.1103/physrevb.86.214103}
  {10.1103/physrevb.86.214103} (\bibinfo {year} {2012})\BibitemShut {NoStop}%
\bibitem [{\citenamefont {Scott}(2008)}]{Scott2008}%
  \BibitemOpen
  \bibfield  {author} {\bibinfo {author} {\bibfnamefont {J.~F.}\ \bibnamefont
  {Scott}},\ }\bibfield  {title} {\bibinfo {title} {{Ferroelectrics go
  bananas}},\ }\href {https://doi.org/10.1088/0953-8984/20/02/021001}
  {\bibfield  {journal} {\bibinfo  {journal} {Journal of Physics Condensed
  Matter}\ }\textbf {\bibinfo {volume} {20}},\ \bibinfo {pages} {18} (\bibinfo
  {year} {2008})}\BibitemShut {NoStop}%
\bibitem [{\citenamefont {Kohn}(1964)}]{Kohn1964}%
  \BibitemOpen
  \bibfield  {author} {\bibinfo {author} {\bibfnamefont {W.}~\bibnamefont
  {Kohn}},\ }\bibfield  {title} {\bibinfo {title} {{Theory of the insulating
  state}},\ }\bibfield  {journal} {\bibinfo  {journal} {Physical Review}\
  }\textbf {\bibinfo {volume} {133}},\ \href
  {https://doi.org/10.1103/PhysRev.133.A171} {10.1103/PhysRev.133.A171}
  (\bibinfo {year} {1964})\BibitemShut {NoStop}%
\bibitem [{\citenamefont {Edwards}\ \emph {et~al.}(2010)\citenamefont
  {Edwards}, \citenamefont {Lodge}, \citenamefont {Hensel},\ and\ \citenamefont
  {Redmer}}]{Edwards2010}%
  \BibitemOpen
  \bibfield  {author} {\bibinfo {author} {\bibfnamefont {P.~P.}\ \bibnamefont
  {Edwards}}, \bibinfo {author} {\bibfnamefont {M.~T.}\ \bibnamefont {Lodge}},
  \bibinfo {author} {\bibfnamefont {F.}~\bibnamefont {Hensel}},\ and\ \bibinfo
  {author} {\bibfnamefont {R.}~\bibnamefont {Redmer}},\ }\bibfield  {title}
  {\bibinfo {title} {{`... a Metal Conducts and a Non-Metal Doesn't'}},\ }\href
  {https://doi.org/10.1098/rsta.2009.0282} {\bibfield  {journal} {\bibinfo
  {journal} {Philosophical Transactions of the Royal Society A: Mathematical,
  Physical and Engineering Sciences}\ }\textbf {\bibinfo {volume} {368}},\
  \bibinfo {pages} {941} (\bibinfo {year} {2010})}\BibitemShut {NoStop}%
\bibitem [{\citenamefont {Edwards}(1998)}]{davis1998nevill}%
  \BibitemOpen
  \bibfield  {author} {\bibinfo {author} {\bibfnamefont {P.~P.}\ \bibnamefont
  {Edwards}},\ }\href@noop {} {\emph {\bibinfo {title} {Nevill Mott:
  reminiscences and appreciations}}},\ edited by\ \bibinfo {editor}
  {\bibfnamefont {E.~A.}\ \bibnamefont {Davis}}\ (\bibinfo  {publisher} {CRC
  Press},\ \bibinfo {year} {1998})\BibitemShut {NoStop}%
\bibitem [{\citenamefont {Takahashi}\ \emph {et~al.}(2017)\citenamefont
  {Takahashi}, \citenamefont {Matsubara}, \citenamefont {Bahramy},
  \citenamefont {Ogawa}, \citenamefont {Hashizume}, \citenamefont {Tokura},\
  and\ \citenamefont {Kawasaki}}]{Takahashi2017}%
  \BibitemOpen
  \bibfield  {author} {\bibinfo {author} {\bibfnamefont {K.~S.}\ \bibnamefont
  {Takahashi}}, \bibinfo {author} {\bibfnamefont {Y.}~\bibnamefont
  {Matsubara}}, \bibinfo {author} {\bibfnamefont {M.~S.}\ \bibnamefont
  {Bahramy}}, \bibinfo {author} {\bibfnamefont {N.}~\bibnamefont {Ogawa}},
  \bibinfo {author} {\bibfnamefont {D.}~\bibnamefont {Hashizume}}, \bibinfo
  {author} {\bibfnamefont {Y.}~\bibnamefont {Tokura}},\ and\ \bibinfo {author}
  {\bibfnamefont {M.}~\bibnamefont {Kawasaki}},\ }\bibfield  {title} {\bibinfo
  {title} {Polar metal phase stabilized in strained {La}-doped {BaTiO}$_3$
  films},\ }\bibfield  {journal} {\bibinfo  {journal} {Scientific Reports}\
  }\textbf {\bibinfo {volume} {7}},\ \href
  {https://doi.org/10.1038/s41598-017-04635-3} {10.1038/s41598-017-04635-3}
  (\bibinfo {year} {2017})\BibitemShut {NoStop}%
\bibitem [{\citenamefont {Fujioka}\ \emph {et~al.}(2015)\citenamefont
  {Fujioka}, \citenamefont {Doi}, \citenamefont {Okuyama}, \citenamefont
  {Morikawa}, \citenamefont {Arima}, \citenamefont {Okada}, \citenamefont
  {Kaneko}, \citenamefont {Fukuda}, \citenamefont {Uchiyama}, \citenamefont
  {Ishikawa}, \citenamefont {Baron}, \citenamefont {Kato}, \citenamefont
  {Takata},\ and\ \citenamefont {Tokura}}]{Fujioka2015}%
  \BibitemOpen
  \bibfield  {author} {\bibinfo {author} {\bibfnamefont {J.}~\bibnamefont
  {Fujioka}}, \bibinfo {author} {\bibfnamefont {A.}~\bibnamefont {Doi}},
  \bibinfo {author} {\bibfnamefont {D.}~\bibnamefont {Okuyama}}, \bibinfo
  {author} {\bibfnamefont {D.}~\bibnamefont {Morikawa}}, \bibinfo {author}
  {\bibfnamefont {T.}~\bibnamefont {Arima}}, \bibinfo {author} {\bibfnamefont
  {K.~N.}\ \bibnamefont {Okada}}, \bibinfo {author} {\bibfnamefont
  {Y.}~\bibnamefont {Kaneko}}, \bibinfo {author} {\bibfnamefont
  {T.}~\bibnamefont {Fukuda}}, \bibinfo {author} {\bibfnamefont
  {H.}~\bibnamefont {Uchiyama}}, \bibinfo {author} {\bibfnamefont
  {D.}~\bibnamefont {Ishikawa}}, \bibinfo {author} {\bibfnamefont {A.~Q.~R.}\
  \bibnamefont {Baron}}, \bibinfo {author} {\bibfnamefont {K.}~\bibnamefont
  {Kato}}, \bibinfo {author} {\bibfnamefont {M.}~\bibnamefont {Takata}},\ and\
  \bibinfo {author} {\bibfnamefont {Y.}~\bibnamefont {Tokura}},\ }\bibfield
  {title} {\bibinfo {title} {{Ferroelectric-like metallic state in electron
  doped BaTiO$_3$}},\ }\href {https://doi.org/10.1038/srep13207} {\bibfield
  {journal} {\bibinfo  {journal} {Scientific Reports}\ }\textbf {\bibinfo
  {volume} {5}},\ \bibinfo {pages} {13207} (\bibinfo {year}
  {2015})}\BibitemShut {NoStop}%
\bibitem [{\citenamefont {Kolodiazhnyi}(2008)}]{Kolodiazhnyi2008}%
  \BibitemOpen
  \bibfield  {author} {\bibinfo {author} {\bibfnamefont {T.}~\bibnamefont
  {Kolodiazhnyi}},\ }\bibfield  {title} {\bibinfo {title} {Insulator-metal
  transition and anomalous sign reversal of the dominant charge carriers in
  perovskite {BaTiO}$_{3-\delta}$},\ }\bibfield  {journal} {\bibinfo  {journal}
  {Physical Review B}\ }\textbf {\bibinfo {volume} {78}},\ \href
  {https://doi.org/10.1103/physrevb.78.045107} {10.1103/physrevb.78.045107}
  (\bibinfo {year} {2008})\BibitemShut {NoStop}%
\bibitem [{\citenamefont {Scalapino}\ \emph {et~al.}(1993)\citenamefont
  {Scalapino}, \citenamefont {White},\ and\ \citenamefont
  {Zhang}}]{Scalapino/White/Zhang1993}%
  \BibitemOpen
  \bibfield  {author} {\bibinfo {author} {\bibfnamefont {D.~J.}\ \bibnamefont
  {Scalapino}}, \bibinfo {author} {\bibfnamefont {S.~R.}\ \bibnamefont
  {White}},\ and\ \bibinfo {author} {\bibfnamefont {S.}~\bibnamefont {Zhang}},\
  }\bibfield  {title} {\bibinfo {title} {Insulator, metal, or superconductor:
  The criteria},\ }\href {https://doi.org/10.1103/PhysRevB.47.7995} {\bibfield
  {journal} {\bibinfo  {journal} {Phys. Rev. B}\ }\textbf {\bibinfo {volume}
  {47}},\ \bibinfo {pages} {7995} (\bibinfo {year} {1993})}\BibitemShut
  {NoStop}%
\bibitem [{\citenamefont {Resta}(2002)}]{Resta2002}%
  \BibitemOpen
  \bibfield  {author} {\bibinfo {author} {\bibfnamefont {R.}~\bibnamefont
  {Resta}},\ }\bibfield  {title} {\bibinfo {title} {Why are insulators
  insulating and metals conducting?},\ }\href
  {https://doi.org/10.1088/0953-8984/14/20/201} {\bibfield  {journal} {\bibinfo
   {journal} {Journal of Physics: Condensed Matter}\ }\textbf {\bibinfo
  {volume} {14}},\ \bibinfo {pages} {R625} (\bibinfo {year}
  {2002})}\BibitemShut {NoStop}%
\bibitem [{\citenamefont {Ashcroft}\ and\ \citenamefont
  {Mermin}(1976)}]{Ashcroft/Mermin:Book}%
  \BibitemOpen
  \bibfield  {author} {\bibinfo {author} {\bibfnamefont {N.~W.}\ \bibnamefont
  {Ashcroft}}\ and\ \bibinfo {author} {\bibfnamefont {N.~D.}\ \bibnamefont
  {Mermin}},\ }\href@noop {} {\emph {\bibinfo {title} {Solid State Physics}}}\
  (\bibinfo  {publisher} {Thomson Learning, Inc.},\ \bibinfo {year}
  {1976})\BibitemShut {NoStop}%
\bibitem [{\citenamefont {Bloch}(1930)}]{Bloch1930}%
  \BibitemOpen
  \bibfield  {author} {\bibinfo {author} {\bibfnamefont {F.}~\bibnamefont
  {Bloch}},\ }\bibfield  {title} {\bibinfo {title} {Zum elektrischen
  widerstandsgesetz bei tiefen temperaturen},\ }\href
  {https://doi.org/https://doi.org/10.1007/BF01341426} {\bibfield  {journal}
  {\bibinfo  {journal} {Zeitschrift für Physik}\ }\textbf {\bibinfo {volume}
  {59}},\ \bibinfo {pages} {208} (\bibinfo {year} {1930})}\BibitemShut
  {NoStop}%
\bibitem [{\citenamefont {Grüneisen}(1933)}]{Grun1933}%
  \BibitemOpen
  \bibfield  {author} {\bibinfo {author} {\bibfnamefont {E.}~\bibnamefont
  {Grüneisen}},\ }\bibfield  {title} {\bibinfo {title} {Die abhängigkeit des
  elektrischen widerstandes reiner metalle von der temperatur},\ }\href
  {https://doi.org/https://doi.org/10.1002/andp.19334080504} {\bibfield
  {journal} {\bibinfo  {journal} {Annalen der Physik}\ }\textbf {\bibinfo
  {volume} {408}},\ \bibinfo {pages} {530} (\bibinfo {year}
  {1933})}\BibitemShut {NoStop}%
\bibitem [{\citenamefont {Yoshida}\ \emph {et~al.}(2004)\citenamefont
  {Yoshida}, \citenamefont {Nagai}, \citenamefont {Ikeda}, \citenamefont
  {Shirakawa}, \citenamefont {Kosaka},\ and\ \citenamefont
  {M\^ori}}]{PhysRevB.69.220411}%
  \BibitemOpen
  \bibfield  {author} {\bibinfo {author} {\bibfnamefont {Y.}~\bibnamefont
  {Yoshida}}, \bibinfo {author} {\bibfnamefont {I.}~\bibnamefont {Nagai}},
  \bibinfo {author} {\bibfnamefont {S.-I.}\ \bibnamefont {Ikeda}}, \bibinfo
  {author} {\bibfnamefont {N.}~\bibnamefont {Shirakawa}}, \bibinfo {author}
  {\bibfnamefont {M.}~\bibnamefont {Kosaka}},\ and\ \bibinfo {author}
  {\bibfnamefont {N.}~\bibnamefont {M\^ori}},\ }\bibfield  {title} {\bibinfo
  {title} {Quasi-two-dimensional metallic ground state of
  {Ca$_3$Ru$_2$O$_7$}},\ }\href {https://doi.org/10.1103/PhysRevB.69.220411}
  {\bibfield  {journal} {\bibinfo  {journal} {Phys. Rev. B}\ }\textbf {\bibinfo
  {volume} {69}},\ \bibinfo {pages} {220411} (\bibinfo {year}
  {2004})}\BibitemShut {NoStop}%
\bibitem [{\citenamefont {Hwang}\ \emph {et~al.}(2010)\citenamefont {Hwang},
  \citenamefont {Kolodiazhnyi}, \citenamefont {Yang},\ and\ \citenamefont
  {Couillard}}]{Hwang2010}%
  \BibitemOpen
  \bibfield  {author} {\bibinfo {author} {\bibfnamefont {J.}~\bibnamefont
  {Hwang}}, \bibinfo {author} {\bibfnamefont {T.}~\bibnamefont {Kolodiazhnyi}},
  \bibinfo {author} {\bibfnamefont {J.}~\bibnamefont {Yang}},\ and\ \bibinfo
  {author} {\bibfnamefont {M.}~\bibnamefont {Couillard}},\ }\bibfield  {title}
  {\bibinfo {title} {{Doping and temperature-dependent optical properties of
  oxygen-reduced BaTiO$_{3-\delta}$}},\ }\href
  {https://doi.org/10.1103/PhysRevB.82.214109} {\bibfield  {journal} {\bibinfo
  {journal} {Physical Review B - Condensed Matter and Materials Physics}\
  }\textbf {\bibinfo {volume} {82}},\ \bibinfo {pages} {1} (\bibinfo {year}
  {2010})},\ \Eprint {https://arxiv.org/abs/1105.1587} {arXiv:1105.1587}
  \BibitemShut {NoStop}%
\bibitem [{\citenamefont {Millis}\ and\ \citenamefont
  {Coppersmith}(1991)}]{Millis1991}%
  \BibitemOpen
  \bibfield  {author} {\bibinfo {author} {\bibfnamefont {A.~J.}\ \bibnamefont
  {Millis}}\ and\ \bibinfo {author} {\bibfnamefont {S.~N.}\ \bibnamefont
  {Coppersmith}},\ }\bibfield  {title} {\bibinfo {title} {{Variational wave
  functions and the Mott transition}},\ }\href@noop {} {\bibfield  {journal}
  {\bibinfo  {journal} {Physical Review B}\ }\textbf {\bibinfo {volume} {43}},\
  \bibinfo {pages} {13770} (\bibinfo {year} {1991})}\BibitemShut {NoStop}%
\bibitem [{\citenamefont {Homes}\ \emph {et~al.}(1993)\citenamefont {Homes},
  \citenamefont {Timusk}, \citenamefont {Liang}, \citenamefont {Bonn},\ and\
  \citenamefont {Hardy}}]{Homes1993}%
  \BibitemOpen
  \bibfield  {author} {\bibinfo {author} {\bibfnamefont {C.~C.}\ \bibnamefont
  {Homes}}, \bibinfo {author} {\bibfnamefont {T.}~\bibnamefont {Timusk}},
  \bibinfo {author} {\bibfnamefont {R.}~\bibnamefont {Liang}}, \bibinfo
  {author} {\bibfnamefont {D.~A.}\ \bibnamefont {Bonn}},\ and\ \bibinfo
  {author} {\bibfnamefont {W.~N.}\ \bibnamefont {Hardy}},\ }\bibfield  {title}
  {\bibinfo {title} {Optical conductivity of c axis oriented
  {YBa$_2$Cu$_3$O$_{6.70}$}: Evidence for a pseudogap},\ }\href
  {https://doi.org/10.1103/physrevlett.71.1645} {\bibfield  {journal} {\bibinfo
   {journal} {Physical Review Letters}\ }\textbf {\bibinfo {volume} {71}},\
  \bibinfo {pages} {1645} (\bibinfo {year} {1993})}\BibitemShut {NoStop}%
\bibitem [{\citenamefont {Dreyer}\ \emph {et~al.}(2022)\citenamefont {Dreyer},
  \citenamefont {Coh},\ and\ \citenamefont {Stengel}}]{dreyer2021nonadiabatic}%
  \BibitemOpen
  \bibfield  {author} {\bibinfo {author} {\bibfnamefont {C.~E.}\ \bibnamefont
  {Dreyer}}, \bibinfo {author} {\bibfnamefont {S.}~\bibnamefont {Coh}},\ and\
  \bibinfo {author} {\bibfnamefont {M.}~\bibnamefont {Stengel}},\ }\bibfield
  {title} {\bibinfo {title} {Nonadiabatic {Born} effective charges in metals
  and the {Drude} weight},\ }\href
  {https://doi.org/10.1103/PhysRevLett.128.095901} {\bibfield  {journal}
  {\bibinfo  {journal} {Phys. Rev. Lett.}\ }\textbf {\bibinfo {volume} {128}},\
  \bibinfo {pages} {095901} (\bibinfo {year} {2022})}\BibitemShut {NoStop}%
\bibitem [{\citenamefont {Anderson}\ \emph {et~al.}(1971)\citenamefont
  {Anderson}, \citenamefont {Alexander},\ and\ \citenamefont
  {Bell}}]{Anderson1971}%
  \BibitemOpen
  \bibfield  {author} {\bibinfo {author} {\bibfnamefont {W.~E.}\ \bibnamefont
  {Anderson}}, \bibinfo {author} {\bibfnamefont {R.~W.}\ \bibnamefont
  {Alexander}},\ and\ \bibinfo {author} {\bibfnamefont {R.~J.}\ \bibnamefont
  {Bell}},\ }\bibfield  {title} {\bibinfo {title} {Surface plasmons and the
  reflectivity of n-type {InSb}},\ }\href
  {https://doi.org/10.1103/physrevlett.27.1057} {\bibfield  {journal} {\bibinfo
   {journal} {Physical Review Letters}\ }\textbf {\bibinfo {volume} {27}},\
  \bibinfo {pages} {1057} (\bibinfo {year} {1971})}\BibitemShut {NoStop}%
\bibitem [{\citenamefont {Bi}\ \emph {et~al.}(2006)\citenamefont {Bi},
  \citenamefont {Ma}, \citenamefont {Yan}, \citenamefont {Fang}, \citenamefont
  {Zhao}, \citenamefont {Yao},\ and\ \citenamefont {Qiu}}]{Bi2006}%
  \BibitemOpen
  \bibfield  {author} {\bibinfo {author} {\bibfnamefont {C.~Z.}\ \bibnamefont
  {Bi}}, \bibinfo {author} {\bibfnamefont {J.~Y.}\ \bibnamefont {Ma}}, \bibinfo
  {author} {\bibfnamefont {J.}~\bibnamefont {Yan}}, \bibinfo {author}
  {\bibfnamefont {X.}~\bibnamefont {Fang}}, \bibinfo {author} {\bibfnamefont
  {B.~R.}\ \bibnamefont {Zhao}}, \bibinfo {author} {\bibfnamefont {D.~Z.}\
  \bibnamefont {Yao}},\ and\ \bibinfo {author} {\bibfnamefont {X.~G.}\
  \bibnamefont {Qiu}},\ }\bibfield  {title} {\bibinfo {title}
  {Electron{\textendash}phonon coupling in {Nb}-doped {SrTiO}$_3$ single
  crystal},\ }\href {https://doi.org/10.1088/0953-8984/18/8/017} {\bibfield
  {journal} {\bibinfo  {journal} {Journal of Physics: Condensed Matter}\
  }\textbf {\bibinfo {volume} {18}},\ \bibinfo {pages} {2553} (\bibinfo {year}
  {2006})}\BibitemShut {NoStop}%
\bibitem [{\citenamefont {Vecchio}\ \emph {et~al.}(2016)\citenamefont
  {Vecchio}, \citenamefont {Giovannetti}, \citenamefont {Autore}, \citenamefont
  {Pietro}, \citenamefont {Perucchi}, \citenamefont {He}, \citenamefont
  {Yamaura}, \citenamefont {Capone},\ and\ \citenamefont
  {Lupi}}]{LoVecchio2016}%
  \BibitemOpen
  \bibfield  {author} {\bibinfo {author} {\bibfnamefont {I.~L.}\ \bibnamefont
  {Vecchio}}, \bibinfo {author} {\bibfnamefont {G.}~\bibnamefont
  {Giovannetti}}, \bibinfo {author} {\bibfnamefont {M.}~\bibnamefont {Autore}},
  \bibinfo {author} {\bibfnamefont {P.~D.}\ \bibnamefont {Pietro}}, \bibinfo
  {author} {\bibfnamefont {A.}~\bibnamefont {Perucchi}}, \bibinfo {author}
  {\bibfnamefont {J.}~\bibnamefont {He}}, \bibinfo {author} {\bibfnamefont
  {K.}~\bibnamefont {Yamaura}}, \bibinfo {author} {\bibfnamefont
  {M.}~\bibnamefont {Capone}},\ and\ \bibinfo {author} {\bibfnamefont
  {S.}~\bibnamefont {Lupi}},\ }\bibfield  {title} {\bibinfo {title} {Electronic
  correlations in the ferroelectric metallic state of {LiOsO}$_3$},\ }\bibfield
   {journal} {\bibinfo  {journal} {Physical Review B}\ }\textbf {\bibinfo
  {volume} {93}},\ \href {https://doi.org/10.1103/physrevb.93.161113}
  {10.1103/physrevb.93.161113} (\bibinfo {year} {2016})\BibitemShut {NoStop}%
\bibitem [{\citenamefont {Bersuker}(2013)}]{Bersuker2013:ChemRev}%
  \BibitemOpen
  \bibfield  {author} {\bibinfo {author} {\bibfnamefont {I.~B.}\ \bibnamefont
  {Bersuker}},\ }\bibfield  {title} {\bibinfo {title}
  {Pseudo-{Jahn{\textendash}Teller} effect{\textemdash}{A} two-state paradigm
  in formation, deformation, and transformation of molecular systems and
  solids},\ }\href {https://doi.org/10.1021/cr300279n} {\bibfield  {journal}
  {\bibinfo  {journal} {Chemical Reviews}\ }\textbf {\bibinfo {volume} {113}},\
  \bibinfo {pages} {1351} (\bibinfo {year} {2013})}\BibitemShut {NoStop}%
\bibitem [{\citenamefont {Hickox-Young}\ \emph {et~al.}(2020)\citenamefont
  {Hickox-Young}, \citenamefont {Puggioni},\ and\ \citenamefont
  {Rondinelli}}]{Hickox-Young_Rondinelli2020:PRB}%
  \BibitemOpen
  \bibfield  {author} {\bibinfo {author} {\bibfnamefont {D.}~\bibnamefont
  {Hickox-Young}}, \bibinfo {author} {\bibfnamefont {D.}~\bibnamefont
  {Puggioni}},\ and\ \bibinfo {author} {\bibfnamefont {J.~M.}\ \bibnamefont
  {Rondinelli}},\ }\bibfield  {title} {\bibinfo {title} {Persistent polar
  distortions from covalent interactions in doped {BaTiO}$_3$},\ }\bibfield
  {journal} {\bibinfo  {journal} {Physical Review B}\ }\textbf {\bibinfo
  {volume} {102}},\ \href {https://doi.org/10.1103/physrevb.102.014108}
  {10.1103/physrevb.102.014108} (\bibinfo {year} {2020})\BibitemShut {NoStop}%
\bibitem [{\citenamefont {Ghosez}\ \emph {et~al.}(1996)\citenamefont {Ghosez},
  \citenamefont {Gonze},\ and\ \citenamefont {Michenaud}}]{Ghosez:1996EPL}%
  \BibitemOpen
  \bibfield  {author} {\bibinfo {author} {\bibfnamefont {P.}~\bibnamefont
  {Ghosez}}, \bibinfo {author} {\bibfnamefont {X.}~\bibnamefont {Gonze}},\ and\
  \bibinfo {author} {\bibfnamefont {J.-P.}\ \bibnamefont {Michenaud}},\
  }\bibfield  {title} {\bibinfo {title} {Coulomb interaction and ferroelectric
  instability of {BaTiO}$_3$},\ }\href
  {https://doi.org/10.1209/epl/i1996-00404-8} {\bibfield  {journal} {\bibinfo
  {journal} {Europhysics Letters ({EPL})}\ }\textbf {\bibinfo {volume} {33}},\
  \bibinfo {pages} {713} (\bibinfo {year} {1996})}\BibitemShut {NoStop}%
\bibitem [{\citenamefont {Salmani-Rezaie}\ \emph
  {et~al.}(2020{\natexlab{a}})\citenamefont {Salmani-Rezaie}, \citenamefont
  {Ahadi}, \citenamefont {Strickland},\ and\ \citenamefont
  {Stemmer}}]{salmanirezaie2020orderdisorder}%
  \BibitemOpen
  \bibfield  {author} {\bibinfo {author} {\bibfnamefont {S.}~\bibnamefont
  {Salmani-Rezaie}}, \bibinfo {author} {\bibfnamefont {K.}~\bibnamefont
  {Ahadi}}, \bibinfo {author} {\bibfnamefont {W.}~\bibnamefont {Strickland}},\
  and\ \bibinfo {author} {\bibfnamefont {S.}~\bibnamefont {Stemmer}},\
  }\bibfield  {title} {\bibinfo {title} {{Order-disorder ferroelectric
  transition of strained SrTiO$_3$}},\ }\bibfield  {journal} {\bibinfo
  {journal} {Physical Review Letters}\ }\textbf {\bibinfo {volume} {125}},\
  \href {https://doi.org/10.1103/physrevlett.125.087601}
  {10.1103/physrevlett.125.087601} (\bibinfo {year}
  {2020}{\natexlab{a}})\BibitemShut {NoStop}%
\bibitem [{\citenamefont {Chepkemboi}\ \emph {et~al.}(2022)\citenamefont
  {Chepkemboi}, \citenamefont {Jorgensen}, \citenamefont {Sato},\ and\
  \citenamefont {Laurita}}]{Chepkemboi2022}%
  \BibitemOpen
  \bibfield  {author} {\bibinfo {author} {\bibfnamefont {C.}~\bibnamefont
  {Chepkemboi}}, \bibinfo {author} {\bibfnamefont {K.}~\bibnamefont
  {Jorgensen}}, \bibinfo {author} {\bibfnamefont {J.}~\bibnamefont {Sato}},\
  and\ \bibinfo {author} {\bibfnamefont {G.}~\bibnamefont {Laurita}},\
  }\bibfield  {title} {\bibinfo {title} {Strategies and considerations for
  least-squares analysis of total scattering data},\ }\href
  {https://doi.org/10.1021/acsomega.2c01285} {\bibfield  {journal} {\bibinfo
  {journal} {{ACS} Omega}\ }\textbf {\bibinfo {volume} {7}},\ \bibinfo {pages}
  {14402} (\bibinfo {year} {2022})}\BibitemShut {NoStop}%
\bibitem [{\citenamefont {Giovannetti}\ \emph {et~al.}(2016)\citenamefont
  {Giovannetti}, \citenamefont {Puggioni}, \citenamefont {Rondinelli},\ and\
  \citenamefont {Capone}}]{Giovannetti2016}%
  \BibitemOpen
  \bibfield  {author} {\bibinfo {author} {\bibfnamefont {G.}~\bibnamefont
  {Giovannetti}}, \bibinfo {author} {\bibfnamefont {D.}~\bibnamefont
  {Puggioni}}, \bibinfo {author} {\bibfnamefont {J.~M.}\ \bibnamefont
  {Rondinelli}},\ and\ \bibinfo {author} {\bibfnamefont {M.}~\bibnamefont
  {Capone}},\ }\bibfield  {title} {\bibinfo {title} {Interplay between electron
  correlations and polar displacements in metallic {SrEuMo$_2$O$_6$}},\
  }\bibfield  {journal} {\bibinfo  {journal} {Physical Review B}\ }\textbf
  {\bibinfo {volume} {93}},\ \href {https://doi.org/10.1103/physrevb.93.115147}
  {10.1103/physrevb.93.115147} (\bibinfo {year} {2016})\BibitemShut {NoStop}%
\bibitem [{\citenamefont {Puggioni}\ \emph {et~al.}(2015)\citenamefont
  {Puggioni}, \citenamefont {Giovannetti}, \citenamefont {Capone},\ and\
  \citenamefont {Rondinelli}}]{Puggioni2015}%
  \BibitemOpen
  \bibfield  {author} {\bibinfo {author} {\bibfnamefont {D.}~\bibnamefont
  {Puggioni}}, \bibinfo {author} {\bibfnamefont {G.}~\bibnamefont
  {Giovannetti}}, \bibinfo {author} {\bibfnamefont {M.}~\bibnamefont
  {Capone}},\ and\ \bibinfo {author} {\bibfnamefont {J.~M.}\ \bibnamefont
  {Rondinelli}},\ }\bibfield  {title} {\bibinfo {title} {Design of a {Mott}
  multiferroic from a nonmagnetic polar metal},\ }\bibfield  {journal}
  {\bibinfo  {journal} {Physical Review Letters}\ }\textbf {\bibinfo {volume}
  {115}},\ \href {https://doi.org/10.1103/physrevlett.115.087202}
  {10.1103/physrevlett.115.087202} (\bibinfo {year} {2015})\BibitemShut
  {NoStop}%
\bibitem [{\citenamefont {M\"uhlbauer}\ \emph {et~al.}(2009)\citenamefont
  {M\"uhlbauer}, \citenamefont {Binz}, \citenamefont {Jonietz}, \citenamefont
  {Pfleiderer}, \citenamefont {Rosch}, \citenamefont {Neubauer}, \citenamefont
  {Georgii},\ and\ \citenamefont {B\"oni}}]{Muhlbauer/etal:2009Science}%
  \BibitemOpen
  \bibfield  {author} {\bibinfo {author} {\bibfnamefont {S.}~\bibnamefont
  {M\"uhlbauer}}, \bibinfo {author} {\bibfnamefont {B.}~\bibnamefont {Binz}},
  \bibinfo {author} {\bibfnamefont {F.}~\bibnamefont {Jonietz}}, \bibinfo
  {author} {\bibfnamefont {C.}~\bibnamefont {Pfleiderer}}, \bibinfo {author}
  {\bibfnamefont {A.}~\bibnamefont {Rosch}}, \bibinfo {author} {\bibfnamefont
  {A.}~\bibnamefont {Neubauer}}, \bibinfo {author} {\bibfnamefont
  {R.}~\bibnamefont {Georgii}},\ and\ \bibinfo {author} {\bibfnamefont
  {P.}~\bibnamefont {B\"oni}},\ }\bibfield  {title} {\bibinfo {title} {Skyrmion
  lattice in a chiral magnet},\ }\href
  {https://doi.org/10.1126/science.1166767} {\bibfield  {journal} {\bibinfo
  {journal} {Science}\ }\textbf {\bibinfo {volume} {323}},\ \bibinfo {pages}
  {915} (\bibinfo {year} {2009})}\BibitemShut {NoStop}%
\bibitem [{\citenamefont {Schulz}\ \emph {et~al.}(2012)\citenamefont {Schulz},
  \citenamefont {Ritz}, \citenamefont {Bauer}, \citenamefont {Halder},
  \citenamefont {Wagner}, \citenamefont {Franz}, \citenamefont {Pfleiderer},
  \citenamefont {Everschor}, \citenamefont {Garst},\ and\ \citenamefont
  {Rosch}}]{Schulz2012}%
  \BibitemOpen
  \bibfield  {author} {\bibinfo {author} {\bibfnamefont {T.}~\bibnamefont
  {Schulz}}, \bibinfo {author} {\bibfnamefont {R.}~\bibnamefont {Ritz}},
  \bibinfo {author} {\bibfnamefont {A.}~\bibnamefont {Bauer}}, \bibinfo
  {author} {\bibfnamefont {M.}~\bibnamefont {Halder}}, \bibinfo {author}
  {\bibfnamefont {M.}~\bibnamefont {Wagner}}, \bibinfo {author} {\bibfnamefont
  {C.}~\bibnamefont {Franz}}, \bibinfo {author} {\bibfnamefont
  {C.}~\bibnamefont {Pfleiderer}}, \bibinfo {author} {\bibfnamefont
  {K.}~\bibnamefont {Everschor}}, \bibinfo {author} {\bibfnamefont
  {M.}~\bibnamefont {Garst}},\ and\ \bibinfo {author} {\bibfnamefont
  {A.}~\bibnamefont {Rosch}},\ }\bibfield  {title} {\bibinfo {title} {Emergent
  electrodynamics of skyrmions in a chiral magnet},\ }\href
  {https://doi.org/10.1038/nphys2231} {\bibfield  {journal} {\bibinfo
  {journal} {Nature Physics}\ }\textbf {\bibinfo {volume} {8}},\ \bibinfo
  {pages} {301} (\bibinfo {year} {2012})}\BibitemShut {NoStop}%
\bibitem [{\citenamefont {Yu}\ \emph {et~al.}(2010)\citenamefont {Yu},
  \citenamefont {Onose}, \citenamefont {Kanazawa}, \citenamefont {Park},
  \citenamefont {Han}, \citenamefont {Matsui}, \citenamefont {Nagaosa},\ and\
  \citenamefont {Tokura}}]{Yu2010}%
  \BibitemOpen
  \bibfield  {author} {\bibinfo {author} {\bibfnamefont {X.~Z.}\ \bibnamefont
  {Yu}}, \bibinfo {author} {\bibfnamefont {Y.}~\bibnamefont {Onose}}, \bibinfo
  {author} {\bibfnamefont {N.}~\bibnamefont {Kanazawa}}, \bibinfo {author}
  {\bibfnamefont {J.~H.}\ \bibnamefont {Park}}, \bibinfo {author}
  {\bibfnamefont {J.~H.}\ \bibnamefont {Han}}, \bibinfo {author} {\bibfnamefont
  {Y.}~\bibnamefont {Matsui}}, \bibinfo {author} {\bibfnamefont
  {N.}~\bibnamefont {Nagaosa}},\ and\ \bibinfo {author} {\bibfnamefont
  {Y.}~\bibnamefont {Tokura}},\ }\bibfield  {title} {\bibinfo {title}
  {Real-space observation of a two-dimensional skyrmion crystal},\ }\href
  {https://doi.org/10.1038/nature09124} {\bibfield  {journal} {\bibinfo
  {journal} {Nature}\ }\textbf {\bibinfo {volume} {465}},\ \bibinfo {pages}
  {901} (\bibinfo {year} {2010})}\BibitemShut {NoStop}%
\bibitem [{\citenamefont {Cheong}\ and\ \citenamefont {Xu}(2022)}]{Cheong2022}%
  \BibitemOpen
  \bibfield  {author} {\bibinfo {author} {\bibfnamefont {S.-W.}\ \bibnamefont
  {Cheong}}\ and\ \bibinfo {author} {\bibfnamefont {X.}~\bibnamefont {Xu}},\
  }\bibfield  {title} {\bibinfo {title} {Magnetic chirality},\ }\bibfield
  {journal} {\bibinfo  {journal} {npj Quantum Materials}\ }\textbf {\bibinfo
  {volume} {7}},\ \href {https://doi.org/10.1038/s41535-022-00447-5}
  {10.1038/s41535-022-00447-5} (\bibinfo {year} {2022})\BibitemShut {NoStop}%
\bibitem [{IEE(2003)}]{IEEE2003}%
  \BibitemOpen
  \bibfield  {title} {\bibinfo {title} {An american national standard {IEEE}
  standard definitions of terms associated with ferroelectric and related
  materials},\ }\href {https://doi.org/10.1109/tuffc.2003.1256302} {\bibfield
  {journal} {\bibinfo  {journal} {{IEEE} Transactions on Ultrasonics,
  Ferroelectrics and Frequency Control}\ }\textbf {\bibinfo {volume} {50}},\
  \bibinfo {pages} {1} (\bibinfo {year} {2003})}\BibitemShut {NoStop}%
\bibitem [{\citenamefont {Aizu}(1969)}]{doi:10.1143/JPSJ.27.387}%
  \BibitemOpen
  \bibfield  {author} {\bibinfo {author} {\bibfnamefont {K.}~\bibnamefont
  {Aizu}},\ }\bibfield  {title} {\bibinfo {title} {Possible species of
  “ferroelastic” crystals and of simultaneously ferroelectric and
  ferroelastic crystals},\ }\href {https://doi.org/10.1143/JPSJ.27.387}
  {\bibfield  {journal} {\bibinfo  {journal} {Journal of the Physical Society
  of Japan}\ }\textbf {\bibinfo {volume} {27}},\ \bibinfo {pages} {387}
  (\bibinfo {year} {1969})}\BibitemShut {NoStop}%
\bibitem [{\citenamefont {Sakai}\ \emph {et~al.}(2016)\citenamefont {Sakai},
  \citenamefont {Ikeura}, \citenamefont {Bahramy}, \citenamefont {Ogawa},
  \citenamefont {Hashizume}, \citenamefont {Fujioka}, \citenamefont {Tokura},\
  and\ \citenamefont {Ishiwata}}]{Sakai2016}%
  \BibitemOpen
  \bibfield  {author} {\bibinfo {author} {\bibfnamefont {H.}~\bibnamefont
  {Sakai}}, \bibinfo {author} {\bibfnamefont {K.}~\bibnamefont {Ikeura}},
  \bibinfo {author} {\bibfnamefont {M.~S.}\ \bibnamefont {Bahramy}}, \bibinfo
  {author} {\bibfnamefont {N.}~\bibnamefont {Ogawa}}, \bibinfo {author}
  {\bibfnamefont {D.}~\bibnamefont {Hashizume}}, \bibinfo {author}
  {\bibfnamefont {J.}~\bibnamefont {Fujioka}}, \bibinfo {author} {\bibfnamefont
  {Y.}~\bibnamefont {Tokura}},\ and\ \bibinfo {author} {\bibfnamefont
  {S.}~\bibnamefont {Ishiwata}},\ }\bibfield  {title} {\bibinfo {title}
  {Critical enhancement of thermopower in a chemically tuned polar semimetal
  {MoTe}$_2$},\ }\href {https://doi.org/10.1126/sciadv.1601378} {\bibfield
  {journal} {\bibinfo  {journal} {Science Advances}\ }\textbf {\bibinfo
  {volume} {2}},\ \bibinfo {pages} {e1601378} (\bibinfo {year}
  {2016})}\BibitemShut {NoStop}%
\bibitem [{\citenamefont {Jin}\ \emph {et~al.}(2019)\citenamefont {Jin},
  \citenamefont {Wang}, \citenamefont {Zhang}, \citenamefont {Ji},
  \citenamefont {Shi}, \citenamefont {Wang}, \citenamefont {Yu}, \citenamefont
  {Zhang}, \citenamefont {Plummer},\ and\ \citenamefont {Zhang}}]{Jin2019}%
  \BibitemOpen
  \bibfield  {author} {\bibinfo {author} {\bibfnamefont {F.}~\bibnamefont
  {Jin}}, \bibinfo {author} {\bibfnamefont {L.}~\bibnamefont {Wang}}, \bibinfo
  {author} {\bibfnamefont {A.}~\bibnamefont {Zhang}}, \bibinfo {author}
  {\bibfnamefont {J.}~\bibnamefont {Ji}}, \bibinfo {author} {\bibfnamefont
  {Y.}~\bibnamefont {Shi}}, \bibinfo {author} {\bibfnamefont {X.}~\bibnamefont
  {Wang}}, \bibinfo {author} {\bibfnamefont {R.}~\bibnamefont {Yu}}, \bibinfo
  {author} {\bibfnamefont {J.}~\bibnamefont {Zhang}}, \bibinfo {author}
  {\bibfnamefont {E.~W.}\ \bibnamefont {Plummer}},\ and\ \bibinfo {author}
  {\bibfnamefont {Q.}~\bibnamefont {Zhang}},\ }\bibfield  {title} {\bibinfo
  {title} {Raman interrogation of the ferroelectric phase transition in polar
  metal {LiOsO}$_3$},\ }\href {https://doi.org/10.1073/pnas.1908956116}
  {\bibfield  {journal} {\bibinfo  {journal} {Proceedings of the National
  Academy of Sciences}\ }\textbf {\bibinfo {volume} {116}},\ \bibinfo {pages}
  {20322} (\bibinfo {year} {2019})}\BibitemShut {NoStop}%
\bibitem [{\citenamefont {Stone}\ \emph {et~al.}(2019)\citenamefont {Stone},
  \citenamefont {Puggioni}, \citenamefont {Lei}, \citenamefont {Gu},
  \citenamefont {Wang}, \citenamefont {Wang}, \citenamefont {Ge}, \citenamefont
  {Lu}, \citenamefont {Mao}, \citenamefont {Rondinelli},\ and\ \citenamefont
  {Gopalan}}]{Stone2019}%
  \BibitemOpen
  \bibfield  {author} {\bibinfo {author} {\bibfnamefont {G.}~\bibnamefont
  {Stone}}, \bibinfo {author} {\bibfnamefont {D.}~\bibnamefont {Puggioni}},
  \bibinfo {author} {\bibfnamefont {S.}~\bibnamefont {Lei}}, \bibinfo {author}
  {\bibfnamefont {M.}~\bibnamefont {Gu}}, \bibinfo {author} {\bibfnamefont
  {K.}~\bibnamefont {Wang}}, \bibinfo {author} {\bibfnamefont {Y.}~\bibnamefont
  {Wang}}, \bibinfo {author} {\bibfnamefont {J.}~\bibnamefont {Ge}}, \bibinfo
  {author} {\bibfnamefont {X.-Z.}\ \bibnamefont {Lu}}, \bibinfo {author}
  {\bibfnamefont {Z.}~\bibnamefont {Mao}}, \bibinfo {author} {\bibfnamefont
  {J.~M.}\ \bibnamefont {Rondinelli}},\ and\ \bibinfo {author} {\bibfnamefont
  {V.}~\bibnamefont {Gopalan}},\ }\bibfield  {title} {\bibinfo {title} {Atomic
  and electronic structure of domains walls in a polar metal},\ }\bibfield
  {journal} {\bibinfo  {journal} {Physical Review B}\ }\textbf {\bibinfo
  {volume} {99}},\ \href {https://doi.org/10.1103/physrevb.99.014105}
  {10.1103/physrevb.99.014105} (\bibinfo {year} {2019})\BibitemShut {NoStop}%
\bibitem [{\citenamefont {Kim}\ \emph {et~al.}(2016)\citenamefont {Kim},
  \citenamefont {Puggioni}, \citenamefont {Yuan}, \citenamefont {Xie},
  \citenamefont {Zhou}, \citenamefont {Campbell}, \citenamefont {Ryan},
  \citenamefont {Choi}, \citenamefont {Kim}, \citenamefont {Patzner},
  \citenamefont {Ryu}, \citenamefont {Podkaminer}, \citenamefont {Irwin},
  \citenamefont {Ma}, \citenamefont {Fennie}, \citenamefont {Rzchowski},
  \citenamefont {Pan}, \citenamefont {Gopalan}, \citenamefont {Rondinelli},\
  and\ \citenamefont {Eom}}]{Kim/Puggioni:2016Nature}%
  \BibitemOpen
  \bibfield  {author} {\bibinfo {author} {\bibfnamefont {T.~H.}\ \bibnamefont
  {Kim}}, \bibinfo {author} {\bibfnamefont {D.}~\bibnamefont {Puggioni}},
  \bibinfo {author} {\bibfnamefont {Y.}~\bibnamefont {Yuan}}, \bibinfo {author}
  {\bibfnamefont {L.}~\bibnamefont {Xie}}, \bibinfo {author} {\bibfnamefont
  {H.}~\bibnamefont {Zhou}}, \bibinfo {author} {\bibfnamefont {N.}~\bibnamefont
  {Campbell}}, \bibinfo {author} {\bibfnamefont {P.~J.}\ \bibnamefont {Ryan}},
  \bibinfo {author} {\bibfnamefont {Y.}~\bibnamefont {Choi}}, \bibinfo {author}
  {\bibfnamefont {J.-W.}\ \bibnamefont {Kim}}, \bibinfo {author} {\bibfnamefont
  {J.~R.}\ \bibnamefont {Patzner}}, \bibinfo {author} {\bibfnamefont
  {S.}~\bibnamefont {Ryu}}, \bibinfo {author} {\bibfnamefont {J.~P.}\
  \bibnamefont {Podkaminer}}, \bibinfo {author} {\bibfnamefont
  {J.}~\bibnamefont {Irwin}}, \bibinfo {author} {\bibfnamefont
  {Y.}~\bibnamefont {Ma}}, \bibinfo {author} {\bibfnamefont {C.~J.}\
  \bibnamefont {Fennie}}, \bibinfo {author} {\bibfnamefont {M.~S.}\
  \bibnamefont {Rzchowski}}, \bibinfo {author} {\bibfnamefont {X.~Q.}\
  \bibnamefont {Pan}}, \bibinfo {author} {\bibfnamefont {V.}~\bibnamefont
  {Gopalan}}, \bibinfo {author} {\bibfnamefont {J.~M.}\ \bibnamefont
  {Rondinelli}},\ and\ \bibinfo {author} {\bibfnamefont {C.~B.}\ \bibnamefont
  {Eom}},\ }\bibfield  {title} {\bibinfo {title} {Polar metals by geometric
  design},\ }\href {https://doi.org/10.1038/nature17628} {\bibfield  {journal}
  {\bibinfo  {journal} {Nature}\ }\textbf {\bibinfo {volume} {533}},\ \bibinfo
  {pages} {68} (\bibinfo {year} {2016})}\BibitemShut {NoStop}%
\bibitem [{\citenamefont {Padmanabhan}\ \emph {et~al.}(2018)\citenamefont
  {Padmanabhan}, \citenamefont {Park}, \citenamefont {Puggioni}, \citenamefont
  {Yuan}, \citenamefont {Cao}, \citenamefont {Gasparov}, \citenamefont {Shi},
  \citenamefont {Chakhalian}, \citenamefont {Rondinelli},\ and\ \citenamefont
  {Gopalan}}]{Padmanabhan2018}%
  \BibitemOpen
  \bibfield  {author} {\bibinfo {author} {\bibfnamefont {H.}~\bibnamefont
  {Padmanabhan}}, \bibinfo {author} {\bibfnamefont {Y.}~\bibnamefont {Park}},
  \bibinfo {author} {\bibfnamefont {D.}~\bibnamefont {Puggioni}}, \bibinfo
  {author} {\bibfnamefont {Y.}~\bibnamefont {Yuan}}, \bibinfo {author}
  {\bibfnamefont {Y.}~\bibnamefont {Cao}}, \bibinfo {author} {\bibfnamefont
  {L.}~\bibnamefont {Gasparov}}, \bibinfo {author} {\bibfnamefont
  {Y.}~\bibnamefont {Shi}}, \bibinfo {author} {\bibfnamefont {J.}~\bibnamefont
  {Chakhalian}}, \bibinfo {author} {\bibfnamefont {J.~M.}\ \bibnamefont
  {Rondinelli}},\ and\ \bibinfo {author} {\bibfnamefont {V.}~\bibnamefont
  {Gopalan}},\ }\bibfield  {title} {\bibinfo {title} {Linear and nonlinear
  optical probe of the ferroelectric-like phase transition in a polar metal,
  {LiOsO}$_3$},\ }\href {https://doi.org/10.1063/1.5042769} {\bibfield
  {journal} {\bibinfo  {journal} {Applied Physics Letters}\ }\textbf {\bibinfo
  {volume} {113}},\ \bibinfo {pages} {122906} (\bibinfo {year}
  {2018})}\BibitemShut {NoStop}%
\bibitem [{\citenamefont {Laurita}\ \emph {et~al.}(2019)\citenamefont
  {Laurita}, \citenamefont {Ron}, \citenamefont {Shan}, \citenamefont
  {Puggioni}, \citenamefont {Koocher}, \citenamefont {Yamaura}, \citenamefont
  {Shi}, \citenamefont {Rondinelli},\ and\ \citenamefont
  {Hsieh}}]{laurita2019evidence}%
  \BibitemOpen
  \bibfield  {author} {\bibinfo {author} {\bibfnamefont {N.~J.}\ \bibnamefont
  {Laurita}}, \bibinfo {author} {\bibfnamefont {A.}~\bibnamefont {Ron}},
  \bibinfo {author} {\bibfnamefont {J.-Y.}\ \bibnamefont {Shan}}, \bibinfo
  {author} {\bibfnamefont {D.}~\bibnamefont {Puggioni}}, \bibinfo {author}
  {\bibfnamefont {N.~Z.}\ \bibnamefont {Koocher}}, \bibinfo {author}
  {\bibfnamefont {K.}~\bibnamefont {Yamaura}}, \bibinfo {author} {\bibfnamefont
  {Y.}~\bibnamefont {Shi}}, \bibinfo {author} {\bibfnamefont {J.~M.}\
  \bibnamefont {Rondinelli}},\ and\ \bibinfo {author} {\bibfnamefont
  {D.}~\bibnamefont {Hsieh}},\ }\bibfield  {title} {\bibinfo {title} {Evidence
  for the weakly coupled electron mechanism in an {Anderson-Blount} polar
  metal},\ }\bibfield  {journal} {\bibinfo  {journal} {Nature Communications}\
  }\textbf {\bibinfo {volume} {10}},\ \href
  {https://doi.org/10.1038/s41467-019-11172-2} {10.1038/s41467-019-11172-2}
  (\bibinfo {year} {2019})\BibitemShut {NoStop}%
\bibitem [{\citenamefont {Wadhawan}(1984)}]{Wadhawan1984}%
  \BibitemOpen
  \bibfield  {author} {\bibinfo {author} {\bibfnamefont {V.~K.}\ \bibnamefont
  {Wadhawan}},\ }\bibfield  {title} {\bibinfo {title} {Ferroelasticity},\
  }\href {https://doi.org/10.1007/bf02744001} {\bibfield  {journal} {\bibinfo
  {journal} {Bulletin of Materials Science}\ }\textbf {\bibinfo {volume} {6}},\
  \bibinfo {pages} {733} (\bibinfo {year} {1984})}\BibitemShut {NoStop}%
\bibitem [{\citenamefont {Slosarek}\ \emph {et~al.}(1989)\citenamefont
  {Slosarek}, \citenamefont {Heuer}, \citenamefont {Zimmermann},\ and\
  \citenamefont {Haeberlen}}]{Slosarek1989}%
  \BibitemOpen
  \bibfield  {author} {\bibinfo {author} {\bibfnamefont {G.}~\bibnamefont
  {Slosarek}}, \bibinfo {author} {\bibfnamefont {A.}~\bibnamefont {Heuer}},
  \bibinfo {author} {\bibfnamefont {H.}~\bibnamefont {Zimmermann}},\ and\
  \bibinfo {author} {\bibfnamefont {U.}~\bibnamefont {Haeberlen}},\ }\bibfield
  {title} {\bibinfo {title} {A study of the paraelectric-ferroelectric phase
  transition of triglycine sulphate by deuteron nuclear magnetic resonance and
  relaxation},\ }\href {https://doi.org/10.1088/0953-8984/1/34/011} {\bibfield
  {journal} {\bibinfo  {journal} {Journal of Physics: Condensed Matter}\
  }\textbf {\bibinfo {volume} {1}},\ \bibinfo {pages} {5931} (\bibinfo {year}
  {1989})}\BibitemShut {NoStop}%
\bibitem [{\citenamefont {Gopalan}\ \emph {et~al.}(2007)\citenamefont
  {Gopalan}, \citenamefont {Dierolf},\ and\ \citenamefont
  {Scrymgeour}}]{Gopalan2007}%
  \BibitemOpen
  \bibfield  {author} {\bibinfo {author} {\bibfnamefont {V.}~\bibnamefont
  {Gopalan}}, \bibinfo {author} {\bibfnamefont {V.}~\bibnamefont {Dierolf}},\
  and\ \bibinfo {author} {\bibfnamefont {D.~A.}\ \bibnamefont {Scrymgeour}},\
  }\bibfield  {title} {\bibinfo {title} {Defect{\textendash}domain wall
  interactions in trigonal ferroelectrics},\ }\href
  {https://doi.org/10.1146/annurev.matsci.37.052506.084247} {\bibfield
  {journal} {\bibinfo  {journal} {Annual Review of Materials Research}\
  }\textbf {\bibinfo {volume} {37}},\ \bibinfo {pages} {449} (\bibinfo {year}
  {2007})}\BibitemShut {NoStop}%
\bibitem [{\citenamefont {Shapovalov}\ \emph {et~al.}(2014)\citenamefont
  {Shapovalov}, \citenamefont {Yudin}, \citenamefont {Tagantsev}, \citenamefont
  {Eliseev}, \citenamefont {Morozovska},\ and\ \citenamefont
  {Setter}}]{Shapovalov2014}%
  \BibitemOpen
  \bibfield  {author} {\bibinfo {author} {\bibfnamefont {K.}~\bibnamefont
  {Shapovalov}}, \bibinfo {author} {\bibfnamefont {P.}~\bibnamefont {Yudin}},
  \bibinfo {author} {\bibfnamefont {A.}~\bibnamefont {Tagantsev}}, \bibinfo
  {author} {\bibfnamefont {E.}~\bibnamefont {Eliseev}}, \bibinfo {author}
  {\bibfnamefont {A.}~\bibnamefont {Morozovska}},\ and\ \bibinfo {author}
  {\bibfnamefont {N.}~\bibnamefont {Setter}},\ }\bibfield  {title} {\bibinfo
  {title} {Elastic coupling between nonferroelastic domain walls},\ }\bibfield
  {journal} {\bibinfo  {journal} {Physical Review Letters}\ }\textbf {\bibinfo
  {volume} {113}},\ \href {https://doi.org/10.1103/physrevlett.113.207601}
  {10.1103/physrevlett.113.207601} (\bibinfo {year} {2014})\BibitemShut
  {NoStop}%
\bibitem [{\citenamefont {Fang}\ and\ \citenamefont {Chen}(2020)}]{Fang2020}%
  \BibitemOpen
  \bibfield  {author} {\bibinfo {author} {\bibfnamefont {Y.-W.}\ \bibnamefont
  {Fang}}\ and\ \bibinfo {author} {\bibfnamefont {H.}~\bibnamefont {Chen}},\
  }\bibfield  {title} {\bibinfo {title} {Design of a multifunctional polar
  metal via first-principles high-throughput structure screening},\ }\bibfield
  {journal} {\bibinfo  {journal} {Communications Materials}\ }\textbf {\bibinfo
  {volume} {1}},\ \href {https://doi.org/10.1038/s43246-019-0005-6}
  {10.1038/s43246-019-0005-6} (\bibinfo {year} {2020})\BibitemShut {NoStop}%
\bibitem [{\citenamefont {Watanabe}\ and\ \citenamefont
  {Yanase}(2017)}]{Watanabe2017}%
  \BibitemOpen
  \bibfield  {author} {\bibinfo {author} {\bibfnamefont {H.}~\bibnamefont
  {Watanabe}}\ and\ \bibinfo {author} {\bibfnamefont {Y.}~\bibnamefont
  {Yanase}},\ }\bibfield  {title} {\bibinfo {title} {Magnetic hexadecapole
  order and magnetopiezoelectric metal state in
  {Ba$_{1-x}$K$_x$Mn$_2$As$_2$}},\ }\bibfield  {journal} {\bibinfo  {journal}
  {Physical Review B}\ }\textbf {\bibinfo {volume} {96}},\ \href
  {https://doi.org/10.1103/physrevb.96.064432} {10.1103/physrevb.96.064432}
  (\bibinfo {year} {2017})\BibitemShut {NoStop}%
\bibitem [{\citenamefont {Bauer}\ \emph {et~al.}(2004)\citenamefont {Bauer},
  \citenamefont {Hilscher}, \citenamefont {Michor}, \citenamefont {Paul},
  \citenamefont {Scheidt}, \citenamefont {Gribanov}, \citenamefont {Seropegin},
  \citenamefont {Noël}, \citenamefont {Sigrist},\ and\ \citenamefont
  {Rogl}}]{Bauer2004}%
  \BibitemOpen
  \bibfield  {author} {\bibinfo {author} {\bibfnamefont {E.}~\bibnamefont
  {Bauer}}, \bibinfo {author} {\bibfnamefont {G.}~\bibnamefont {Hilscher}},
  \bibinfo {author} {\bibfnamefont {H.}~\bibnamefont {Michor}}, \bibinfo
  {author} {\bibfnamefont {C.}~\bibnamefont {Paul}}, \bibinfo {author}
  {\bibfnamefont {E.~W.}\ \bibnamefont {Scheidt}}, \bibinfo {author}
  {\bibfnamefont {A.}~\bibnamefont {Gribanov}}, \bibinfo {author}
  {\bibfnamefont {Y.}~\bibnamefont {Seropegin}}, \bibinfo {author}
  {\bibfnamefont {H.}~\bibnamefont {Noël}}, \bibinfo {author} {\bibfnamefont
  {M.}~\bibnamefont {Sigrist}},\ and\ \bibinfo {author} {\bibfnamefont
  {P.}~\bibnamefont {Rogl}},\ }\bibfield  {title} {\bibinfo {title} {Heavy
  fermion superconductivity and magnetic order in noncentrosymmetric
  {CePt$_3$Si}},\ }\bibfield  {journal} {\bibinfo  {journal} {Physical Review
  Letters}\ }\textbf {\bibinfo {volume} {92}},\ \href
  {https://doi.org/10.1103/physrevlett.92.027003}
  {10.1103/physrevlett.92.027003} (\bibinfo {year} {2004})\BibitemShut
  {NoStop}%
\bibitem [{\citenamefont {Jiao}\ \emph {et~al.}(2020)\citenamefont {Jiao},
  \citenamefont {Fang}, \citenamefont {Sun}, \citenamefont {Shan},
  \citenamefont {Yu}, \citenamefont {Feng}, \citenamefont {Wang}, \citenamefont
  {Ma}, \citenamefont {Uwatoko}, \citenamefont {Yamaura}, \citenamefont {Guo},
  \citenamefont {Chen},\ and\ \citenamefont {Cheng}}]{Jiao2020}%
  \BibitemOpen
  \bibfield  {author} {\bibinfo {author} {\bibfnamefont {Y.}~\bibnamefont
  {Jiao}}, \bibinfo {author} {\bibfnamefont {Y.-W.}\ \bibnamefont {Fang}},
  \bibinfo {author} {\bibfnamefont {J.}~\bibnamefont {Sun}}, \bibinfo {author}
  {\bibfnamefont {P.}~\bibnamefont {Shan}}, \bibinfo {author} {\bibfnamefont
  {Z.}~\bibnamefont {Yu}}, \bibinfo {author} {\bibfnamefont {H.~L.}\
  \bibnamefont {Feng}}, \bibinfo {author} {\bibfnamefont {B.}~\bibnamefont
  {Wang}}, \bibinfo {author} {\bibfnamefont {H.}~\bibnamefont {Ma}}, \bibinfo
  {author} {\bibfnamefont {Y.}~\bibnamefont {Uwatoko}}, \bibinfo {author}
  {\bibfnamefont {K.}~\bibnamefont {Yamaura}}, \bibinfo {author} {\bibfnamefont
  {Y.}~\bibnamefont {Guo}}, \bibinfo {author} {\bibfnamefont {H.}~\bibnamefont
  {Chen}},\ and\ \bibinfo {author} {\bibfnamefont {J.}~\bibnamefont {Cheng}},\
  }\bibfield  {title} {\bibinfo {title} {Coupled magnetic and structural phase
  transitions in the antiferromagnetic polar metal {Pb$_2$CoOsO$_6$} under
  pressure},\ }\bibfield  {journal} {\bibinfo  {journal} {Physical Review B}\
  }\textbf {\bibinfo {volume} {102}},\ \href
  {https://doi.org/10.1103/physrevb.102.144418} {10.1103/physrevb.102.144418}
  (\bibinfo {year} {2020})\BibitemShut {NoStop}%
\bibitem [{\citenamefont {Rischau}\ \emph {et~al.}(2017)\citenamefont
  {Rischau}, \citenamefont {Lin}, \citenamefont {Grams}, \citenamefont {Finck},
  \citenamefont {Harms}, \citenamefont {Engelmayer}, \citenamefont {Lorenz},
  \citenamefont {Gallais}, \citenamefont {Fauqu{\'{e}}}, \citenamefont
  {Hemberger},\ and\ \citenamefont {Behnia}}]{Rischau2017}%
  \BibitemOpen
  \bibfield  {author} {\bibinfo {author} {\bibfnamefont {C.~W.}\ \bibnamefont
  {Rischau}}, \bibinfo {author} {\bibfnamefont {X.}~\bibnamefont {Lin}},
  \bibinfo {author} {\bibfnamefont {C.~P.}\ \bibnamefont {Grams}}, \bibinfo
  {author} {\bibfnamefont {D.}~\bibnamefont {Finck}}, \bibinfo {author}
  {\bibfnamefont {S.}~\bibnamefont {Harms}}, \bibinfo {author} {\bibfnamefont
  {J.}~\bibnamefont {Engelmayer}}, \bibinfo {author} {\bibfnamefont
  {T.}~\bibnamefont {Lorenz}}, \bibinfo {author} {\bibfnamefont
  {Y.}~\bibnamefont {Gallais}}, \bibinfo {author} {\bibfnamefont
  {B.}~\bibnamefont {Fauqu{\'{e}}}}, \bibinfo {author} {\bibfnamefont
  {J.}~\bibnamefont {Hemberger}},\ and\ \bibinfo {author} {\bibfnamefont
  {K.}~\bibnamefont {Behnia}},\ }\bibfield  {title} {\bibinfo {title} {A
  ferroelectric quantum phase transition inside the superconducting dome of
  {Sr$_{1-x}$Ca$_x$TiO$_{3-\delta}$}},\ }\href
  {https://doi.org/10.1038/nphys4085} {\bibfield  {journal} {\bibinfo
  {journal} {Nature Physics}\ }\textbf {\bibinfo {volume} {13}},\ \bibinfo
  {pages} {643} (\bibinfo {year} {2017})}\BibitemShut {NoStop}%
\bibitem [{\citenamefont {Berger}\ \emph {et~al.}(2021)\citenamefont {Berger},
  \citenamefont {Jamnuch}, \citenamefont {Uzundal}, \citenamefont {Woodahl},
  \citenamefont {Padmanabhan}, \citenamefont {Amado}, \citenamefont {Manset},
  \citenamefont {Hirata}, \citenamefont {Kubota}, \citenamefont {Owada},
  \citenamefont {Tono}, \citenamefont {Yabashi}, \citenamefont {Wang},
  \citenamefont {Shi}, \citenamefont {Gopalan}, \citenamefont {Schwartz},
  \citenamefont {Drisdell}, \citenamefont {Matsuda}, \citenamefont {Freeland},
  \citenamefont {Pascal},\ and\ \citenamefont {Zuerch}}]{Berger2021}%
  \BibitemOpen
  \bibfield  {author} {\bibinfo {author} {\bibfnamefont {E.}~\bibnamefont
  {Berger}}, \bibinfo {author} {\bibfnamefont {S.}~\bibnamefont {Jamnuch}},
  \bibinfo {author} {\bibfnamefont {C.~B.}\ \bibnamefont {Uzundal}}, \bibinfo
  {author} {\bibfnamefont {C.}~\bibnamefont {Woodahl}}, \bibinfo {author}
  {\bibfnamefont {H.}~\bibnamefont {Padmanabhan}}, \bibinfo {author}
  {\bibfnamefont {A.}~\bibnamefont {Amado}}, \bibinfo {author} {\bibfnamefont
  {P.}~\bibnamefont {Manset}}, \bibinfo {author} {\bibfnamefont
  {Y.}~\bibnamefont {Hirata}}, \bibinfo {author} {\bibfnamefont
  {Y.}~\bibnamefont {Kubota}}, \bibinfo {author} {\bibfnamefont
  {S.}~\bibnamefont {Owada}}, \bibinfo {author} {\bibfnamefont
  {K.}~\bibnamefont {Tono}}, \bibinfo {author} {\bibfnamefont {M.}~\bibnamefont
  {Yabashi}}, \bibinfo {author} {\bibfnamefont {C.}~\bibnamefont {Wang}},
  \bibinfo {author} {\bibfnamefont {Y.}~\bibnamefont {Shi}}, \bibinfo {author}
  {\bibfnamefont {V.}~\bibnamefont {Gopalan}}, \bibinfo {author} {\bibfnamefont
  {C.~P.}\ \bibnamefont {Schwartz}}, \bibinfo {author} {\bibfnamefont {W.~S.}\
  \bibnamefont {Drisdell}}, \bibinfo {author} {\bibfnamefont {I.}~\bibnamefont
  {Matsuda}}, \bibinfo {author} {\bibfnamefont {J.~W.}\ \bibnamefont
  {Freeland}}, \bibinfo {author} {\bibfnamefont {T.~A.}\ \bibnamefont
  {Pascal}},\ and\ \bibinfo {author} {\bibfnamefont {M.}~\bibnamefont
  {Zuerch}},\ }\bibfield  {title} {\bibinfo {title} {Extreme ultraviolet second
  harmonic generation spectroscopy in a polar metal},\ }\href
  {https://doi.org/10.1021/acs.nanolett.1c01502} {\bibfield  {journal}
  {\bibinfo  {journal} {Nano Letters}\ }\textbf {\bibinfo {volume} {21}},\
  \bibinfo {pages} {6095} (\bibinfo {year} {2021})}\BibitemShut {NoStop}%
\bibitem [{\citenamefont {Pan}\ \emph {et~al.}(2013)\citenamefont {Pan},
  \citenamefont {Nikitin}, \citenamefont {Bay}, \citenamefont {Huang},
  \citenamefont {Paulsen}, \citenamefont {Yan},\ and\ \citenamefont
  {de~Visser}}]{Pan2013}%
  \BibitemOpen
  \bibfield  {author} {\bibinfo {author} {\bibfnamefont {Y.}~\bibnamefont
  {Pan}}, \bibinfo {author} {\bibfnamefont {A.~M.}\ \bibnamefont {Nikitin}},
  \bibinfo {author} {\bibfnamefont {T.~V.}\ \bibnamefont {Bay}}, \bibinfo
  {author} {\bibfnamefont {Y.~K.}\ \bibnamefont {Huang}}, \bibinfo {author}
  {\bibfnamefont {C.}~\bibnamefont {Paulsen}}, \bibinfo {author} {\bibfnamefont
  {B.~H.}\ \bibnamefont {Yan}},\ and\ \bibinfo {author} {\bibfnamefont
  {A.}~\bibnamefont {de~Visser}},\ }\bibfield  {title} {\bibinfo {title}
  {Superconductivity and magnetic order in the noncentrosymmetric
  half-{Heusler} compound {ErPdBi}},\ }\href
  {https://doi.org/10.1209/0295-5075/104/27001} {\bibfield  {journal} {\bibinfo
   {journal} {{EPL} (Europhysics Letters)}\ }\textbf {\bibinfo {volume}
  {104}},\ \bibinfo {pages} {27001} (\bibinfo {year} {2013})}\BibitemShut
  {NoStop}%
\bibitem [{\citenamefont {Jiang}\ \emph {et~al.}(2017)\citenamefont {Jiang},
  \citenamefont {Bai}, \citenamefont {Chen}, \citenamefont {He}, \citenamefont
  {Zhang}, \citenamefont {Zhang}, \citenamefont {Shi}, \citenamefont {Park},
  \citenamefont {Scott}, \citenamefont {Hwang},\ and\ \citenamefont
  {Jiang}}]{Jiang2017}%
  \BibitemOpen
  \bibfield  {author} {\bibinfo {author} {\bibfnamefont {J.}~\bibnamefont
  {Jiang}}, \bibinfo {author} {\bibfnamefont {Z.~L.}\ \bibnamefont {Bai}},
  \bibinfo {author} {\bibfnamefont {Z.~H.}\ \bibnamefont {Chen}}, \bibinfo
  {author} {\bibfnamefont {L.}~\bibnamefont {He}}, \bibinfo {author}
  {\bibfnamefont {D.~W.}\ \bibnamefont {Zhang}}, \bibinfo {author}
  {\bibfnamefont {Q.~H.}\ \bibnamefont {Zhang}}, \bibinfo {author}
  {\bibfnamefont {J.~A.}\ \bibnamefont {Shi}}, \bibinfo {author} {\bibfnamefont
  {M.~H.}\ \bibnamefont {Park}}, \bibinfo {author} {\bibfnamefont {J.~F.}\
  \bibnamefont {Scott}}, \bibinfo {author} {\bibfnamefont {C.~S.}\ \bibnamefont
  {Hwang}},\ and\ \bibinfo {author} {\bibfnamefont {A.~Q.}\ \bibnamefont
  {Jiang}},\ }\bibfield  {title} {\bibinfo {title} {Temporary formation of
  highly conducting domain walls for non-destructive read-out of ferroelectric
  domain-wall resistance switching memories},\ }\href
  {https://doi.org/10.1038/nmat5028} {\bibfield  {journal} {\bibinfo  {journal}
  {Nature Materials}\ }\textbf {\bibinfo {volume} {17}},\ \bibinfo {pages} {49}
  (\bibinfo {year} {2017})}\BibitemShut {NoStop}%
\bibitem [{\citenamefont {Choi}\ \emph {et~al.}(2010)\citenamefont {Choi},
  \citenamefont {Horibe}, \citenamefont {Yi}, \citenamefont {Choi},
  \citenamefont {Wu},\ and\ \citenamefont {Cheong}}]{Choi2010}%
  \BibitemOpen
  \bibfield  {author} {\bibinfo {author} {\bibfnamefont {T.}~\bibnamefont
  {Choi}}, \bibinfo {author} {\bibfnamefont {Y.}~\bibnamefont {Horibe}},
  \bibinfo {author} {\bibfnamefont {H.~T.}\ \bibnamefont {Yi}}, \bibinfo
  {author} {\bibfnamefont {Y.~J.}\ \bibnamefont {Choi}}, \bibinfo {author}
  {\bibfnamefont {W.}~\bibnamefont {Wu}},\ and\ \bibinfo {author}
  {\bibfnamefont {S.-W.}\ \bibnamefont {Cheong}},\ }\bibfield  {title}
  {\bibinfo {title} {Insulating interlocked ferroelectric and structural
  antiphase domain walls in multiferroic {YMnO}3},\ }\href
  {https://doi.org/10.1038/nmat2632} {\bibfield  {journal} {\bibinfo  {journal}
  {Nature Materials}\ }\textbf {\bibinfo {volume} {9}},\ \bibinfo {pages} {253}
  (\bibinfo {year} {2010})}\BibitemShut {NoStop}%
\bibitem [{\citenamefont {Jiang}\ \emph {et~al.}(2015)\citenamefont {Jiang},
  \citenamefont {Zhao}, \citenamefont {Zhang}, \citenamefont {Deng},
  \citenamefont {Zhang}, \citenamefont {Li}, \citenamefont {Hu}, \citenamefont
  {Yang},\ and\ \citenamefont {Chu}}]{jiang2015strain}%
  \BibitemOpen
  \bibfield  {author} {\bibinfo {author} {\bibfnamefont {K.}~\bibnamefont
  {Jiang}}, \bibinfo {author} {\bibfnamefont {R.}~\bibnamefont {Zhao}},
  \bibinfo {author} {\bibfnamefont {P.}~\bibnamefont {Zhang}}, \bibinfo
  {author} {\bibfnamefont {Q.}~\bibnamefont {Deng}}, \bibinfo {author}
  {\bibfnamefont {J.}~\bibnamefont {Zhang}}, \bibinfo {author} {\bibfnamefont
  {W.}~\bibnamefont {Li}}, \bibinfo {author} {\bibfnamefont {Z.}~\bibnamefont
  {Hu}}, \bibinfo {author} {\bibfnamefont {H.}~\bibnamefont {Yang}},\ and\
  \bibinfo {author} {\bibfnamefont {J.}~\bibnamefont {Chu}},\ }\bibfield
  {title} {\bibinfo {title} {Strain and temperature dependent absorption
  spectra studies for identifying the phase structure and band gap of
  {EuTiO$_3$} perovskite films},\ }\href@noop {} {\bibfield  {journal}
  {\bibinfo  {journal} {Physical Chemistry Chemical Physics}\ }\textbf
  {\bibinfo {volume} {17}},\ \bibinfo {pages} {31618} (\bibinfo {year}
  {2015})}\BibitemShut {NoStop}%
\bibitem [{\citenamefont {Holstad}\ \emph {et~al.}(2020)\citenamefont
  {Holstad}, \citenamefont {R{\ae}der}, \citenamefont {Evans}, \citenamefont
  {Sm{\aa}br{\aa}ten}, \citenamefont {Krohns}, \citenamefont {Schaab},
  \citenamefont {Yan}, \citenamefont {Bourret}, \citenamefont {van Helvoort},
  \citenamefont {Grande}, \citenamefont {Selbach}, \citenamefont {Agar},\ and\
  \citenamefont {Meier}}]{Holstad2020}%
  \BibitemOpen
  \bibfield  {author} {\bibinfo {author} {\bibfnamefont {T.~S.}\ \bibnamefont
  {Holstad}}, \bibinfo {author} {\bibfnamefont {T.~M.}\ \bibnamefont
  {R{\ae}der}}, \bibinfo {author} {\bibfnamefont {D.~M.}\ \bibnamefont
  {Evans}}, \bibinfo {author} {\bibfnamefont {D.~R.}\ \bibnamefont
  {Sm{\aa}br{\aa}ten}}, \bibinfo {author} {\bibfnamefont {S.}~\bibnamefont
  {Krohns}}, \bibinfo {author} {\bibfnamefont {J.}~\bibnamefont {Schaab}},
  \bibinfo {author} {\bibfnamefont {Z.}~\bibnamefont {Yan}}, \bibinfo {author}
  {\bibfnamefont {E.}~\bibnamefont {Bourret}}, \bibinfo {author} {\bibfnamefont
  {A.~T.~J.}\ \bibnamefont {van Helvoort}}, \bibinfo {author} {\bibfnamefont
  {T.}~\bibnamefont {Grande}}, \bibinfo {author} {\bibfnamefont {S.~M.}\
  \bibnamefont {Selbach}}, \bibinfo {author} {\bibfnamefont {J.~C.}\
  \bibnamefont {Agar}},\ and\ \bibinfo {author} {\bibfnamefont
  {D.}~\bibnamefont {Meier}},\ }\bibfield  {title} {\bibinfo {title}
  {Application of a long short-term memory for deconvoluting conductance
  contributions at charged ferroelectric domain walls},\ }\bibfield  {journal}
  {\bibinfo  {journal} {npj Computational Materials}\ }\textbf {\bibinfo
  {volume} {6}},\ \href {https://doi.org/10.1038/s41524-020-00426-z}
  {10.1038/s41524-020-00426-z} (\bibinfo {year} {2020})\BibitemShut {NoStop}%
\bibitem [{\citenamefont {Cao}\ \emph {et~al.}(2018)\citenamefont {Cao},
  \citenamefont {Fatemi}, \citenamefont {Fang}, \citenamefont {Watanabe},
  \citenamefont {Taniguchi}, \citenamefont {Kaxiras},\ and\ \citenamefont
  {Jarillo-Herrero}}]{Cao2018}%
  \BibitemOpen
  \bibfield  {author} {\bibinfo {author} {\bibfnamefont {Y.}~\bibnamefont
  {Cao}}, \bibinfo {author} {\bibfnamefont {V.}~\bibnamefont {Fatemi}},
  \bibinfo {author} {\bibfnamefont {S.}~\bibnamefont {Fang}}, \bibinfo {author}
  {\bibfnamefont {K.}~\bibnamefont {Watanabe}}, \bibinfo {author}
  {\bibfnamefont {T.}~\bibnamefont {Taniguchi}}, \bibinfo {author}
  {\bibfnamefont {E.}~\bibnamefont {Kaxiras}},\ and\ \bibinfo {author}
  {\bibfnamefont {P.}~\bibnamefont {Jarillo-Herrero}},\ }\bibfield  {title}
  {\bibinfo {title} {Unconventional superconductivity in magic-angle graphene
  superlattices},\ }\href {https://doi.org/10.1038/nature26160} {\bibfield
  {journal} {\bibinfo  {journal} {Nature}\ }\textbf {\bibinfo {volume} {556}},\
  \bibinfo {pages} {43} (\bibinfo {year} {2018})}\BibitemShut {NoStop}%
\bibitem [{\citenamefont {Zhou}\ \emph {et~al.}(2019)\citenamefont {Zhou},
  \citenamefont {Wu}, \citenamefont {Zhou}, \citenamefont {Zeng}, \citenamefont
  {Li}, \citenamefont {Li}, \citenamefont {Guo}, \citenamefont {Xiao},
  \citenamefont {Huang}, \citenamefont {Lv}, \citenamefont {Han}, \citenamefont
  {Yang}, \citenamefont {Li}, \citenamefont {Lim}, \citenamefont {Wang},
  \citenamefont {Zhang}, \citenamefont {Chua}, \citenamefont {Zeng},
  \citenamefont {Venkatesan}, \citenamefont {Chen}, \citenamefont {Feng},
  \citenamefont {Pennycook},\ and\ \citenamefont {Ariando}}]{Zhou2019}%
  \BibitemOpen
  \bibfield  {author} {\bibinfo {author} {\bibfnamefont {W.~X.}\ \bibnamefont
  {Zhou}}, \bibinfo {author} {\bibfnamefont {H.~J.}\ \bibnamefont {Wu}},
  \bibinfo {author} {\bibfnamefont {J.}~\bibnamefont {Zhou}}, \bibinfo {author}
  {\bibfnamefont {S.~W.}\ \bibnamefont {Zeng}}, \bibinfo {author}
  {\bibfnamefont {C.~J.}\ \bibnamefont {Li}}, \bibinfo {author} {\bibfnamefont
  {M.~S.}\ \bibnamefont {Li}}, \bibinfo {author} {\bibfnamefont
  {R.}~\bibnamefont {Guo}}, \bibinfo {author} {\bibfnamefont {J.~X.}\
  \bibnamefont {Xiao}}, \bibinfo {author} {\bibfnamefont {Z.}~\bibnamefont
  {Huang}}, \bibinfo {author} {\bibfnamefont {W.~M.}\ \bibnamefont {Lv}},
  \bibinfo {author} {\bibfnamefont {K.}~\bibnamefont {Han}}, \bibinfo {author}
  {\bibfnamefont {P.}~\bibnamefont {Yang}}, \bibinfo {author} {\bibfnamefont
  {C.~G.}\ \bibnamefont {Li}}, \bibinfo {author} {\bibfnamefont {Z.~S.}\
  \bibnamefont {Lim}}, \bibinfo {author} {\bibfnamefont {H.}~\bibnamefont
  {Wang}}, \bibinfo {author} {\bibfnamefont {Y.}~\bibnamefont {Zhang}},
  \bibinfo {author} {\bibfnamefont {S.~J.}\ \bibnamefont {Chua}}, \bibinfo
  {author} {\bibfnamefont {K.~Y.}\ \bibnamefont {Zeng}}, \bibinfo {author}
  {\bibfnamefont {T.}~\bibnamefont {Venkatesan}}, \bibinfo {author}
  {\bibfnamefont {J.~S.}\ \bibnamefont {Chen}}, \bibinfo {author}
  {\bibfnamefont {Y.~P.}\ \bibnamefont {Feng}}, \bibinfo {author}
  {\bibfnamefont {S.~J.}\ \bibnamefont {Pennycook}},\ and\ \bibinfo {author}
  {\bibfnamefont {A.}~\bibnamefont {Ariando}},\ }\bibfield  {title} {\bibinfo
  {title} {Artificial two-dimensional polar metal by charge transfer to a
  ferroelectric insulator},\ }\bibfield  {journal} {\bibinfo  {journal}
  {Communications Physics}\ }\textbf {\bibinfo {volume} {2}},\ \href
  {https://doi.org/10.1038/s42005-019-0227-4} {10.1038/s42005-019-0227-4}
  (\bibinfo {year} {2019})\BibitemShut {NoStop}%
\bibitem [{\citenamefont {Zhao}\ \emph {et~al.}(2018)\citenamefont {Zhao},
  \citenamefont {Filippetti}, \citenamefont {Escorihuela-Sayalero},
  \citenamefont {Delugas}, \citenamefont {Canadell}, \citenamefont {Bellaiche},
  \citenamefont {Fiorentini},\ and\ \citenamefont
  {{\'{I}}{\~{n}}iguez}}]{Zhao_Iniguez2018:PRB}%
  \BibitemOpen
  \bibfield  {author} {\bibinfo {author} {\bibfnamefont {H.~J.}\ \bibnamefont
  {Zhao}}, \bibinfo {author} {\bibfnamefont {A.}~\bibnamefont {Filippetti}},
  \bibinfo {author} {\bibfnamefont {C.}~\bibnamefont {Escorihuela-Sayalero}},
  \bibinfo {author} {\bibfnamefont {P.}~\bibnamefont {Delugas}}, \bibinfo
  {author} {\bibfnamefont {E.}~\bibnamefont {Canadell}}, \bibinfo {author}
  {\bibfnamefont {L.}~\bibnamefont {Bellaiche}}, \bibinfo {author}
  {\bibfnamefont {V.}~\bibnamefont {Fiorentini}},\ and\ \bibinfo {author}
  {\bibfnamefont {J.}~\bibnamefont {{\'{I}}{\~{n}}iguez}},\ }\bibfield  {title}
  {\bibinfo {title} {Meta-screening and permanence of polar distortion in
  metallized ferroelectrics},\ }\bibfield  {journal} {\bibinfo  {journal}
  {Physical Review B}\ }\textbf {\bibinfo {volume} {97}},\ \href
  {https://doi.org/10.1103/physrevb.97.054107} {10.1103/physrevb.97.054107}
  (\bibinfo {year} {2018})\BibitemShut {NoStop}%
\bibitem [{\citenamefont {Lee}\ \emph {et~al.}(2010)\citenamefont {Lee},
  \citenamefont {Wilke}, \citenamefont {Trolier-McKinstry}, \citenamefont
  {Zhang},\ and\ \citenamefont {Randall}}]{Lee2010}%
  \BibitemOpen
  \bibfield  {author} {\bibinfo {author} {\bibfnamefont {S.}~\bibnamefont
  {Lee}}, \bibinfo {author} {\bibfnamefont {R.~H.~T.}\ \bibnamefont {Wilke}},
  \bibinfo {author} {\bibfnamefont {S.}~\bibnamefont {Trolier-McKinstry}},
  \bibinfo {author} {\bibfnamefont {S.}~\bibnamefont {Zhang}},\ and\ \bibinfo
  {author} {\bibfnamefont {C.~A.}\ \bibnamefont {Randall}},\ }\bibfield
  {title} {\bibinfo {title} {{Sr$_x$Ba$_{1-x}$Nb$_2$O$_{6-\delta}$}
  ferroelectric-thermoelectrics: Crystal anisotropy, conduction mechanism, and
  power factor},\ }\href {https://doi.org/10.1063/1.3291563} {\bibfield
  {journal} {\bibinfo  {journal} {Applied Physics Letters}\ }\textbf {\bibinfo
  {volume} {96}},\ \bibinfo {pages} {031910} (\bibinfo {year}
  {2010})}\BibitemShut {NoStop}%
\bibitem [{\citenamefont {Lee}\ \emph {et~al.}(2012)\citenamefont {Lee},
  \citenamefont {Bock}, \citenamefont {Trolier-McKinstry},\ and\ \citenamefont
  {Randall}}]{Lee2012}%
  \BibitemOpen
  \bibfield  {author} {\bibinfo {author} {\bibfnamefont {S.}~\bibnamefont
  {Lee}}, \bibinfo {author} {\bibfnamefont {J.~A.}\ \bibnamefont {Bock}},
  \bibinfo {author} {\bibfnamefont {S.}~\bibnamefont {Trolier-McKinstry}},\
  and\ \bibinfo {author} {\bibfnamefont {C.~A.}\ \bibnamefont {Randall}},\
  }\bibfield  {title} {\bibinfo {title} {Ferroelectric-thermoelectricity and
  {Mott} transition of ferroelectric oxides with high electronic
  conductivity},\ }\href {https://doi.org/10.1016/j.jeurceramsoc.2012.06.007}
  {\bibfield  {journal} {\bibinfo  {journal} {Journal of the European Ceramic
  Society}\ }\textbf {\bibinfo {volume} {32}},\ \bibinfo {pages} {3971}
  (\bibinfo {year} {2012})}\BibitemShut {NoStop}%
\bibitem [{\citenamefont {Banik}\ \emph {et~al.}(2019)\citenamefont {Banik},
  \citenamefont {Ghosh}, \citenamefont {Arora}, \citenamefont {Dutta},
  \citenamefont {Pandey}, \citenamefont {Acharya}, \citenamefont {Soni},
  \citenamefont {Waghmare},\ and\ \citenamefont {Biswas}}]{Banik2019}%
  \BibitemOpen
  \bibfield  {author} {\bibinfo {author} {\bibfnamefont {A.}~\bibnamefont
  {Banik}}, \bibinfo {author} {\bibfnamefont {T.}~\bibnamefont {Ghosh}},
  \bibinfo {author} {\bibfnamefont {R.}~\bibnamefont {Arora}}, \bibinfo
  {author} {\bibfnamefont {M.}~\bibnamefont {Dutta}}, \bibinfo {author}
  {\bibfnamefont {J.}~\bibnamefont {Pandey}}, \bibinfo {author} {\bibfnamefont
  {S.}~\bibnamefont {Acharya}}, \bibinfo {author} {\bibfnamefont
  {A.}~\bibnamefont {Soni}}, \bibinfo {author} {\bibfnamefont {U.~V.}\
  \bibnamefont {Waghmare}},\ and\ \bibinfo {author} {\bibfnamefont
  {K.}~\bibnamefont {Biswas}},\ }\bibfield  {title} {\bibinfo {title}
  {Engineering ferroelectric instability to achieve ultralow thermal
  conductivity and high thermoelectric performance in {Sn$_{1-x}$Ge$_x$Te}},\
  }\href {https://doi.org/10.1039/c8ee03162b} {\bibfield  {journal} {\bibinfo
  {journal} {Energy {\&} Environmental Science}\ }\textbf {\bibinfo {volume}
  {12}},\ \bibinfo {pages} {589} (\bibinfo {year} {2019})}\BibitemShut
  {NoStop}%
\bibitem [{\citenamefont {Dangi{\'{c}}}\ \emph {et~al.}(2020)\citenamefont
  {Dangi{\'{c}}}, \citenamefont {Fahy},\ and\ \citenamefont
  {Savi{\'{c}}}}]{Dangi2020}%
  \BibitemOpen
  \bibfield  {author} {\bibinfo {author} {\bibfnamefont {D.}~\bibnamefont
  {Dangi{\'{c}}}}, \bibinfo {author} {\bibfnamefont {S.}~\bibnamefont {Fahy}},\
  and\ \bibinfo {author} {\bibfnamefont {I.}~\bibnamefont {Savi{\'{c}}}},\
  }\bibfield  {title} {\bibinfo {title} {Giant thermoelectric power factor in
  charged ferroelectric domain walls of {GeTe} with van {Hove} singularities},\
  }\bibfield  {journal} {\bibinfo  {journal} {npj Computational Materials}\
  }\textbf {\bibinfo {volume} {6}},\ \href
  {https://doi.org/10.1038/s41524-020-00468-3} {10.1038/s41524-020-00468-3}
  (\bibinfo {year} {2020})\BibitemShut {NoStop}%
\bibitem [{\citenamefont {Zhou}\ and\ \citenamefont
  {Rabe}(2015)}]{PhysRevLett.115.106401}%
  \BibitemOpen
  \bibfield  {author} {\bibinfo {author} {\bibfnamefont {Y.}~\bibnamefont
  {Zhou}}\ and\ \bibinfo {author} {\bibfnamefont {K.~M.}\ \bibnamefont
  {Rabe}},\ }\bibfield  {title} {\bibinfo {title} {Coupled nonpolar-polar
  metal-insulator transition in {${\mathrm{SrCrO}}_{3}/{\mathrm{SrTiO}}_{3}$}
  superlattices: A first-principles study},\ }\href
  {https://doi.org/10.1103/PhysRevLett.115.106401} {\bibfield  {journal}
  {\bibinfo  {journal} {Phys. Rev. Lett.}\ }\textbf {\bibinfo {volume} {115}},\
  \bibinfo {pages} {106401} (\bibinfo {year} {2015})}\BibitemShut {NoStop}%
\bibitem [{\citenamefont {Zhang}\ \emph {et~al.}(2019)\citenamefont {Zhang},
  \citenamefont {Gong}, \citenamefont {Li}, \citenamefont {Lin}, \citenamefont
  {Yan}, \citenamefont {Dong},\ and\ \citenamefont
  {Liu}}]{Zhang_pssr.201900436}%
  \BibitemOpen
  \bibfield  {author} {\bibinfo {author} {\bibfnamefont {Y.}~\bibnamefont
  {Zhang}}, \bibinfo {author} {\bibfnamefont {J.-J.}\ \bibnamefont {Gong}},
  \bibinfo {author} {\bibfnamefont {C.-F.}\ \bibnamefont {Li}}, \bibinfo
  {author} {\bibfnamefont {L.}~\bibnamefont {Lin}}, \bibinfo {author}
  {\bibfnamefont {Z.-B.}\ \bibnamefont {Yan}}, \bibinfo {author} {\bibfnamefont
  {S.}~\bibnamefont {Dong}},\ and\ \bibinfo {author} {\bibfnamefont {J.-M.}\
  \bibnamefont {Liu}},\ }\bibfield  {title} {\bibinfo {title} {{Strain-Induced
  Slater Transition in Polar Metal LiOsO$_3$}},\ }\href
  {https://doi.org/10.1002/pssr.201900436} {\bibfield  {journal} {\bibinfo
  {journal} {physica status solidi (RRL) -- Rapid Research Letters}\ }\textbf
  {\bibinfo {volume} {13}},\ \bibinfo {pages} {900436} (\bibinfo {year}
  {2019})}\BibitemShut {NoStop}%
\bibitem [{\citenamefont {Li}\ and\ \citenamefont {Birol}(2021)}]{li2021free}%
  \BibitemOpen
  \bibfield  {author} {\bibinfo {author} {\bibfnamefont {S.}~\bibnamefont
  {Li}}\ and\ \bibinfo {author} {\bibfnamefont {T.}~\bibnamefont {Birol}},\
  }\bibfield  {title} {\bibinfo {title} {Free-carrier-induced ferroelectricity
  in layered perovskites},\ }\bibfield  {journal} {\bibinfo  {journal}
  {Physical Review Letters}\ }\textbf {\bibinfo {volume} {127}},\ \href
  {https://doi.org/10.1103/physrevlett.127.087601}
  {10.1103/physrevlett.127.087601} (\bibinfo {year} {2021})\BibitemShut
  {NoStop}%
\bibitem [{\citenamefont {Cao}\ \emph {et~al.}(2022)\citenamefont {Cao},
  \citenamefont {Ren}, \citenamefont {Shao}, \citenamefont {Tsymbal},\ and\
  \citenamefont {Mishra}}]{Cao2022arxiv}%
  \BibitemOpen
  \bibfield  {author} {\bibinfo {author} {\bibfnamefont {T.}~\bibnamefont
  {Cao}}, \bibinfo {author} {\bibfnamefont {G.}~\bibnamefont {Ren}}, \bibinfo
  {author} {\bibfnamefont {D.-F.}\ \bibnamefont {Shao}}, \bibinfo {author}
  {\bibfnamefont {E.~Y.}\ \bibnamefont {Tsymbal}},\ and\ \bibinfo {author}
  {\bibfnamefont {R.}~\bibnamefont {Mishra}},\ }\href
  {https://doi.org/10.48550/ARXIV.2209.09436} {\bibinfo {title} {Stabilizing
  polar phases in binary metal oxides by hole doping}} (\bibinfo {year}
  {2022})\BibitemShut {NoStop}%
\bibitem [{\citenamefont {Nisi}\ \emph {et~al.}(2020)\citenamefont {Nisi},
  \citenamefont {Subramanian}, \citenamefont {He}, \citenamefont {Ulman},
  \citenamefont {El-Sherif}, \citenamefont {Sigger}, \citenamefont
  {Lassauni{\`{e}}re}, \citenamefont {Wetherington}, \citenamefont {Briggs},
  \citenamefont {Gray}, \citenamefont {Holleitner}, \citenamefont {Bassim},
  \citenamefont {Quek}, \citenamefont {Robinson},\ and\ \citenamefont
  {Wurstbauer}}]{Nisi2020}%
  \BibitemOpen
  \bibfield  {author} {\bibinfo {author} {\bibfnamefont {K.}~\bibnamefont
  {Nisi}}, \bibinfo {author} {\bibfnamefont {S.}~\bibnamefont {Subramanian}},
  \bibinfo {author} {\bibfnamefont {W.}~\bibnamefont {He}}, \bibinfo {author}
  {\bibfnamefont {K.~A.}\ \bibnamefont {Ulman}}, \bibinfo {author}
  {\bibfnamefont {H.}~\bibnamefont {El-Sherif}}, \bibinfo {author}
  {\bibfnamefont {F.}~\bibnamefont {Sigger}}, \bibinfo {author} {\bibfnamefont
  {M.}~\bibnamefont {Lassauni{\`{e}}re}}, \bibinfo {author} {\bibfnamefont
  {M.~T.}\ \bibnamefont {Wetherington}}, \bibinfo {author} {\bibfnamefont
  {N.}~\bibnamefont {Briggs}}, \bibinfo {author} {\bibfnamefont
  {J.}~\bibnamefont {Gray}}, \bibinfo {author} {\bibfnamefont {A.~W.}\
  \bibnamefont {Holleitner}}, \bibinfo {author} {\bibfnamefont
  {N.}~\bibnamefont {Bassim}}, \bibinfo {author} {\bibfnamefont {S.~Y.}\
  \bibnamefont {Quek}}, \bibinfo {author} {\bibfnamefont {J.~A.}\ \bibnamefont
  {Robinson}},\ and\ \bibinfo {author} {\bibfnamefont {U.}~\bibnamefont
  {Wurstbauer}},\ }\bibfield  {title} {\bibinfo {title}
  {Light{\textendash}matter interaction in quantum confined {2D} polar
  metals},\ }\href {https://doi.org/10.1002/adfm.202005977} {\bibfield
  {journal} {\bibinfo  {journal} {Advanced Functional Materials}\ }\textbf
  {\bibinfo {volume} {31}},\ \bibinfo {pages} {2005977} (\bibinfo {year}
  {2020})}\BibitemShut {NoStop}%
\bibitem [{\citenamefont {Steves}\ \emph {et~al.}(2020)\citenamefont {Steves},
  \citenamefont {Wang}, \citenamefont {Briggs}, \citenamefont {Zhao},
  \citenamefont {El-Sherif}, \citenamefont {Bersch}, \citenamefont
  {Subramanian}, \citenamefont {Dong}, \citenamefont {Bowen}, \citenamefont
  {Duran}, \citenamefont {Nisi}, \citenamefont {Lassauni{\`{e}}re},
  \citenamefont {Wurstbauer}, \citenamefont {Bassim}, \citenamefont {Fonseca},
  \citenamefont {Robinson}, \citenamefont {Crespi}, \citenamefont {Robinson},\
  and\ \citenamefont {Jr}}]{Steves2020}%
  \BibitemOpen
  \bibfield  {author} {\bibinfo {author} {\bibfnamefont {M.~A.}\ \bibnamefont
  {Steves}}, \bibinfo {author} {\bibfnamefont {Y.}~\bibnamefont {Wang}},
  \bibinfo {author} {\bibfnamefont {N.}~\bibnamefont {Briggs}}, \bibinfo
  {author} {\bibfnamefont {T.}~\bibnamefont {Zhao}}, \bibinfo {author}
  {\bibfnamefont {H.}~\bibnamefont {El-Sherif}}, \bibinfo {author}
  {\bibfnamefont {B.~M.}\ \bibnamefont {Bersch}}, \bibinfo {author}
  {\bibfnamefont {S.}~\bibnamefont {Subramanian}}, \bibinfo {author}
  {\bibfnamefont {C.}~\bibnamefont {Dong}}, \bibinfo {author} {\bibfnamefont
  {T.}~\bibnamefont {Bowen}}, \bibinfo {author} {\bibfnamefont {A.~D. L.~F.}\
  \bibnamefont {Duran}}, \bibinfo {author} {\bibfnamefont {K.}~\bibnamefont
  {Nisi}}, \bibinfo {author} {\bibfnamefont {M.}~\bibnamefont
  {Lassauni{\`{e}}re}}, \bibinfo {author} {\bibfnamefont {U.}~\bibnamefont
  {Wurstbauer}}, \bibinfo {author} {\bibfnamefont {N.~D.}\ \bibnamefont
  {Bassim}}, \bibinfo {author} {\bibfnamefont {J.}~\bibnamefont {Fonseca}},
  \bibinfo {author} {\bibfnamefont {J.~T.}\ \bibnamefont {Robinson}}, \bibinfo
  {author} {\bibfnamefont {V.~H.}\ \bibnamefont {Crespi}}, \bibinfo {author}
  {\bibfnamefont {J.}~\bibnamefont {Robinson}},\ and\ \bibinfo {author}
  {\bibfnamefont {K.~L.~K.}\ \bibnamefont {Jr}},\ }\bibfield  {title} {\bibinfo
  {title} {Unexpected near-infrared to visible nonlinear optical properties
  from {2-D} polar metals},\ }\href
  {https://doi.org/10.1021/acs.nanolett.0c03481} {\bibfield  {journal}
  {\bibinfo  {journal} {Nano Letters}\ }\textbf {\bibinfo {volume} {20}},\
  \bibinfo {pages} {8312} (\bibinfo {year} {2020})}\BibitemShut {NoStop}%
\bibitem [{\citenamefont {Feng}\ \emph {et~al.}(2021)\citenamefont {Feng},
  \citenamefont {Kang}, \citenamefont {Manuel}, \citenamefont {Orlandi},
  \citenamefont {Su}, \citenamefont {Chen}, \citenamefont {Tsujimoto},
  \citenamefont {Hadermann}, \citenamefont {Kotliar}, \citenamefont {Yamaura},
  \citenamefont {McCabe},\ and\ \citenamefont {Greenblatt}}]{Feng2021}%
  \BibitemOpen
  \bibfield  {author} {\bibinfo {author} {\bibfnamefont {H.~L.}\ \bibnamefont
  {Feng}}, \bibinfo {author} {\bibfnamefont {C.-J.}\ \bibnamefont {Kang}},
  \bibinfo {author} {\bibfnamefont {P.}~\bibnamefont {Manuel}}, \bibinfo
  {author} {\bibfnamefont {F.}~\bibnamefont {Orlandi}}, \bibinfo {author}
  {\bibfnamefont {Y.}~\bibnamefont {Su}}, \bibinfo {author} {\bibfnamefont
  {J.}~\bibnamefont {Chen}}, \bibinfo {author} {\bibfnamefont {Y.}~\bibnamefont
  {Tsujimoto}}, \bibinfo {author} {\bibfnamefont {J.}~\bibnamefont
  {Hadermann}}, \bibinfo {author} {\bibfnamefont {G.}~\bibnamefont {Kotliar}},
  \bibinfo {author} {\bibfnamefont {K.}~\bibnamefont {Yamaura}}, \bibinfo
  {author} {\bibfnamefont {E.~E.}\ \bibnamefont {McCabe}},\ and\ \bibinfo
  {author} {\bibfnamefont {M.}~\bibnamefont {Greenblatt}},\ }\bibfield  {title}
  {\bibinfo {title} {Antiferromagnetic order breaks inversion symmetry in a
  metallic double perovskite, {Pb$_2$NiOsO$_6$}},\ }\href
  {https://doi.org/10.1021/acs.chemmater.1c01032} {\bibfield  {journal}
  {\bibinfo  {journal} {Chemistry of Materials}\ }\textbf {\bibinfo {volume}
  {33}},\ \bibinfo {pages} {4188} (\bibinfo {year} {2021})}\BibitemShut
  {NoStop}%
\bibitem [{\citenamefont {Salmani-Rezaie}\ \emph {et~al.}(2021)\citenamefont
  {Salmani-Rezaie}, \citenamefont {Galletti}, \citenamefont {Schumann},
  \citenamefont {Russell}, \citenamefont {Jeong}, \citenamefont {Li},
  \citenamefont {Harter},\ and\ \citenamefont {Stemmer}}]{SalmaniRezaie2021}%
  \BibitemOpen
  \bibfield  {author} {\bibinfo {author} {\bibfnamefont {S.}~\bibnamefont
  {Salmani-Rezaie}}, \bibinfo {author} {\bibfnamefont {L.}~\bibnamefont
  {Galletti}}, \bibinfo {author} {\bibfnamefont {T.}~\bibnamefont {Schumann}},
  \bibinfo {author} {\bibfnamefont {R.}~\bibnamefont {Russell}}, \bibinfo
  {author} {\bibfnamefont {H.}~\bibnamefont {Jeong}}, \bibinfo {author}
  {\bibfnamefont {Y.}~\bibnamefont {Li}}, \bibinfo {author} {\bibfnamefont
  {J.~W.}\ \bibnamefont {Harter}},\ and\ \bibinfo {author} {\bibfnamefont
  {S.}~\bibnamefont {Stemmer}},\ }\bibfield  {title} {\bibinfo {title}
  {{Superconductivity in magnetically doped SrTiO$_3$}},\ }\href
  {https://doi.org/10.1063/5.0052319} {\bibfield  {journal} {\bibinfo
  {journal} {Applied Physics Letters}\ }\textbf {\bibinfo {volume} {118}},\
  \bibinfo {pages} {202602} (\bibinfo {year} {2021})}\BibitemShut {NoStop}%
\bibitem [{\citenamefont {Jacobsen}\ \emph {et~al.}(2020)\citenamefont
  {Jacobsen}, \citenamefont {Feng}, \citenamefont {Princep}, \citenamefont
  {Rahn}, \citenamefont {Guo}, \citenamefont {Chen}, \citenamefont
  {Matsushita}, \citenamefont {Tsujimoto}, \citenamefont {Nagao}, \citenamefont
  {Khalyavin}, \citenamefont {Manuel}, \citenamefont {Murray}, \citenamefont
  {Donnerer}, \citenamefont {Vale}, \citenamefont {Sala}, \citenamefont
  {Yamaura},\ and\ \citenamefont {Boothroyd}}]{Jacobsen2020}%
  \BibitemOpen
  \bibfield  {author} {\bibinfo {author} {\bibfnamefont {H.}~\bibnamefont
  {Jacobsen}}, \bibinfo {author} {\bibfnamefont {H.~L.}\ \bibnamefont {Feng}},
  \bibinfo {author} {\bibfnamefont {A.~J.}\ \bibnamefont {Princep}}, \bibinfo
  {author} {\bibfnamefont {M.~C.}\ \bibnamefont {Rahn}}, \bibinfo {author}
  {\bibfnamefont {Y.}~\bibnamefont {Guo}}, \bibinfo {author} {\bibfnamefont
  {J.}~\bibnamefont {Chen}}, \bibinfo {author} {\bibfnamefont {Y.}~\bibnamefont
  {Matsushita}}, \bibinfo {author} {\bibfnamefont {Y.}~\bibnamefont
  {Tsujimoto}}, \bibinfo {author} {\bibfnamefont {M.}~\bibnamefont {Nagao}},
  \bibinfo {author} {\bibfnamefont {D.}~\bibnamefont {Khalyavin}}, \bibinfo
  {author} {\bibfnamefont {P.}~\bibnamefont {Manuel}}, \bibinfo {author}
  {\bibfnamefont {C.~A.}\ \bibnamefont {Murray}}, \bibinfo {author}
  {\bibfnamefont {C.}~\bibnamefont {Donnerer}}, \bibinfo {author}
  {\bibfnamefont {J.~G.}\ \bibnamefont {Vale}}, \bibinfo {author}
  {\bibfnamefont {M.~M.}\ \bibnamefont {Sala}}, \bibinfo {author}
  {\bibfnamefont {K.}~\bibnamefont {Yamaura}},\ and\ \bibinfo {author}
  {\bibfnamefont {A.~T.}\ \bibnamefont {Boothroyd}},\ }\bibfield  {title}
  {\bibinfo {title} {Magnetically induced metal-insulator transition in
  {Pb$_2$CaOsO$_6$}},\ }\bibfield  {journal} {\bibinfo  {journal} {Physical
  Review B}\ }\textbf {\bibinfo {volume} {102}},\ \href
  {https://doi.org/10.1103/physrevb.102.214409} {10.1103/physrevb.102.214409}
  (\bibinfo {year} {2020})\BibitemShut {NoStop}%
\bibitem [{\citenamefont {Zhu}\ \emph {et~al.}(2017)\citenamefont {Zhu},
  \citenamefont {Peng}, \citenamefont {Tian}, \citenamefont {Hong},
  \citenamefont {Mao},\ and\ \citenamefont {Ke}}]{Zhu2017}%
  \BibitemOpen
  \bibfield  {author} {\bibinfo {author} {\bibfnamefont {M.}~\bibnamefont
  {Zhu}}, \bibinfo {author} {\bibfnamefont {J.}~\bibnamefont {Peng}}, \bibinfo
  {author} {\bibfnamefont {W.}~\bibnamefont {Tian}}, \bibinfo {author}
  {\bibfnamefont {T.}~\bibnamefont {Hong}}, \bibinfo {author} {\bibfnamefont
  {Z.~Q.}\ \bibnamefont {Mao}},\ and\ \bibinfo {author} {\bibfnamefont
  {X.}~\bibnamefont {Ke}},\ }\bibfield  {title} {\bibinfo {title} {Tuning the
  competing phases of bilayer ruthenate {Ca$_3$Ru$_2$O$_7$} via dilute {Mn}
  impurities and magnetic field},\ }\bibfield  {journal} {\bibinfo  {journal}
  {Physical Review B}\ }\textbf {\bibinfo {volume} {95}},\ \href
  {https://doi.org/10.1103/physrevb.95.144426} {10.1103/physrevb.95.144426}
  (\bibinfo {year} {2017})\BibitemShut {NoStop}%
\bibitem [{\citenamefont {Lei}\ \emph {et~al.}(2019)\citenamefont {Lei},
  \citenamefont {Chikara}, \citenamefont {Puggioni}, \citenamefont {Peng},
  \citenamefont {Zhu}, \citenamefont {Gu}, \citenamefont {Zhao}, \citenamefont
  {Wang}, \citenamefont {Yuan}, \citenamefont {Akamatsu}, \citenamefont {Chan},
  \citenamefont {Ke}, \citenamefont {Mao}, \citenamefont {Rondinelli},
  \citenamefont {Jaime}, \citenamefont {Singleton}, \citenamefont {Weickert},
  \citenamefont {Zapf},\ and\ \citenamefont {Gopalan}}]{Lei2019:phasediagram}%
  \BibitemOpen
  \bibfield  {author} {\bibinfo {author} {\bibfnamefont {S.}~\bibnamefont
  {Lei}}, \bibinfo {author} {\bibfnamefont {S.}~\bibnamefont {Chikara}},
  \bibinfo {author} {\bibfnamefont {D.}~\bibnamefont {Puggioni}}, \bibinfo
  {author} {\bibfnamefont {J.}~\bibnamefont {Peng}}, \bibinfo {author}
  {\bibfnamefont {M.}~\bibnamefont {Zhu}}, \bibinfo {author} {\bibfnamefont
  {M.}~\bibnamefont {Gu}}, \bibinfo {author} {\bibfnamefont {W.}~\bibnamefont
  {Zhao}}, \bibinfo {author} {\bibfnamefont {Y.}~\bibnamefont {Wang}}, \bibinfo
  {author} {\bibfnamefont {Y.}~\bibnamefont {Yuan}}, \bibinfo {author}
  {\bibfnamefont {H.}~\bibnamefont {Akamatsu}}, \bibinfo {author}
  {\bibfnamefont {M.~H.~W.}\ \bibnamefont {Chan}}, \bibinfo {author}
  {\bibfnamefont {X.}~\bibnamefont {Ke}}, \bibinfo {author} {\bibfnamefont
  {Z.}~\bibnamefont {Mao}}, \bibinfo {author} {\bibfnamefont {J.~M.}\
  \bibnamefont {Rondinelli}}, \bibinfo {author} {\bibfnamefont
  {M.}~\bibnamefont {Jaime}}, \bibinfo {author} {\bibfnamefont
  {J.}~\bibnamefont {Singleton}}, \bibinfo {author} {\bibfnamefont
  {F.}~\bibnamefont {Weickert}}, \bibinfo {author} {\bibfnamefont {V.~S.}\
  \bibnamefont {Zapf}},\ and\ \bibinfo {author} {\bibfnamefont
  {V.}~\bibnamefont {Gopalan}},\ }\bibfield  {title} {\bibinfo {title}
  {Comprehensive magnetic phase diagrams of the polar metal
  {Ca$_3$(Ru$_{0.95}$Fe$_{0.05}$)$_2$O$_7$}},\ }\bibfield  {journal} {\bibinfo
  {journal} {Physical Review B}\ }\textbf {\bibinfo {volume} {99}},\ \href
  {https://doi.org/10.1103/physrevb.99.224411} {10.1103/physrevb.99.224411}
  (\bibinfo {year} {2019})\BibitemShut {NoStop}%
\bibitem [{\citenamefont {Tsuda}\ \emph {et~al.}(2013)\citenamefont {Tsuda},
  \citenamefont {Kikugawa}, \citenamefont {Sugii}, \citenamefont {Uji},
  \citenamefont {Ueda}, \citenamefont {Nishio},\ and\ \citenamefont
  {Maeno}}]{Tsuda2013}%
  \BibitemOpen
  \bibfield  {author} {\bibinfo {author} {\bibfnamefont {S.}~\bibnamefont
  {Tsuda}}, \bibinfo {author} {\bibfnamefont {N.}~\bibnamefont {Kikugawa}},
  \bibinfo {author} {\bibfnamefont {K.}~\bibnamefont {Sugii}}, \bibinfo
  {author} {\bibfnamefont {S.}~\bibnamefont {Uji}}, \bibinfo {author}
  {\bibfnamefont {S.}~\bibnamefont {Ueda}}, \bibinfo {author} {\bibfnamefont
  {M.}~\bibnamefont {Nishio}},\ and\ \bibinfo {author} {\bibfnamefont
  {Y.}~\bibnamefont {Maeno}},\ }\bibfield  {title} {\bibinfo {title} {Mott
  transition extremely sensitive to impurities in {Ca$_3$Ru$_2$O$_7$} revealed
  by hard x-ray photoemission studies},\ }\bibfield  {journal} {\bibinfo
  {journal} {Physical Review B}\ }\textbf {\bibinfo {volume} {87}},\ \href
  {https://doi.org/10.1103/physrevb.87.241107} {10.1103/physrevb.87.241107}
  (\bibinfo {year} {2013})\BibitemShut {NoStop}%
\bibitem [{\citenamefont {Malyi}\ \emph {et~al.}(2020)\citenamefont {Malyi},
  \citenamefont {Dalpian}, \citenamefont {Zhao}, \citenamefont {Wang},\ and\
  \citenamefont {Zunger}}]{Malyi2020}%
  \BibitemOpen
  \bibfield  {author} {\bibinfo {author} {\bibfnamefont {O.~I.}\ \bibnamefont
  {Malyi}}, \bibinfo {author} {\bibfnamefont {G.~M.}\ \bibnamefont {Dalpian}},
  \bibinfo {author} {\bibfnamefont {X.-G.}\ \bibnamefont {Zhao}}, \bibinfo
  {author} {\bibfnamefont {Z.}~\bibnamefont {Wang}},\ and\ \bibinfo {author}
  {\bibfnamefont {A.}~\bibnamefont {Zunger}},\ }\bibfield  {title} {\bibinfo
  {title} {Realization of predicted exotic materials: The burden of proof},\
  }\href {https://doi.org/10.1016/j.mattod.2019.08.003} {\bibfield  {journal}
  {\bibinfo  {journal} {Materials Today}\ }\textbf {\bibinfo {volume} {32}},\
  \bibinfo {pages} {35} (\bibinfo {year} {2020})}\BibitemShut {NoStop}%
\bibitem [{\citenamefont {Ma}\ \emph {et~al.}(2021)\citenamefont {Ma},
  \citenamefont {Yang},\ and\ \citenamefont {Chen}}]{Ma2021}%
  \BibitemOpen
  \bibfield  {author} {\bibinfo {author} {\bibfnamefont {J.}~\bibnamefont
  {Ma}}, \bibinfo {author} {\bibfnamefont {R.}~\bibnamefont {Yang}},\ and\
  \bibinfo {author} {\bibfnamefont {H.}~\bibnamefont {Chen}},\ }\bibfield
  {title} {\bibinfo {title} {A large modulation of electron-phonon coupling and
  an emergent superconducting dome in doped strong ferroelectrics},\ }\bibfield
   {journal} {\bibinfo  {journal} {Nature Communications}\ }\textbf {\bibinfo
  {volume} {12}},\ \href {https://doi.org/10.1038/s41467-021-22541-1}
  {10.1038/s41467-021-22541-1} (\bibinfo {year} {2021})\BibitemShut {NoStop}%
\bibitem [{\citenamefont {Kennes}\ \emph {et~al.}(2021)\citenamefont {Kennes},
  \citenamefont {Claassen}, \citenamefont {Xian}, \citenamefont {Georges},
  \citenamefont {Millis}, \citenamefont {Hone}, \citenamefont {Dean},
  \citenamefont {Basov}, \citenamefont {Pasupathy},\ and\ \citenamefont
  {Rubio}}]{Kennes2021}%
  \BibitemOpen
  \bibfield  {author} {\bibinfo {author} {\bibfnamefont {D.~M.}\ \bibnamefont
  {Kennes}}, \bibinfo {author} {\bibfnamefont {M.}~\bibnamefont {Claassen}},
  \bibinfo {author} {\bibfnamefont {L.}~\bibnamefont {Xian}}, \bibinfo {author}
  {\bibfnamefont {A.}~\bibnamefont {Georges}}, \bibinfo {author} {\bibfnamefont
  {A.~J.}\ \bibnamefont {Millis}}, \bibinfo {author} {\bibfnamefont
  {J.}~\bibnamefont {Hone}}, \bibinfo {author} {\bibfnamefont {C.~R.}\
  \bibnamefont {Dean}}, \bibinfo {author} {\bibfnamefont {D.~N.}\ \bibnamefont
  {Basov}}, \bibinfo {author} {\bibfnamefont {A.~N.}\ \bibnamefont
  {Pasupathy}},\ and\ \bibinfo {author} {\bibfnamefont {A.}~\bibnamefont
  {Rubio}},\ }\bibfield  {title} {\bibinfo {title} {Moir{\'{e}}
  heterostructures as a condensed-matter quantum simulator},\ }\href
  {https://doi.org/10.1038/s41567-020-01154-3} {\bibfield  {journal} {\bibinfo
  {journal} {Nature Physics}\ }\textbf {\bibinfo {volume} {17}},\ \bibinfo
  {pages} {155} (\bibinfo {year} {2021})}\BibitemShut {NoStop}%
\bibitem [{\citenamefont {Susner}\ \emph {et~al.}(2017)\citenamefont {Susner},
  \citenamefont {Chyasnavichyus}, \citenamefont {McGuire}, \citenamefont
  {Ganesh},\ and\ \citenamefont {Maksymovych}}]{Susner2017}%
  \BibitemOpen
  \bibfield  {author} {\bibinfo {author} {\bibfnamefont {M.~A.}\ \bibnamefont
  {Susner}}, \bibinfo {author} {\bibfnamefont {M.}~\bibnamefont
  {Chyasnavichyus}}, \bibinfo {author} {\bibfnamefont {M.~A.}\ \bibnamefont
  {McGuire}}, \bibinfo {author} {\bibfnamefont {P.}~\bibnamefont {Ganesh}},\
  and\ \bibinfo {author} {\bibfnamefont {P.}~\bibnamefont {Maksymovych}},\
  }\bibfield  {title} {\bibinfo {title} {Metal thio- and selenophosphates as
  multifunctional {van der Waals} layered materials},\ }\href
  {https://doi.org/10.1002/adma.201602852} {\bibfield  {journal} {\bibinfo
  {journal} {Advanced Materials}\ }\textbf {\bibinfo {volume} {29}},\ \bibinfo
  {pages} {1602852} (\bibinfo {year} {2017})}\BibitemShut {NoStop}%
\bibitem [{\citenamefont {Mak}\ \emph {et~al.}(2019)\citenamefont {Mak},
  \citenamefont {Shan},\ and\ \citenamefont {Ralph}}]{Mak2019}%
  \BibitemOpen
  \bibfield  {author} {\bibinfo {author} {\bibfnamefont {K.~F.}\ \bibnamefont
  {Mak}}, \bibinfo {author} {\bibfnamefont {J.}~\bibnamefont {Shan}},\ and\
  \bibinfo {author} {\bibfnamefont {D.~C.}\ \bibnamefont {Ralph}},\ }\bibfield
  {title} {\bibinfo {title} {Probing and controlling magnetic states in {2D}
  layered magnetic materials},\ }\href
  {https://doi.org/10.1038/s42254-019-0110-y} {\bibfield  {journal} {\bibinfo
  {journal} {Nature Reviews Physics}\ }\textbf {\bibinfo {volume} {1}},\
  \bibinfo {pages} {646} (\bibinfo {year} {2019})}\BibitemShut {NoStop}%
\bibitem [{\citenamefont {Gong}\ and\ \citenamefont {Zhang}(2019)}]{Gong2019}%
  \BibitemOpen
  \bibfield  {author} {\bibinfo {author} {\bibfnamefont {C.}~\bibnamefont
  {Gong}}\ and\ \bibinfo {author} {\bibfnamefont {X.}~\bibnamefont {Zhang}},\
  }\bibfield  {title} {\bibinfo {title} {Two-dimensional magnetic crystals and
  emergent heterostructure devices},\ }\href
  {https://doi.org/10.1126/science.aav4450} {\bibfield  {journal} {\bibinfo
  {journal} {Science}\ }\textbf {\bibinfo {volume} {363}},\ \bibinfo {pages}
  {eaav4450} (\bibinfo {year} {2019})}\BibitemShut {NoStop}%
\bibitem [{\citenamefont {Xu}\ \emph {et~al.}(2020)\citenamefont {Xu},
  \citenamefont {Zhang}, \citenamefont {Zhu}, \citenamefont {Wang},
  \citenamefont {Shimada}, \citenamefont {Kitamura},\ and\ \citenamefont
  {Zhang}}]{Xu2020}%
  \BibitemOpen
  \bibfield  {author} {\bibinfo {author} {\bibfnamefont {T.}~\bibnamefont
  {Xu}}, \bibinfo {author} {\bibfnamefont {J.}~\bibnamefont {Zhang}}, \bibinfo
  {author} {\bibfnamefont {Y.}~\bibnamefont {Zhu}}, \bibinfo {author}
  {\bibfnamefont {J.}~\bibnamefont {Wang}}, \bibinfo {author} {\bibfnamefont
  {T.}~\bibnamefont {Shimada}}, \bibinfo {author} {\bibfnamefont
  {T.}~\bibnamefont {Kitamura}},\ and\ \bibinfo {author} {\bibfnamefont
  {T.-Y.}\ \bibnamefont {Zhang}},\ }\bibfield  {title} {\bibinfo {title}
  {Two-dimensional polar metal of a {PbTe} monolayer by electrostatic doping},\
  }\href {https://doi.org/10.1039/d0nh00188k} {\bibfield  {journal} {\bibinfo
  {journal} {Nanoscale Horizons}\ }\textbf {\bibinfo {volume} {5}},\ \bibinfo
  {pages} {1400} (\bibinfo {year} {2020})}\BibitemShut {NoStop}%
\bibitem [{\citenamefont {Ye}\ \emph {et~al.}(2019)\citenamefont {Ye},
  \citenamefont {Hu}, \citenamefont {Ke}, \citenamefont {Zhu}, \citenamefont
  {Zhang}, \citenamefont {Xie}, \citenamefont {Zhang}, \citenamefont {Zhang},
  \citenamefont {Luo}, \citenamefont {Gu}, \citenamefont {He}, \citenamefont
  {Zhang}, \citenamefont {Zhang},\ and\ \citenamefont
  {Chen}}]{ye2019observation}%
  \BibitemOpen
  \bibfield  {author} {\bibinfo {author} {\bibfnamefont {M.}~\bibnamefont
  {Ye}}, \bibinfo {author} {\bibfnamefont {S.}~\bibnamefont {Hu}}, \bibinfo
  {author} {\bibfnamefont {S.}~\bibnamefont {Ke}}, \bibinfo {author}
  {\bibfnamefont {Y.}~\bibnamefont {Zhu}}, \bibinfo {author} {\bibfnamefont
  {Y.}~\bibnamefont {Zhang}}, \bibinfo {author} {\bibfnamefont
  {L.}~\bibnamefont {Xie}}, \bibinfo {author} {\bibfnamefont {Y.}~\bibnamefont
  {Zhang}}, \bibinfo {author} {\bibfnamefont {D.}~\bibnamefont {Zhang}},
  \bibinfo {author} {\bibfnamefont {Z.}~\bibnamefont {Luo}}, \bibinfo {author}
  {\bibfnamefont {M.}~\bibnamefont {Gu}}, \bibinfo {author} {\bibfnamefont
  {J.}~\bibnamefont {He}}, \bibinfo {author} {\bibfnamefont {P.}~\bibnamefont
  {Zhang}}, \bibinfo {author} {\bibfnamefont {W.}~\bibnamefont {Zhang}},\ and\
  \bibinfo {author} {\bibfnamefont {L.}~\bibnamefont {Chen}},\ }\href
  {https://doi.org/10.48550/ARXIV.1908.08726} {\bibinfo {title} {Observation of
  a room temperature two-dimensional ferroelectric metal}} (\bibinfo {year}
  {2019})\BibitemShut {NoStop}%
\bibitem [{\citenamefont {Luo}\ \emph {et~al.}(2017)\citenamefont {Luo},
  \citenamefont {Xu},\ and\ \citenamefont {Xiang}}]{Luo2017}%
  \BibitemOpen
  \bibfield  {author} {\bibinfo {author} {\bibfnamefont {W.}~\bibnamefont
  {Luo}}, \bibinfo {author} {\bibfnamefont {K.}~\bibnamefont {Xu}},\ and\
  \bibinfo {author} {\bibfnamefont {H.}~\bibnamefont {Xiang}},\ }\bibfield
  {title} {\bibinfo {title} {Two-dimensional hyperferroelectric metals: A
  different route to ferromagnetic-ferroelectric multiferroics},\ }\bibfield
  {journal} {\bibinfo  {journal} {Physical Review B}\ }\textbf {\bibinfo
  {volume} {96}},\ \href {https://doi.org/10.1103/physrevb.96.235415}
  {10.1103/physrevb.96.235415} (\bibinfo {year} {2017})\BibitemShut {NoStop}%
\bibitem [{\citenamefont {Lu}\ \emph {et~al.}(2019)\citenamefont {Lu},
  \citenamefont {Chen}, \citenamefont {Luo}, \citenamefont
  {{\'{I}}{\~{n}}iguez}, \citenamefont {Bellaiche},\ and\ \citenamefont
  {Xiang}}]{Lu2019}%
  \BibitemOpen
  \bibfield  {author} {\bibinfo {author} {\bibfnamefont {J.}~\bibnamefont
  {Lu}}, \bibinfo {author} {\bibfnamefont {G.}~\bibnamefont {Chen}}, \bibinfo
  {author} {\bibfnamefont {W.}~\bibnamefont {Luo}}, \bibinfo {author}
  {\bibfnamefont {J.}~\bibnamefont {{\'{I}}{\~{n}}iguez}}, \bibinfo {author}
  {\bibfnamefont {L.}~\bibnamefont {Bellaiche}},\ and\ \bibinfo {author}
  {\bibfnamefont {H.}~\bibnamefont {Xiang}},\ }\bibfield  {title} {\bibinfo
  {title} {Ferroelectricity with asymmetric hysteresis in metallic {LiOsO}$_3$
  ultrathin films},\ }\bibfield  {journal} {\bibinfo  {journal} {Physical
  Review Letters}\ }\textbf {\bibinfo {volume} {122}},\ \href
  {https://doi.org/10.1103/physrevlett.122.227601}
  {10.1103/physrevlett.122.227601} (\bibinfo {year} {2019})\BibitemShut
  {NoStop}%
\bibitem [{\citenamefont {Bostr\"{o}m}\ \emph {et~al.}(2018)\citenamefont
  {Bostr\"{o}m}, \citenamefont {Senn},\ and\ \citenamefont
  {Goodwin}}]{Bostrm2018}%
  \BibitemOpen
  \bibfield  {author} {\bibinfo {author} {\bibfnamefont {H.~L.~B.}\
  \bibnamefont {Bostr\"{o}m}}, \bibinfo {author} {\bibfnamefont {M.~S.}\
  \bibnamefont {Senn}},\ and\ \bibinfo {author} {\bibfnamefont {A.~L.}\
  \bibnamefont {Goodwin}},\ }\bibfield  {title} {\bibinfo {title} {Recipes for
  improper ferroelectricity in molecular perovskites},\ }\bibfield  {journal}
  {\bibinfo  {journal} {Nature Communications}\ }\textbf {\bibinfo {volume}
  {9}},\ \href {https://doi.org/10.1038/s41467-018-04764-x}
  {10.1038/s41467-018-04764-x} (\bibinfo {year} {2018})\BibitemShut {NoStop}%
\bibitem [{\citenamefont {Aubrey}\ \emph {et~al.}(2021)\citenamefont {Aubrey},
  \citenamefont {Valdes}, \citenamefont {Filip}, \citenamefont {Connor},
  \citenamefont {Lindquist}, \citenamefont {Neaton},\ and\ \citenamefont
  {Karunadasa}}]{Aubrey2021}%
  \BibitemOpen
  \bibfield  {author} {\bibinfo {author} {\bibfnamefont {M.~L.}\ \bibnamefont
  {Aubrey}}, \bibinfo {author} {\bibfnamefont {A.~S.}\ \bibnamefont {Valdes}},
  \bibinfo {author} {\bibfnamefont {M.~R.}\ \bibnamefont {Filip}}, \bibinfo
  {author} {\bibfnamefont {B.~A.}\ \bibnamefont {Connor}}, \bibinfo {author}
  {\bibfnamefont {K.~P.}\ \bibnamefont {Lindquist}}, \bibinfo {author}
  {\bibfnamefont {J.~B.}\ \bibnamefont {Neaton}},\ and\ \bibinfo {author}
  {\bibfnamefont {H.~I.}\ \bibnamefont {Karunadasa}},\ }\bibfield  {title}
  {\bibinfo {title} {Directed assembly of layered perovskite heterostructures
  as single crystals},\ }\href {https://doi.org/10.1038/s41586-021-03810-x}
  {\bibfield  {journal} {\bibinfo  {journal} {Nature}\ }\textbf {\bibinfo
  {volume} {597}},\ \bibinfo {pages} {355} (\bibinfo {year}
  {2021})}\BibitemShut {NoStop}%
\bibitem [{\citenamefont {Peng}\ \emph {et~al.}(2020)\citenamefont {Peng},
  \citenamefont {Hu}, \citenamefont {Murakami}, \citenamefont {Zhang},\ and\
  \citenamefont {Monserrat}}]{Peng2020}%
  \BibitemOpen
  \bibfield  {author} {\bibinfo {author} {\bibfnamefont {B.}~\bibnamefont
  {Peng}}, \bibinfo {author} {\bibfnamefont {Y.}~\bibnamefont {Hu}}, \bibinfo
  {author} {\bibfnamefont {S.}~\bibnamefont {Murakami}}, \bibinfo {author}
  {\bibfnamefont {T.}~\bibnamefont {Zhang}},\ and\ \bibinfo {author}
  {\bibfnamefont {B.}~\bibnamefont {Monserrat}},\ }\bibfield  {title} {\bibinfo
  {title} {Topological phonons in oxide perovskites controlled by light},\
  }\href {https://doi.org/10.1126/sciadv.abd1618} {\bibfield  {journal}
  {\bibinfo  {journal} {Science Advances}\ }\textbf {\bibinfo {volume} {6}},\
  \bibinfo {pages} {eabd1618} (\bibinfo {year} {2020})}\BibitemShut {NoStop}%
\bibitem [{\citenamefont {Wijethunge}\ \emph {et~al.}(2021)\citenamefont
  {Wijethunge}, \citenamefont {Zhang},\ and\ \citenamefont {Du}}]{D1TC02213J}%
  \BibitemOpen
  \bibfield  {author} {\bibinfo {author} {\bibfnamefont {D.}~\bibnamefont
  {Wijethunge}}, \bibinfo {author} {\bibfnamefont {L.}~\bibnamefont {Zhang}},\
  and\ \bibinfo {author} {\bibfnamefont {A.}~\bibnamefont {Du}},\ }\bibfield
  {title} {\bibinfo {title} {{Prediction of two-dimensional ferroelectric metal
  Mxenes}},\ }\href {https://doi.org/10.1039/D1TC02213J} {\bibfield  {journal}
  {\bibinfo  {journal} {J. Mater. Chem. C}\ }\textbf {\bibinfo {volume} {9}},\
  \bibinfo {pages} {11343} (\bibinfo {year} {2021})}\BibitemShut {NoStop}%
\bibitem [{\citenamefont {Yang}\ \emph {et~al.}(2018)\citenamefont {Yang},
  \citenamefont {Wu},\ and\ \citenamefont
  {Li}}]{doi:10.1021/acs.jpclett.8b03654}%
  \BibitemOpen
  \bibfield  {author} {\bibinfo {author} {\bibfnamefont {Q.}~\bibnamefont
  {Yang}}, \bibinfo {author} {\bibfnamefont {M.}~\bibnamefont {Wu}},\ and\
  \bibinfo {author} {\bibfnamefont {J.}~\bibnamefont {Li}},\ }\bibfield
  {title} {\bibinfo {title} {Origin of two-dimensional vertical
  ferroelectricity in {WTe$_2$} bilayer and multilayer},\ }\href
  {https://doi.org/10.1021/acs.jpclett.8b03654} {\bibfield  {journal} {\bibinfo
   {journal} {The Journal of Physical Chemistry Letters}\ }\textbf {\bibinfo
  {volume} {9}},\ \bibinfo {pages} {7160} (\bibinfo {year} {2018})},\ \Eprint
  {https://arxiv.org/abs/https://doi.org/10.1021/acs.jpclett.8b03654}
  {https://doi.org/10.1021/acs.jpclett.8b03654} \BibitemShut {NoStop}%
\bibitem [{\citenamefont {Li}\ and\ \citenamefont
  {Wu}(2017)}]{doi:10.1021/acsnano.7b02756}%
  \BibitemOpen
  \bibfield  {author} {\bibinfo {author} {\bibfnamefont {L.}~\bibnamefont
  {Li}}\ and\ \bibinfo {author} {\bibfnamefont {M.}~\bibnamefont {Wu}},\
  }\bibfield  {title} {\bibinfo {title} {Binary compound bilayer and multilayer
  with vertical polarizations: Two-dimensional ferroelectrics, multiferroics,
  and nanogenerators},\ }\href {https://doi.org/10.1021/acsnano.7b02756}
  {\bibfield  {journal} {\bibinfo  {journal} {ACS Nano}\ }\textbf {\bibinfo
  {volume} {11}},\ \bibinfo {pages} {6382} (\bibinfo {year} {2017})},\ \bibinfo
  {note} {pMID: 28602074},\ \Eprint
  {https://arxiv.org/abs/https://doi.org/10.1021/acsnano.7b02756}
  {https://doi.org/10.1021/acsnano.7b02756} \BibitemShut {NoStop}%
\bibitem [{\citenamefont {Mann}\ \emph {et~al.}(2022)\citenamefont {Mann},
  \citenamefont {D{\'{\i}}ez}, \citenamefont {Xu}, \citenamefont {Lebedev},
  \citenamefont {Kolen'ko},\ and\ \citenamefont {Shatruk}}]{Mann_2022}%
  \BibitemOpen
  \bibfield  {author} {\bibinfo {author} {\bibfnamefont {D.~K.}\ \bibnamefont
  {Mann}}, \bibinfo {author} {\bibfnamefont {A.~M.}\ \bibnamefont
  {D{\'{\i}}ez}}, \bibinfo {author} {\bibfnamefont {J.}~\bibnamefont {Xu}},
  \bibinfo {author} {\bibfnamefont {O.~I.}\ \bibnamefont {Lebedev}}, \bibinfo
  {author} {\bibfnamefont {Y.~V.}\ \bibnamefont {Kolen'ko}},\ and\ \bibinfo
  {author} {\bibfnamefont {M.}~\bibnamefont {Shatruk}},\ }\bibfield  {title}
  {\bibinfo {title} {{Polar Layered Intermetallic {LaCo}$_2$P$_2$ as a Water
  Oxidation Electrocatalyst}},\ }\href {https://doi.org/10.1021/acsami.1c19858}
  {\bibfield  {journal} {\bibinfo  {journal} {{ACS} Applied Materials {\&}
  Interfaces}\ }\textbf {\bibinfo {volume} {14}},\ \bibinfo {pages} {14120}
  (\bibinfo {year} {2022})}\BibitemShut {NoStop}%
\bibitem [{\citenamefont {Puggioni}\ \emph {et~al.}(2018)\citenamefont
  {Puggioni}, \citenamefont {Giovannetti},\ and\ \citenamefont
  {Rondinelli}}]{Puggioni_electrodes}%
  \BibitemOpen
  \bibfield  {author} {\bibinfo {author} {\bibfnamefont {D.}~\bibnamefont
  {Puggioni}}, \bibinfo {author} {\bibfnamefont {G.}~\bibnamefont
  {Giovannetti}},\ and\ \bibinfo {author} {\bibfnamefont {J.~M.}\ \bibnamefont
  {Rondinelli}},\ }\bibfield  {title} {\bibinfo {title} {Polar metals as
  electrodes to suppress the critical-thickness limit in ferroelectric
  nanocapacitors},\ }\href {https://doi.org/10.1063/1.5049607} {\bibfield
  {journal} {\bibinfo  {journal} {Journal of Applied Physics}\ }\textbf
  {\bibinfo {volume} {124}},\ \bibinfo {pages} {174102} (\bibinfo {year}
  {2018})},\ \Eprint {https://arxiv.org/abs/https://doi.org/10.1063/1.5049607}
  {https://doi.org/10.1063/1.5049607} \BibitemShut {NoStop}%
\bibitem [{\citenamefont {Puggioni}\ and\ \citenamefont
  {Rondinelli}(2019)}]{puggioni_rondinelli_2019_pat}%
  \BibitemOpen
  \bibfield  {author} {\bibinfo {author} {\bibfnamefont {D.}~\bibnamefont
  {Puggioni}}\ and\ \bibinfo {author} {\bibfnamefont {J.~M.}\ \bibnamefont
  {Rondinelli}},\ }\href {https://patents.google.com/patent/US20180040711A1/en}
  {\bibinfo {title} {Noncentrosymmetric metal electrodes for ferroic devices}}
  (\bibinfo {year} {2019})\BibitemShut {NoStop}%
\bibitem [{\citenamefont {Liu}\ \emph {et~al.}(2019)\citenamefont {Liu},
  \citenamefont {Yang}, \citenamefont {Hu}, \citenamefont {Zhao}, \citenamefont
  {Chen},\ and\ \citenamefont {Ren}}]{C9NR05404A}%
  \BibitemOpen
  \bibfield  {author} {\bibinfo {author} {\bibfnamefont {X.}~\bibnamefont
  {Liu}}, \bibinfo {author} {\bibfnamefont {Y.}~\bibnamefont {Yang}}, \bibinfo
  {author} {\bibfnamefont {T.}~\bibnamefont {Hu}}, \bibinfo {author}
  {\bibfnamefont {G.}~\bibnamefont {Zhao}}, \bibinfo {author} {\bibfnamefont
  {C.}~\bibnamefont {Chen}},\ and\ \bibinfo {author} {\bibfnamefont
  {W.}~\bibnamefont {Ren}},\ }\bibfield  {title} {\bibinfo {title} {{Vertical
  ferroelectric switching by in-plane sliding of two-dimensional bilayer
  WTe$_2$}},\ }\href {https://doi.org/10.1039/C9NR05404A} {\bibfield  {journal}
  {\bibinfo  {journal} {Nanoscale}\ }\textbf {\bibinfo {volume} {11}},\
  \bibinfo {pages} {18575} (\bibinfo {year} {2019})}\BibitemShut {NoStop}%
\bibitem [{\citenamefont {Moaied~M.}(2015)}]{plasmonics1}%
  \BibitemOpen
  \bibfield  {author} {\bibinfo {author} {\bibfnamefont {O.~K.}\ \bibnamefont
  {Moaied~M.}, \bibfnamefont {Yajadda~M.M.A.}},\ }\bibfield  {title} {\bibinfo
  {title} {Quantum effects of nonlocal plasmons in epsilon-near-zero properties
  of a thin gold film slab},\ }\href
  {https://doi.org/https://doi.org/10.1007/s11468-015-9951-0} {\bibfield
  {journal} {\bibinfo  {journal} {Plasmonics}\ }\textbf {\bibinfo {volume}
  {10}},\ \bibinfo {pages} {1615} (\bibinfo {year} {2015})}\BibitemShut
  {NoStop}%
\bibitem [{\citenamefont {Raza}\ \emph {et~al.}(2013)\citenamefont {Raza},
  \citenamefont {Christensen}, \citenamefont {Wubs}, \citenamefont
  {Bozhevolnyi},\ and\ \citenamefont {Mortensen}}]{plasmonics2}%
  \BibitemOpen
  \bibfield  {author} {\bibinfo {author} {\bibfnamefont {S.}~\bibnamefont
  {Raza}}, \bibinfo {author} {\bibfnamefont {T.}~\bibnamefont {Christensen}},
  \bibinfo {author} {\bibfnamefont {M.}~\bibnamefont {Wubs}}, \bibinfo {author}
  {\bibfnamefont {S.~I.}\ \bibnamefont {Bozhevolnyi}},\ and\ \bibinfo {author}
  {\bibfnamefont {N.~A.}\ \bibnamefont {Mortensen}},\ }\bibfield  {title}
  {\bibinfo {title} {Nonlocal response in thin-film waveguides: Loss versus
  nonlocality and breaking of complementarity},\ }\href
  {https://doi.org/10.1103/PhysRevB.88.115401} {\bibfield  {journal} {\bibinfo
  {journal} {Phys. Rev. B}\ }\textbf {\bibinfo {volume} {88}},\ \bibinfo
  {pages} {115401} (\bibinfo {year} {2013})}\BibitemShut {NoStop}%
\bibitem [{\citenamefont {Chalabi}\ \emph {et~al.}(2014)\citenamefont
  {Chalabi}, \citenamefont {Schoen},\ and\ \citenamefont
  {Brongersma}}]{hotelectron}%
  \BibitemOpen
  \bibfield  {author} {\bibinfo {author} {\bibfnamefont {H.}~\bibnamefont
  {Chalabi}}, \bibinfo {author} {\bibfnamefont {D.}~\bibnamefont {Schoen}},\
  and\ \bibinfo {author} {\bibfnamefont {M.~L.}\ \bibnamefont {Brongersma}},\
  }\bibfield  {title} {\bibinfo {title} {Hot-electron photodetection with a
  plasmonic nanostripe antenna},\ }\href {https://doi.org/10.1021/nl4044373}
  {\bibfield  {journal} {\bibinfo  {journal} {Nano Letters}\ }\textbf {\bibinfo
  {volume} {14}},\ \bibinfo {pages} {1374} (\bibinfo {year} {2014})},\ \bibinfo
  {note} {pMID: 24502677},\ \Eprint
  {https://arxiv.org/abs/https://doi.org/10.1021/nl4044373}
  {https://doi.org/10.1021/nl4044373} \BibitemShut {NoStop}%
\bibitem [{\citenamefont {Hou}\ \emph {et~al.}(2017)\citenamefont {Hou},
  \citenamefont {Shen}, \citenamefont {Shi}, \citenamefont {Kapadia},\ and\
  \citenamefont {Cronin}}]{photocat}%
  \BibitemOpen
  \bibfield  {author} {\bibinfo {author} {\bibfnamefont {B.}~\bibnamefont
  {Hou}}, \bibinfo {author} {\bibfnamefont {L.}~\bibnamefont {Shen}}, \bibinfo
  {author} {\bibfnamefont {H.}~\bibnamefont {Shi}}, \bibinfo {author}
  {\bibfnamefont {R.}~\bibnamefont {Kapadia}},\ and\ \bibinfo {author}
  {\bibfnamefont {S.~B.}\ \bibnamefont {Cronin}},\ }\bibfield  {title}
  {\bibinfo {title} {Hot electron-driven photocatalytic water splitting},\
  }\href {https://doi.org/10.1039/C6CP07542H} {\bibfield  {journal} {\bibinfo
  {journal} {Phys. Chem. Chem. Phys.}\ }\textbf {\bibinfo {volume} {19}},\
  \bibinfo {pages} {2877} (\bibinfo {year} {2017})}\BibitemShut {NoStop}%
\bibitem [{\citenamefont {Juan}\ \emph {et~al.}(2009)\citenamefont {Juan},
  \citenamefont {Gordon}, \citenamefont {Pang}, \citenamefont {Eftekhari},\
  and\ \citenamefont {Quidant}}]{Juan2009}%
  \BibitemOpen
  \bibfield  {author} {\bibinfo {author} {\bibfnamefont {M.~L.}\ \bibnamefont
  {Juan}}, \bibinfo {author} {\bibfnamefont {R.}~\bibnamefont {Gordon}},
  \bibinfo {author} {\bibfnamefont {Y.}~\bibnamefont {Pang}}, \bibinfo {author}
  {\bibfnamefont {F.}~\bibnamefont {Eftekhari}},\ and\ \bibinfo {author}
  {\bibfnamefont {R.}~\bibnamefont {Quidant}},\ }\bibfield  {title} {\bibinfo
  {title} {Self-induced back-action optical trapping of dielectric
  nanoparticles},\ }\href {https://doi.org/doi.org/10.1038/nphys1422}
  {\bibfield  {journal} {\bibinfo  {journal} {Nature Physics}\ }\textbf
  {\bibinfo {volume} {5}},\ \bibinfo {pages} {915} (\bibinfo {year}
  {2009})}\BibitemShut {NoStop}%
\bibitem [{\citenamefont {Juan~M.}(2011)}]{Juan:2011}%
  \BibitemOpen
  \bibfield  {author} {\bibinfo {author} {\bibfnamefont {Q.~R.}\ \bibnamefont
  {Juan~M.}, \bibfnamefont {Righini~M.}},\ }\bibfield  {title} {\bibinfo
  {title} {Plasmon nano-optical tweezers},\ }\href
  {https://doi.org/10.1038/nphoton.2011.56} {\bibfield  {journal} {\bibinfo
  {journal} {Nature Photon}\ }\textbf {\bibinfo {volume} {5}},\ \bibinfo
  {pages} {349} (\bibinfo {year} {2011})}\BibitemShut {NoStop}%
\bibitem [{\citenamefont {Ndukaife}\ \emph {et~al.}(2016)\citenamefont
  {Ndukaife}, \citenamefont {Kildishev}, \citenamefont {Nnanna}, \citenamefont
  {Shalaev}, \citenamefont {Wereley},\ and\ \citenamefont
  {Boltasseva}}]{Ndukaife:2016}%
  \BibitemOpen
  \bibfield  {author} {\bibinfo {author} {\bibfnamefont {J.~C.}\ \bibnamefont
  {Ndukaife}}, \bibinfo {author} {\bibfnamefont {A.~V.}\ \bibnamefont
  {Kildishev}}, \bibinfo {author} {\bibfnamefont {A.~G.~A.}\ \bibnamefont
  {Nnanna}}, \bibinfo {author} {\bibfnamefont {V.~M.}\ \bibnamefont {Shalaev}},
  \bibinfo {author} {\bibfnamefont {S.~T.}\ \bibnamefont {Wereley}},\ and\
  \bibinfo {author} {\bibfnamefont {A.}~\bibnamefont {Boltasseva}},\ }\bibfield
   {title} {\bibinfo {title} {Long-range and rapid transport of individual
  nano-objects by a hybrid electrothermoplasmonic nanotweezer},\ }\href
  {https://doi.org/doi.org/10.1038/nnano.2015.248} {\bibfield  {journal}
  {\bibinfo  {journal} {Nature Nanotech}\ }\textbf {\bibinfo {volume} {11}},\
  \bibinfo {pages} {53} (\bibinfo {year} {2016})}\BibitemShut {NoStop}%
\bibitem [{\citenamefont {Basov}\ \emph {et~al.}(2017)\citenamefont {Basov},
  \citenamefont {Averitt},\ and\ \citenamefont {Hsieh}}]{Basov2017}%
  \BibitemOpen
  \bibfield  {author} {\bibinfo {author} {\bibfnamefont {D.~N.}\ \bibnamefont
  {Basov}}, \bibinfo {author} {\bibfnamefont {R.~D.}\ \bibnamefont {Averitt}},\
  and\ \bibinfo {author} {\bibfnamefont {D.}~\bibnamefont {Hsieh}},\ }\bibfield
   {title} {\bibinfo {title} {Towards properties on demand in quantum
  materials},\ }\href {https://doi.org/10.1038/nmat5017} {\bibfield  {journal}
  {\bibinfo  {journal} {Nature Materials}\ }\textbf {\bibinfo {volume} {16}},\
  \bibinfo {pages} {1077} (\bibinfo {year} {2017})}\BibitemShut {NoStop}%
\bibitem [{\citenamefont {Wang}\ \emph {et~al.}(2012)\citenamefont {Wang},
  \citenamefont {Liu}, \citenamefont {Burton}, \citenamefont {Jaswal},\ and\
  \citenamefont {Tsymbal}}]{Wang/Tsymbal:2012PRL}%
  \BibitemOpen
  \bibfield  {author} {\bibinfo {author} {\bibfnamefont {Y.}~\bibnamefont
  {Wang}}, \bibinfo {author} {\bibfnamefont {X.}~\bibnamefont {Liu}}, \bibinfo
  {author} {\bibfnamefont {J.~D.}\ \bibnamefont {Burton}}, \bibinfo {author}
  {\bibfnamefont {S.~S.}\ \bibnamefont {Jaswal}},\ and\ \bibinfo {author}
  {\bibfnamefont {E.~Y.}\ \bibnamefont {Tsymbal}},\ }\bibfield  {title}
  {\bibinfo {title} {Ferroelectric instability under screened {Coulomb}
  interactions},\ }\bibfield  {journal} {\bibinfo  {journal} {Physical Review
  Letters}\ }\textbf {\bibinfo {volume} {109}},\ \href
  {https://doi.org/10.1103/physrevlett.109.247601}
  {10.1103/physrevlett.109.247601} (\bibinfo {year} {2012})\BibitemShut
  {NoStop}%
\bibitem [{\citenamefont {Michel}\ \emph {et~al.}(2021)\citenamefont {Michel},
  \citenamefont {Esswein},\ and\ \citenamefont {Spaldin}}]{Michel2021}%
  \BibitemOpen
  \bibfield  {author} {\bibinfo {author} {\bibfnamefont {V.~F.}\ \bibnamefont
  {Michel}}, \bibinfo {author} {\bibfnamefont {T.}~\bibnamefont {Esswein}},\
  and\ \bibinfo {author} {\bibfnamefont {N.~A.}\ \bibnamefont {Spaldin}},\
  }\bibfield  {title} {\bibinfo {title} {Interplay between ferroelectricity and
  metallicity in {BaTiO}$_3$},\ }\bibfield  {journal} {\bibinfo  {journal}
  {Journal of Materials Chemistry C}\ }\href
  {https://doi.org/10.1039/d1tc01868j} {10.1039/d1tc01868j} (\bibinfo {year}
  {2021})\BibitemShut {NoStop}%
\bibitem [{\citenamefont {Serre}\ and\ \citenamefont
  {Ghazali}(1983)}]{Serre1983}%
  \BibitemOpen
  \bibfield  {author} {\bibinfo {author} {\bibfnamefont {J.}~\bibnamefont
  {Serre}}\ and\ \bibinfo {author} {\bibfnamefont {A.}~\bibnamefont
  {Ghazali}},\ }\bibfield  {title} {\bibinfo {title} {From band tailing to
  impurity-band formation and discussion of localization in doped
  semiconductors: A multiple-scattering approach},\ }\href
  {https://doi.org/10.1103/physrevb.28.4704} {\bibfield  {journal} {\bibinfo
  {journal} {Physical Review B}\ }\textbf {\bibinfo {volume} {28}},\ \bibinfo
  {pages} {4704} (\bibinfo {year} {1983})}\BibitemShut {NoStop}%
\bibitem [{\citenamefont {Efros}\ \emph {et~al.}(1979)\citenamefont {Efros},
  \citenamefont {Lien},\ and\ \citenamefont {Shklovskii}}]{Efros1979}%
  \BibitemOpen
  \bibfield  {author} {\bibinfo {author} {\bibfnamefont {A.~L.}\ \bibnamefont
  {Efros}}, \bibinfo {author} {\bibfnamefont {N.~V.}\ \bibnamefont {Lien}},\
  and\ \bibinfo {author} {\bibfnamefont {B.~I.}\ \bibnamefont {Shklovskii}},\
  }\bibfield  {title} {\bibinfo {title} {Impurity band structure in lightly
  doped semiconductors},\ }\href {https://doi.org/10.1088/0022-3719/12/10/018}
  {\bibfield  {journal} {\bibinfo  {journal} {Journal of Physics C: Solid State
  Physics}\ }\textbf {\bibinfo {volume} {12}},\ \bibinfo {pages} {1869}
  (\bibinfo {year} {1979})}\BibitemShut {NoStop}%
\bibitem [{\citenamefont {Raghavan}\ \emph {et~al.}(2016)\citenamefont
  {Raghavan}, \citenamefont {Zhang}, \citenamefont {Shoron},\ and\
  \citenamefont {Stemmer}}]{Raghavan2016}%
  \BibitemOpen
  \bibfield  {author} {\bibinfo {author} {\bibfnamefont {S.}~\bibnamefont
  {Raghavan}}, \bibinfo {author} {\bibfnamefont {J.~Y.}\ \bibnamefont {Zhang}},
  \bibinfo {author} {\bibfnamefont {O.~F.}\ \bibnamefont {Shoron}},\ and\
  \bibinfo {author} {\bibfnamefont {S.}~\bibnamefont {Stemmer}},\ }\bibfield
  {title} {\bibinfo {title} {Probing the metal-insulator transition in
  {BaTiO}$_3$ by electrostatic doping},\ }\bibfield  {journal} {\bibinfo
  {journal} {Physical Review Letters}\ }\textbf {\bibinfo {volume} {117}},\
  \href {https://doi.org/10.1103/physrevlett.117.037602}
  {10.1103/physrevlett.117.037602} (\bibinfo {year} {2016})\BibitemShut
  {NoStop}%
\bibitem [{\citenamefont {Page}\ \emph {et~al.}(2008)\citenamefont {Page},
  \citenamefont {Kolodiazhnyi}, \citenamefont {Proffen}, \citenamefont
  {Cheetham},\ and\ \citenamefont {Seshadri}}]{Page2008}%
  \BibitemOpen
  \bibfield  {author} {\bibinfo {author} {\bibfnamefont {K.}~\bibnamefont
  {Page}}, \bibinfo {author} {\bibfnamefont {T.}~\bibnamefont {Kolodiazhnyi}},
  \bibinfo {author} {\bibfnamefont {T.}~\bibnamefont {Proffen}}, \bibinfo
  {author} {\bibfnamefont {A.~K.}\ \bibnamefont {Cheetham}},\ and\ \bibinfo
  {author} {\bibfnamefont {R.}~\bibnamefont {Seshadri}},\ }\bibfield  {title}
  {\bibinfo {title} {Local structural origins of the distinct electronic
  properties of {Nb-substituted SrTiO$_3$ and BaTiO$_3$}},\ }\bibfield
  {journal} {\bibinfo  {journal} {Physical Review Letters}\ }\textbf {\bibinfo
  {volume} {101}},\ \href {https://doi.org/10.1103/physrevlett.101.205502}
  {10.1103/physrevlett.101.205502} (\bibinfo {year} {2008})\BibitemShut
  {NoStop}%
\bibitem [{\citenamefont {Walsh}\ and\ \citenamefont
  {Zunger}(2017)}]{Walsh2017}%
  \BibitemOpen
  \bibfield  {author} {\bibinfo {author} {\bibfnamefont {A.}~\bibnamefont
  {Walsh}}\ and\ \bibinfo {author} {\bibfnamefont {A.}~\bibnamefont {Zunger}},\
  }\bibfield  {title} {\bibinfo {title} {Instilling defect tolerance in new
  compounds},\ }\href {https://doi.org/10.1038/nmat4973} {\bibfield  {journal}
  {\bibinfo  {journal} {Nature Materials}\ }\textbf {\bibinfo {volume} {16}},\
  \bibinfo {pages} {964} (\bibinfo {year} {2017})}\BibitemShut {NoStop}%
\bibitem [{\citenamefont {Marcelli}\ \emph {et~al.}(2017)\citenamefont
  {Marcelli}, \citenamefont {Coreno}, \citenamefont {Stredansky}, \citenamefont
  {Xu}, \citenamefont {Zou}, \citenamefont {Fan}, \citenamefont {Chu},
  \citenamefont {Wei}, \citenamefont {Cossaro}, \citenamefont {Ricci},
  \citenamefont {Bianconi},\ and\ \citenamefont {D'Elia}}]{Marcelli2017}%
  \BibitemOpen
  \bibfield  {author} {\bibinfo {author} {\bibfnamefont {A.}~\bibnamefont
  {Marcelli}}, \bibinfo {author} {\bibfnamefont {M.}~\bibnamefont {Coreno}},
  \bibinfo {author} {\bibfnamefont {M.}~\bibnamefont {Stredansky}}, \bibinfo
  {author} {\bibfnamefont {W.}~\bibnamefont {Xu}}, \bibinfo {author}
  {\bibfnamefont {C.}~\bibnamefont {Zou}}, \bibinfo {author} {\bibfnamefont
  {L.}~\bibnamefont {Fan}}, \bibinfo {author} {\bibfnamefont {W.}~\bibnamefont
  {Chu}}, \bibinfo {author} {\bibfnamefont {S.}~\bibnamefont {Wei}}, \bibinfo
  {author} {\bibfnamefont {A.}~\bibnamefont {Cossaro}}, \bibinfo {author}
  {\bibfnamefont {A.}~\bibnamefont {Ricci}}, \bibinfo {author} {\bibfnamefont
  {A.}~\bibnamefont {Bianconi}},\ and\ \bibinfo {author} {\bibfnamefont
  {A.}~\bibnamefont {D'Elia}},\ }\bibfield  {title} {\bibinfo {title}
  {Nanoscale phase separation and lattice complexity in {VO}$_2$: The
  metal{\textendash}insulator transition investigated by {XANES} via auger
  electron yield at the vanadium {L}$_{23}$-edge and resonant photoemission},\
  }\href {https://doi.org/10.3390/condmat2040038} {\bibfield  {journal}
  {\bibinfo  {journal} {Condensed Matter}\ }\textbf {\bibinfo {volume} {2}},\
  \bibinfo {pages} {38} (\bibinfo {year} {2017})}\BibitemShut {NoStop}%
\bibitem [{\citenamefont {Salmani-Rezaie}\ \emph
  {et~al.}(2020{\natexlab{b}})\citenamefont {Salmani-Rezaie}, \citenamefont
  {Ahadi},\ and\ \citenamefont {Stemmer}}]{SalmaniRezaie2020}%
  \BibitemOpen
  \bibfield  {author} {\bibinfo {author} {\bibfnamefont {S.}~\bibnamefont
  {Salmani-Rezaie}}, \bibinfo {author} {\bibfnamefont {K.}~\bibnamefont
  {Ahadi}},\ and\ \bibinfo {author} {\bibfnamefont {S.}~\bibnamefont
  {Stemmer}},\ }\bibfield  {title} {\bibinfo {title} {Polar nanodomains in a
  ferroelectric superconductor},\ }\href
  {https://doi.org/10.1021/acs.nanolett.0c02285} {\bibfield  {journal}
  {\bibinfo  {journal} {Nano Letters}\ }\textbf {\bibinfo {volume} {20}},\
  \bibinfo {pages} {6542} (\bibinfo {year} {2020}{\natexlab{b}})}\BibitemShut
  {NoStop}%
\end{thebibliography}%

\end{document}